\documentclass[10pt]{article}
\usepackage{tgpagella} 

\usepackage{moresize}
\usepackage{tabularx}
\usepackage{tablefootnote}
\usepackage[dvips]{graphics}
\usepackage[utf8]{inputenc}
\usepackage{amsmath}
\usepackage{amsfonts}
\usepackage{amssymb,gensymb}
\usepackage{graphicx}
\usepackage{fancyhdr}
\usepackage{multirow}
\usepackage{wrapfig, framed}
\usepackage{algorithm}
\usepackage{algorithmic}
\usepackage{epsfig,endnotes,color,url,paralist, multirow,float}

\usepackage{subfig}
\usepackage[export]{adjustbox}

\usepackage[left=2cm,right=2cm,top=2cm,bottom=2cm]{geometry}
\usepackage[inline,shortlabels]{enumitem}
\usepackage{mathtools}
\usepackage{mathrsfs}
\usepackage[table]{xcolor}

\definecolor{lightgray}{gray}{0.9}
\definecolor{LightCyan}{rgb}{0.88,1,1}

\usepackage{lipsum}
\usepackage{comment}
\usepackage{soul}
\usepackage{tikz}
\usepackage{xcolor}
\usepackage[format=plain,
            labelfont={bf,it},
            textfont=it]{caption}
\usepackage{xspace}
\usepackage{hyperref}
\usepackage{booktabs}
\newtheorem{definition}{Definition}
\usepackage{cite}
\usepackage{url}

\usepackage[compact]{titlesec}

\newcommand{\add}{ADD\xspace}
\newcommand{\mdd}{MDD\xspace}
\newcommand{\pwue}{pWUE\xspace}

\newcommand{\ouralg}{\texttt{WCI}\xspace}

\makeatletter
\newcommand\mysize{\@setfontsize\notsotiny\@vipt\@viipt}
\makeatother

\begin{document}

\newpage
\pagenumbering{arabic}
\pagestyle{plain}
\setcounter{page}{1}
\setcounter{section}{0}

\begin{center}{{\bf \Large
Small Bottle, Big Pipe: Quantifying and Addressing the Impact of Data Centers on Public Water Systems}}
\end{center}

\vspace{0.5cm}

\begin{table}[!h]
\centering
\begin{tabular}{m{0.2\textwidth}m{0.2\textwidth}m{0.2\textwidth} m{0.2\textwidth}}
\centering Yuelin Han\\ \emph{UC Riverside}
& \centering Pengfei Li\\ \emph{RIT}
& \centering Adam Wierman\\ \emph{Caltech}
& \centering
Shaolei Ren\tablefootnote{Corresponding authors: Adam Wierman (adamw@caltech.edu) and Shaolei Ren (shaolei@ucr.edu)}\\ \emph{UC Riverside}
\end{tabular}
\end{table}

\begin{center}
    \textbf{Abstract}
    \end{center}
Water is a critical resource for data centers and an efficient means of cooling. However, meeting the growing water demand of data centers requires substantial peak water withdrawals, which many  communities in the United States do not have the available capacity to supply, especially during the hottest days of the year.
This largely overlooked peak water capacity constraint is emerging as a bottleneck for data centers and can force operators to rely on waterless, less efficient dry cooling, thereby increasing electricity demand and further stressing the power grid during summer peaks.
 In this paper, we focus on the direct water withdrawal of U.S. data centers for cooling and examine their impacts on public water systems. Based on public sources including
government records and water utility data,
our analysis indicates that, if the 2024 water use intensity persists, U.S. data centers could collectively require 697--1,451~million gallons per day (MGD) of new water capacity through 2030, comparable to New York City's average daily supply of approximately 1,000~MGD. Under an optimistic scenario with an industry-wide compound annual water use intensity reduction by 10\%, the water capacity demand decreases to 227--604~MGD, although the case of high-growth IT loads could still require enough capacity to hypothetically supply about half of New York City's daily water demand for most of the year, excluding peak hot days.
The total valuation of the new water capacity need is on the order of \$10~billion, reaching up to \$58~billion in the high-growth case.
These impacts are highly concentrated on individual communities hosting data centers, placing additional stress on already capacity-constrained public water systems without substantial infrastructure upgrades.
Finally, we provide recommendations to address the growing water capacity demand of U.S. data centers, including reporting peak water use, developing corporate-community partnerships to benefit community well-being,
adopting a Water Capacity Neutral approach (colloquially ``Pipe Neutral'') to allow host communities to retain limited water capacity resources, and implementing coordinated water-power planning to responsibly
leverage water  for peak power reduction and opportunistically
utilize surplus power to mitigate peak water impacts on public water systems.

\vspace{0.3cm}

\section{Introduction}\label{sec:introduction}

The advancement of artificial intelligence (AI) and computing technologies holds the promise of reshaping industries, accelerating scientific discovery, and offering powerful tools to address pressing societal challenges, from  controlling plasma for nuclear fusion to predicting extreme river-related events for watershed resilience \cite{AI_Benefits_Google_DavidPatterson_CACM_2025_10.1145/3746132}. 
Yet this rapid progress is being powered by an unprecedented buildout of warehouse-scale data centers filled with power-hungry servers \cite{DoE_DataCenter_EnergyReport_US_2024}.
As these data center facilities proliferate, 
they may pose significant local infrastructure challenges in the communities where they operate,
 raising public concerns that residents may shoulder disproportionate infrastructure costs and face sustained obligations to address associated financial, environmental, and health risks
\cite{Microsoft_CommunityFirst_AI_Infrastructure_Blog_2026,AI_Health_Public_Concerns_AirPollution_TheLancet_2026_mclellan2026ai}.

A well-recognized challenge of data centers' rapid expansion stems from their increasing electricity usage, with a single AI data center campus capable of drawing hundreds of megawatts, enough to power a mid-sized city. 
In the United States, the surging demand for AI
is projected to drive
the data center electricity use to 6.7 to 12.0\% of the national total in 2028, up from 4.4\% in 2023  \cite{DoE_DataCenter_EnergyReport_US_2024}. 
As a result of the electricity consumption, the climate impacts of data centers' carbon footprint \cite{Water_Carbon_AI_Projection_2030_FengqiYou_Cornell_NatureSustainability_2025_xiao2025environmental,AI_Water_Carbon_Mistral_Large2_Blog_2025,ML_Carbon_Bloom_176B_Sasha_Luccioni_arXiv_2022_luccioni2022estimating,ML_CarbonFoorptint_Google_JeffDean_Journal_2022_9810097,AI_Water_Carbon_Transparancy_Alex_deVriesGao_Pattern_Journal_2025_DEVRIESGAO2026101430} and the public health burden from their associated air pollution \cite{Shaolei_Health_Impact_2030_arXiv_2024,Health_DataCenter_CallforResearch_GaoPeng_Harvard_Journal_2025_TAO2025100157,AI_Health_Public_Concerns_AirPollution_TheLancet_2026_mclellan2026ai} have received heightened attention. 
Additionally, an immediate physical constraint often lies in the  already strained local power infrastructure \cite{DoE_AI_DataCenter_EnergyDemand_Recommendation_2024,DataCenter_LoadGrowth_Flexibility_Rethinking_Duke_Report_2025_norris2025rethinking}.

Alongside electricity, water is not just a basic necessity
for human-being but also ``a critical input supporting the data center and AI revolution,'' as noted in a recent roundtable discussion on data centers convened by the U.S. Environmental Protection Agency (EPA) \cite{EPA_RoundtableDiscussion_Air_Water_Energy_Jan_2026_PressRelease}.
Indeed, even setting aside off-site water use for electricity generation and chip manufacturing (Section~\ref{sec:datacenter_type}),
on-site evaporative cooling for data centers is widely recognized as a power-efficient approach~\cite{Equinix_WaterEnergy_Tradeoff_split_20_80_Blog_2025,Amazon_Water_400m_25_35per_PeakPowerReduction_AI_Cloud_Louisiana_PressRelease_2026}
and even ``the most efficient means of cooling in many places''~\cite{Google_SustainabilityReport_2025}.
For example, compared to waterless dry cooling, evaporative cooling for a large technology company's new data center campuses in Louisiana
``reduces electricity demand by 25--35\% at the same time when the grid experiences peak summer loads and regional power demand is at its highest''~\cite{Amazon_Water_400m_25_35per_PeakPowerReduction_AI_Cloud_Louisiana_PressRelease_2026}.
Notably, multiple leading technology companies have signed agreements with local water utilities to secure substantial water  allocation for cooling their future data center campuses across various states \cite{Water_Lebanon_LEAP_DataCenter_Meta_Final_2025,Meta_Lebanon_Water_AI_NoWaterMajority_100million_2026,Meta_Water_Lousiana_DataCenter_News_Feb_2026,Amazon_Water_400m_25_35per_PeakPowerReduction_AI_Cloud_Louisiana_PressRelease_2026,Google_Water_8MGD_Vrginia_WaterUtilityAgreement_2026}.

However, despite the significant benefits of water and the often relatively modest \emph{annual} total, 
data center water use is highly concentrated during the hottest days of the year, resulting in substantial 
peak daily water withdrawals. 
For example, peak daily water demand for a large state-of-the-art data center 
using evaporative cooling during the summer can often exceed 1~million gallons per day (MGD) and, for some planned facilities, may even reach as high as 8~MGD~\cite{Google_Water_8MGD_Vrginia_WaterUtilityAgreement_2026,Water_Lebanon_LEAP_DataCenter_Meta_Final_2025,Meta_Lebanon_Water_AI_NoWaterMajority_100million_2026}.

There is no ``national reservoir'' that data centers can tap into. Instead, data centers predominantly rely on treated water---primarily potable---supplied by local public water systems.\footnote{As detailed in Section~\ref{sec:us_public_water}, we use the term ``\emph{public water system}'' to refer broadly to community water systems and related water infrastructure \cite{EPA_PublicWaterSystem_website}, including treated reclaimed water and wastewater treatment systems.}
Unfortunately, many public water systems in the United States are aging, fragmented, and/or financially constrained~\cite{EPA_Water_InfrastructureResilience_website},
thus lacking sufficient available capacity to meet the additional peak water demands of large data centers~\cite{Water_DataCenter_FAQ_NoSurplus_SRBC_2025}.
For example, when interviewed about a new data center requesting
6 million gallons per day (MGD) of water capacity---an amount that could potentially exhaust the available surplus supply in Newton County, Georgia---a representative of the county's water authority stated that ``we just don't have the water''~\cite{AI_Water_Georgia_CountyDailyUse_News_NYTimes_2025}.
Without substantial upgrades, even large water utilities, such as Loudoun Water that serves 
over 300,000 residents and
many tens of data centers in Northern Virginia \cite{Water_LoudounWater_Future90MGD_Current40MGD_Website}, could be unable to meet the total water capacity needs if all data centers were to employ evaporative cooling during the summer (Appendix~\ref{appendix:water_loudoun_hypothetical}).

The high capital expenditure required to expand water capacity to meet the additional demand from data centers---as reflected by three recent water infrastructure upgrades for large technology companies costing a total of nearly \$1 billion ~\cite{Amazon_Water_400m_25_35per_PeakPowerReduction_AI_Cloud_Louisiana_PressRelease_2026,Google_Water_8MGD_300million_30x_BottlingCompany_Virginia_News_2025,Water_Lebanon_LEAP_DataCenter_Meta_Final_2025}---is frequently beyond the capability of many public water systems (Table~\ref{table:surplus_capacity_epa}), particularly small and medium-sized systems which are often under-resourced and together constitute nearly 99\% of community water infrastructure in the United States~\cite{EPA_WaterDrinking_625billion_2021_7th_Report_Congress_2023}.
Like power grids, public water systems are designed to safely and reliably meet the maximum demand across all pressure zones, with margins to handle extreme conditions such as prolonged heatwaves and droughts. However, fulfilling this mission is becoming increasingly challenging due to longstanding underinvestment and limited technical resources.
According to the EPA surveys on drinking water and wastewater systems \cite{EPA_WaterDrinking_625billion_2021_7th_Report_Congress_2023,EPA_WaterWaste_CleanNeeds_630billion_2022_Report_Congress_2024} and a recent third-party analysis \cite{Water_ValueCompaign_3point4_trillion_Report_2025}, 
U.S. public water and wastewater systems now face a total infrastructure funding need of \$1.3–3.4 trillion over the next 20 years to improve service reliability, safety, and resilience. Moreover, as highlighted in the EPA's report to Congress in 2024  \cite{EPA_WaterAffordability_NeedsAssessment_Report_Congress_2024}, water affordability has become a significant challenge nationwide, with an estimate of 12.1 to 19.2 million U.S. households lacking affordable access to water services. These factors can further constrain the ability of public water systems to expand capacity for new industry-scale demands.

To address water capacity shortages without increasing the burden on local ratepayers, state laws often require large water users to make ``pro rata'' fair contributions to fund infrastructure upgrades and system expansion \cite{Google_Water_8MGD_2MGDinitial_CountyWebsite_2026,Google_Water_8MGD_Vrginia_New_WaterSource_Agreement_2026}, sometimes in combination with corporate-community partnerships. For example, a large technology company has recently committed up to \$400 million to invest in local public water infrastructure to support its new Louisiana data centers hosting AI and cloud services \cite{Amazon_Water_400m_25_35per_PeakPowerReduction_AI_Cloud_Louisiana_PressRelease_2026}. 
In another case, a water capacity request of 1.2~MGD from a data center developer
(first reportedly for cooling \cite{OpenAI_Water_1point2_MGD_Peak_News_2025} and later allegedly for fire suppression \cite{OpenAI_Water_1point2_MGD_Peak_forFire_ResponseToPublicRecords_WI_Dept_NaturalResources_Jan_2026})
triggered major water infrastructure upgrades costing the developer over \$100 million \cite{OpenAI_Community_175million_Investment_VantageDC_News_2025}. Without these upgrades, the host town in Wisconsin, which had less than 2~MGD of remaining water capacity according to the water utility's 2024 annual report~\cite{Water_PortWashington_WaterUtility_4MGD_LessThan2MGD_Available_Website}, would likely see operational safety margin significantly reduced for its water system.

Even when financial resources are available, however, public water infrastructure projects can take multiple years to complete and may face further delays due to additional constraints, including engineering water rights allocation and multi-agency approvals~\cite{Water_DataCenter_FAQ_NoSurplus_SRBC_2025,Water_WaterRights_ConsumptiveConstraintMitigation_SRBC_website,Water_Google_TheDalles_ReserviorCapacity_Forest_OregonPublicBroadcasting_2026}. 
In a recent instance, an Indiana water project, partly financed by the state, to add 25~MGD of new water supply capacity and expand wastewater treatment capacity by 15~MGD for an economic development district is expected to take approximately six years to complete and cost more than \$1~billion, including \$560~million for new water supply~\cite{Water_Lebanon_LEAP_WaterOverview_560million_2025}, \$225~million for upgrades to existing infrastructure, and \$350~million for wastewater treatment expansion~\cite{Water_Lebanon_LEAP_DataCenter_Meta_Final_2025}.
As a result, a leading technology company' data center located in that district to host AI and core products may not be able to operate at full capacity until its allocated water capacity of 8~MGD becomes available in 2031~\cite{Water_Lebanon_LEAP_DataCenter_Meta_Final_2025,Meta_Lebanon_Water_AI_NoWaterMajority_100million_2026}.

In some cases, the availability of local and regional water sources can become a key and complex bottleneck. For example, expanding the capacity of a reservoir located on federal land for an Oregon town would require federal-level review and approval beyond the town's jurisdiction~\cite{Water_Google_TheDalles_ReserviorCapacity_Forest_OregonPublicBroadcasting_2026}, which can significantly constrain the water supply available to new large users or expansions of existing demand.

In another instance involving a data center project in Virginia, the county record~\cite{Google_Water_8MGD_Vrginia_New_WaterSource_Agreement_2026} indicates that the future expansion of water capacity from 2 MGD to 8 MGD to serve the data center ``will limit the water supply available [to the water authority's service area] and will impair economic and other development opportunities in those localities that are dependent on water supply.'' The county further recognizes
that ``new water supply sources must be identified, accessed, and
developed'' for the water and wastewater system~\cite{Google_Water_8MGD_Vrginia_New_WaterSource_Agreement_2026}.
This requirement effectively accelerates the water authority's long-range water source planning by decades~\cite{Google_Water_8MGD_New_WaterSource_Accelerated_2060_to_2030_News_2026}.
As a result, future expansion of the data center project may face substantial delays while new water supply sources are developed, a process which itself is a ``lengthy and costly project''~\cite{Google_Water_8MGD_Vrginia_New_WaterSource_Agreement_2026}.

Compared with power grid expansion, public water system upgrades are often equally complex and time-consuming, and in some cases even more so, as they are heavily constrained by source availability, local hydrology, water rights, regulatory oversight, and/or environmental considerations \cite{EPA_WaterInfrastructure_ResilienceFiance_website,Water_DrinkingWaterRegulation_California_Website}. 
For example, it can take roughly 20 years or more to search for a new water source and have it in service \cite{Google_Water_8MGD_2MGDinitial_CountyWebsite_2026}. 
Moreover, many communities rely on reservoirs naturally replenished by rainfall or snowpack, which are highly sensitive to drought conditions and further constrain the development of additional reliable water supplies \cite{Water_DataCenter_FAQ_NoSurplus_SRBC_2025}. In addition, unlike cross-state electricity transmission, transporting water over long distances is considerably more challenging and generally not a viable option.

Although data centers have increasingly constructed on-site (typically gas-fired) power plants for electricity to address grid capacity constraints~\cite{DataCenter_Meta_Gas_Louisiana_CNBC_2025,DataCenter_OpenAI_Stargate_Texas_GasFired_AP_Sep_2025}, sourcing water from on-site groundwater wells or nearby rivers for cooling is rare (Table~\ref{tab:dc_metrics_2024}) and requires comprehensive environmental reviews, constrained site selection, and water rights approvals, thereby entailing significant business risks~\cite{Water_DataCenter_FAQ_NoSurplus_SRBC_2025}.

Consequently, the limited available capacity of public water systems is  emerging as a critical, yet frequently underestimated, bottleneck to the rapid growth of data centers. Insufficient
water capacity often compels data centers to rely on waterless, less-efficient dry cooling, which increases electricity demand and can further strain the power grid during summer peaks~\cite{Microsoft_Water_Zero_MoreEnergy_DataCenter_2024,Equinix_WaterEnergy_Tradeoff_split_20_80_Blog_2025,Amazon_AI_Water_EnergyWater_Tradeoff_Video,Amazon_Water_400m_25_35per_PeakPowerReduction_AI_Cloud_Louisiana_PressRelease_2026}. Moreover, such ``unmet'' water demands are often treated as ``zero'' water demands in public discourse, effectively overlooking or deferring the challenges of limited available capacity in many public water systems rather than proactively addressing them. Indeed, an Energy and Environmental Economics (E3) study released in December~2025 projects that water availability will become an even more crucial factor than the well-recognized power access for data center siting within the next five years~\cite{Water_CommunityOpposition_Rising_E3_WhitePaper_2025}.

\paragraph{Contributions.} In this paper, we focus on data centers' direct water withdrawal (also referred to as water \emph{use} for clarity) in the United States, which hosts the largest number of data centers in the world~\cite{IEA_AI_Energy_Report_2025}, and examine the impacts of data centers on public water systems. 

We first show that data centers use water differently than other users in two key aspects: compared to typical public water users such as residential buildings and offices, data centers have   higher \emph{consumptive} ratios (i.e., the ratio of evaporated or consumed water to total water withdrawn) and exhibit substantially higher daily \emph{peaking} factors (i.e., the ratio of maximum daily use to average daily use). While the high consumptive ratio could affect regional water availability 
\cite{Water_Planning_WashingtonMetro_2050Prediction_DataCenter_PeakingFactor_3point62_Report_2025}, the high peaking factor, often reaching
6 and sometimes exceeding 10 (Appendix~\ref{appendix:peaking_factor}), poses an even greater challenge for public water systems, many of which already struggle to accommodate new water demands during the summer. Thus, even though some data centers use evaporative cooling only for a limited number of days and have low annual water usage, their capacity impacts on public water systems remain significant \cite{Water_WestDesMoines_IA_Website,Google_Water_8MGD_Vrginia_WaterUtilityAgreement_2026,Water_Lebanon_LEAP_DataCenter_Meta_Final_2025}. Importantly, this challenge is largely hidden from both total annual water figures and the commonly-reported annualized water usage effectiveness (WUE, defined as the ratio of total direct/on-site water consumption to total IT energy).

Next, we study U.S. data center water use for the period of 2024-2030, including total annual water withdrawal, water capacity need, and the valuation of water capacity. To substantiate our projections, we examine the 2024 sustainability reports of five leading hyperscale data center operators and seven large colocation providers (Table~\ref{tab:dc_metrics_2024}), collectively representing approximately 86\% and 22\% of U.S.\ hyperscale and colocation IT loads in 2024, respectively. We use this data to establish industry-wide baseline values for water use and consumptive ratios. While data center operators rarely disclose their peak water requirements, we review public reporting, including government records, water utility data, and planning documents, to conservatively set peaking factors for peak water demand and to inform the valuation of water capacity.
In addition to the baseline scenario, we incorporate annual industry-wide WUE reductions 
by assuming compound annual rates of 5\% and 10\% 
under moderate and optimistic scenarios, respectively,
reflecting both efficiency improvement and tightening water capacity constraints in many U.S. public water systems.

Our analysis reveals the following key findings.

\begin{itemize}

\item \textbf{Annual water withdrawal and consumption:} 
In 2030, under the baseline scenario, data centers could withdraw 80--150~billion gallons and consume 60--110~billion gallons annually, representing roughly 0.6--1.1\% of total annual withdrawals and 4--7\% of total annual consumptive use in U.S. public water systems. The higher share of consumption reflects that data centers
evaporate approximately 75\% of the water they withdraw, compared to roughly 12\% for other public water users on average \cite{Water_US_Public_WaterAnalysis_USGS_Report_2025_Medalie2025}. 
These figures could decrease by half under an optimistic scenario with substantial water use reductions.
While aggregated annual totals are relatively modest, a pressing challenge arises from data centers' large and rapidly growing peak daily water demand, which many public water systems are unable to accommodate given longstanding resource constraints and the difficulty of developing new water supplies.

\item  \textbf{New water capacity demand:} Meeting the growing demand of U.S. data centers requires substantial new peak water capacity.
From 2024 to 2030, U.S. data centers are projected to require 697–1,451~MGD of new water capacity under the baseline scenario, comparable in scale to New York City's average daily supply of roughly 1,000~MGD \cite{Water_NYC_1512MGDin1979_1billionNow_2025}. 
Under an optimistic scenario with substantial water reductions, the total water capacity need drops to 227–604~MGD, although high-growth IT loads could still require enough capacity to hypothetically supply about half of New York City's daily water demand for most of the year, except during the hottest peak days.
Further, the water capacity demand is highly concentrated on individual communities hosting data centers, placing additional stress on already capacity-constrained public water systems in the absence of substantial infrastructure upgrades.

\item  \textbf{Valuation of new water capacity demand:}
The total valuation of new water capacity needed to support U.S. data centers through 2030 is generally on the order of \$10~billion. Under the baseline scenario, valuations range from \$7–28~billion in the low-growth case and \$15–58~billion in the high-growth case, while the optimistic scenario reduces these figures to \$2–9~billion and \$6–24~billion, respectively. Although modest relative to the overall investment in the AI industry, such new water capacity expansion can be challenging for financially constrained public water systems \cite{EPA_WaterDrinking_625billion_2021_7th_Report_Congress_2023,EPA_WaterWaste_CleanNeeds_630billion_2022_Report_Congress_2024,EPA_WaterAffordability_NeedsAssessment_Report_Congress_2024}, highlighting the critical importance of corporate–community partnerships to expand infrastructure without raising local water rates \cite{Microsoft_CommunityFirst_AI_Infrastructure_Blog_2026,Amazon_Water_400m_25_35per_PeakPowerReduction_AI_Cloud_Louisiana_PressRelease_2026}.

\end{itemize}

Finally, to address the growing water demand of U.S. data centers, water capacity resources must be planned proactively through multi-stakeholder collaboration rather than treated as an afterthought. Our key recommendations include: 
reporting peak water use to inform public water systems and decision-makers; 
developing corporate-community partnerships to
aligning
corporate responsibility with community well-being;
adopting a Water Capacity Neutral approach, or colloquially ``{Pipe Neutral},'' that allows host communities to retain limited available water capacity for future development; 
and implementing coordinated water-power planning to
responsibly use water as a ``water battery'' for peak power reduction during peak hours, while opportunistically utilizing surplus power to alleviate peak water stress on public water systems during heatwaves. 
Together, these strategies can enhance operational efficiency, protect local water systems, and ensure long-term community resilience while supporting data center growth and technological advancement.

\vspace{.15in}

\noindent \emph{Disclaimer: The goal of this study is to quantify and assess the potential impact of data centers on public water systems in the United States. This study is not intended to advocate for or against the construction of data centers.
We do not take a position on decisions pertaining to any specific data center, including decisions that may be directly or indirectly related to water use. In addition, our estimates of water capacity valuation are intended to illustrate the scale of new water infrastructure needed to meet data center demand and should not be interpreted as the actual financial obligations of data center operators, which may vary depending on contractual arrangements, accounting practices, local rate structures, and other context specific factors.}

\section{Preliminaries on U.S. Data Centers and Public Water Systems}

This section provides background on data centers and their water use, followed by a brief overview of public water systems in the United States and discussion on the emerging water capacity bottleneck for data centers.

\subsection{Overview of U.S. Data Centers and Water Use}\label{sec:datacenter_type}

We broadly classify data centers into the following three types according
to a recent report from LBNL \cite{DoE_DataCenter_EnergyReport_US_2024}. 

\begin{itemize}

\item \textit{Hyperscaler}: Hyperscaler data centers typically refer to large-scale facilities owned and operated by major technology companies to support massive computing workloads, often spanning multiple buildings or campuses. 

\item \textit{Colocation}: Colocation data centers are operated by third-party providers that lease space, power, and cooling to multiple tenants (or a single large tenant), enabling organizations to outsource the physical infrastructure while retaining full control over their IT hardware.
Some colocation data centers have IT loads comparable to, or even exceeding, those of hyperscale facilities.

\item \textit{Others}: Other types of data centers include
enterprise or in-house data centers, telecommunications data centers, and small server rooms, which are often owned and operated by individual organizations to meet internal computing and networking needs. They are typically smaller and more geographically dispersed than hyperscalers.

\end{itemize}

In the United States, as of 2024, colocation and hyperscaler data centers constitute the dominant total data center energy consumption \cite{DoE_DataCenter_EnergyReport_US_2024}. Specifically, colocation data centers account for approximately 48\%,  hyperscalers represent 36-40\%, and the other data centers make up the remaining small share. 
Fueled by accelerating AI demand, the total U.S. data center energy consumption is expected to grow at an annual rate of 13–27\%, reaching 325–580 TWh in 2028 \cite{DoE_DataCenter_EnergyReport_US_2024}.
Colocation and hyperscaler data centers continue to constitute the two largest segments, with colocation accounting for a slightly higher share.
In contrast, other types of data centers, including enterprise and in-house facilities, are expected to comprise a declining share of total data center energy consumption, falling to approximately 4-6\% by 2028 \cite{DoE_DataCenter_EnergyReport_US_2024}.

\paragraph{Data center water use.}
There are three related concepts in water-related studies: water \emph{withdrawal}, water \emph{consumption}, and water 
\emph{discharge}. 
Water withdrawal refers to freshwater taken from surface or groundwater sources for use, regardless of whether it is later returned, and reflects competition for shared water resources \cite{Water_WaterWithdrawal_OECD}. Water consumption denotes the portion of withdrawn water that is not returned because it is evaporated or otherwise removed, thereby directly affecting downstream water availability \cite{Water_Consumption_Withdrawal_WorldResourceInstitute}. 
Water discharge is defined as the difference between withdrawal and consumption and refers to the portion of withdrawn water that is immediately returned. 

Data centers use water both directly (Scope 1) and indirectly (Scopes 2 and 3)  \cite{Shaolei_Water_AI_Thirsty_CACM}.  We discuss each in the remainder of this section.

\paragraph{Scope 1.} The direct water use of data centers is primarily associated with cooling, although humidification control can also require 
water. For a large data center, humidification demand alone may reach on the order of millions of gallons per year
\cite{Water_ProjectWashington_DoubleUsage_10million_Gallon_Humidification_DataCenter_Delaware_2025}.
Critically, IT equipment, including servers, storage systems, and networking devices, generates heat that must be dissipated to the surrounding environment to ensure reliable operation. The cooling process typically occurs in two stages \cite{Equinix_WaterEnergy_Tradeoff_YouTube_2025,Equinix_LiquidCooling_MythsDebunked_Blog_June_2024}. 

The first stage is \emph{server}-level cooling, which transfers heat from the IT equipment to an intermediate facility-level heat exchanger through either air-based or closed-loop liquid-based cooling, without direct water consumption. Notably, the high power density of GPU-powered AI servers often necessitates liquid cooling 
(e.g., direct-to-chip and immersion cooling) because of its superior heat removal capability \cite{Water_HybridCooling_LiquidCooling_NREL_TechReport_2018_sickinger2018thermosyphon,Equinix_WaterEnergy_Tradeoff_YouTube_2025}, whereas
liquid-cooled AI racks still have approximately  20\% or more loads cooled
by air \cite{Equinix_WaterEnergy_Tradeoff_split_20_80_Blog_2025,PUE_32_45degree_LiquidAir_nVidia_Paper_2022_heydari2022power}.

The second stage is \emph{facility}-level cooling, which rejects heat from the facility to the outside environment and may involve water use depending on the cooling technology employed. Common heat rejection mechanisms include: (1) wet cooling towers that primarily rely on water evaporation;  and (2) air-cooled systems (dry coolers), often supplemented by direct evaporative or adiabatic cooling to reduce the peak power demand during the hottest days of the year if applicable.

There is a fundamental tradeoff between power and water use for facility-level cooling \cite{Equinix_WaterEnergy_Tradeoff_YouTube_2025}, which we discuss in detail in Appendix~\ref{appendix:water_power_tradeoff}. Briefly speaking, 
evaporative cooling uses substantially more on-site water than waterless dry cooling,  ``but it also consumes much less energy'' as noted by a large colocation provider \cite{Equinix_WaterEnergy_Tradeoff_split_20_80_Blog_2025}.
Likewise, a leading technology company acknowledges in its sustainability report that ``water is the most efficient means of cooling in many places'' \cite{Google_SustainabilityReport_2025}.
Notably, multiple large technology companies have signed agreements with local water utilities to secure substantial water capacity (e.g., up to 8~MGD at full buildout) for cooling their newest AI and cloud data center campuses across various states \cite{Water_Lebanon_LEAP_DataCenter_Meta_Final_2025,Meta_Lebanon_Water_AI_NoWaterMajority_100million_2026,Meta_Water_Lousiana_DataCenter_News_Feb_2026,Amazon_Water_400m_25_35per_PeakPowerReduction_AI_Cloud_Louisiana_PressRelease_2026,Google_Water_8MGD_Vrginia_WaterUtilityAgreement_2026},
with some water capacity expected to be delivered in 2031 due to required infrastructure upgrades \cite{Water_Lebanon_LEAP_DataCenter_Meta_Final_2025}.

\begin{figure}[!t]
    \centering
    \subfloat[On-site WUE]{
    \includegraphics[width=0.315\textwidth,valign=b]{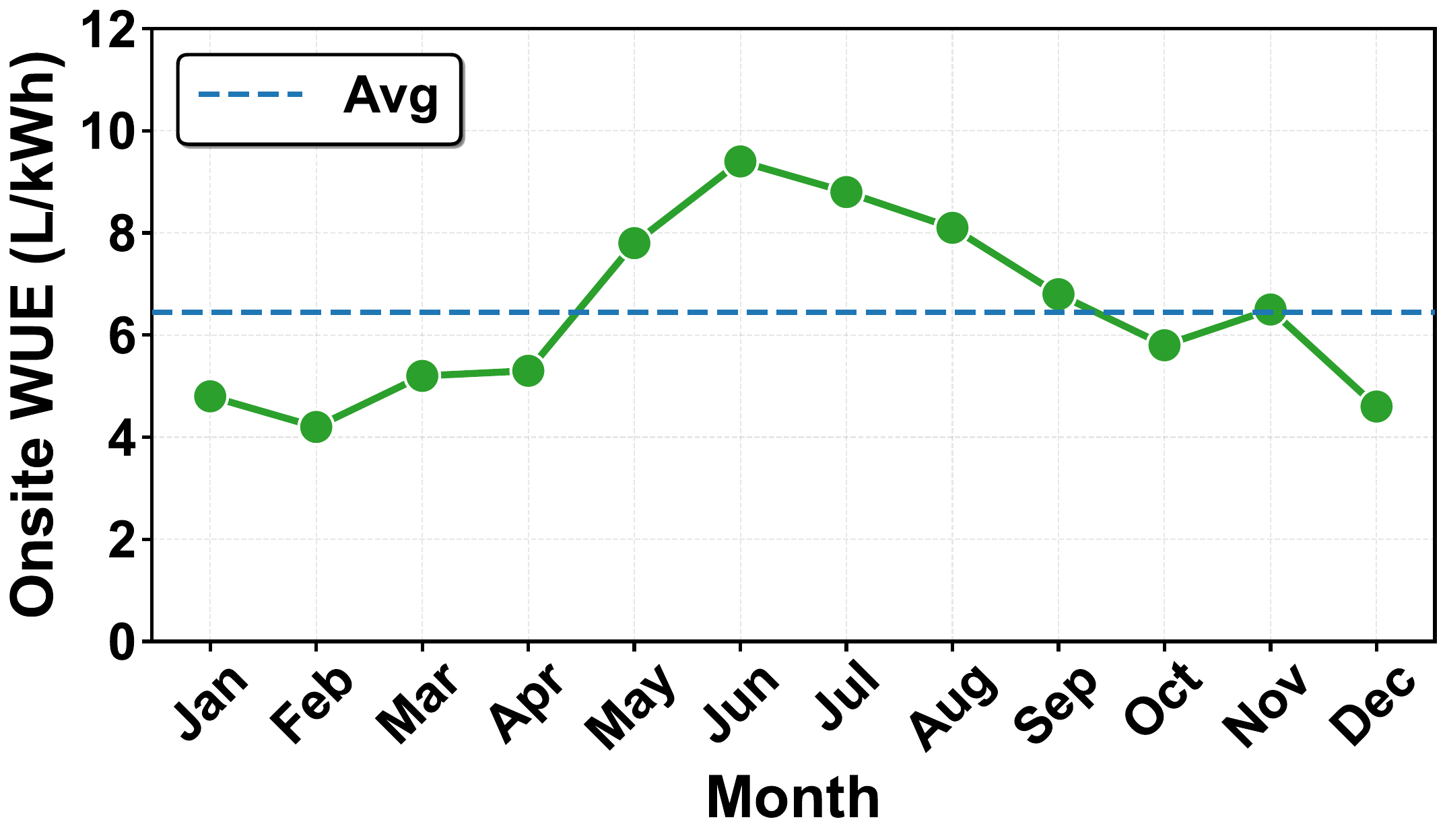}\label{fig:datacenter_wue_arizona} 
    }
    \subfloat[PUE Comparison]{
    \includegraphics[width=0.315\textwidth,valign=b]{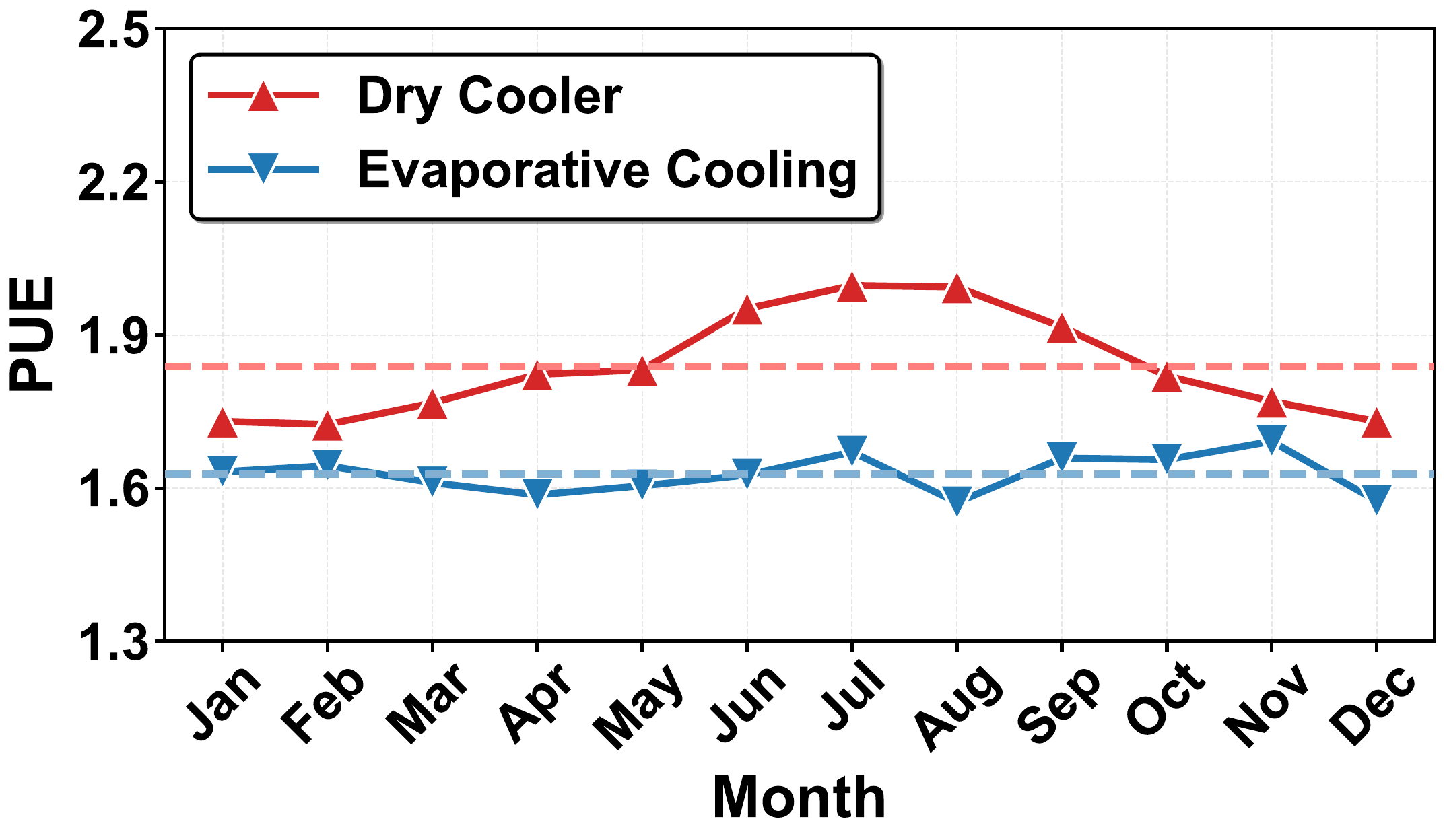}\label{fig:datacenter_pue_consumption_arizona} 
    }
    \subfloat[Consumptive Ratio]{
    \includegraphics[width=0.315\textwidth,valign=b]{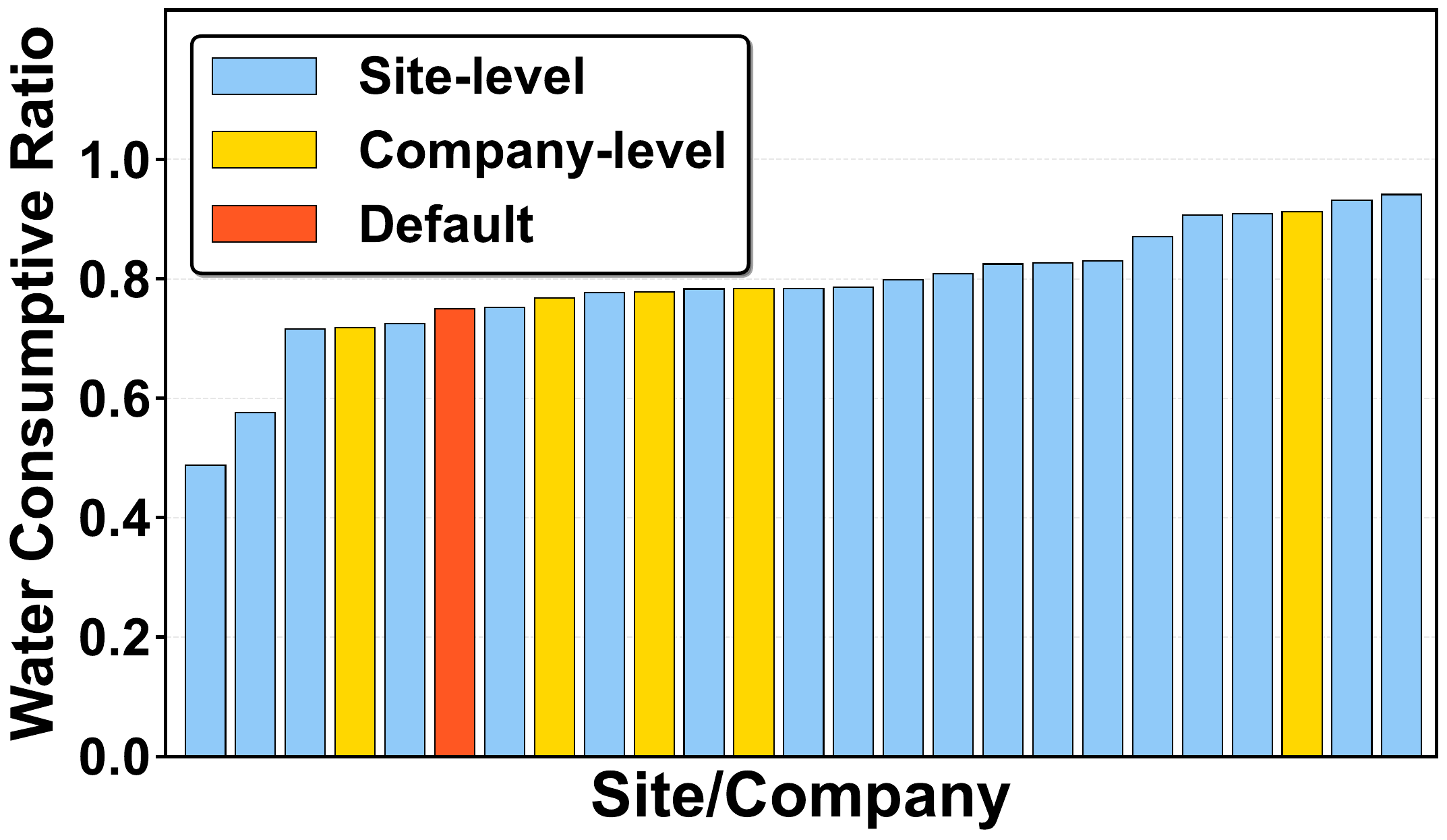}\label{fig:datacenter_comsumptive_ratio} 
    }
    \caption{(a) Monthly WUE of a data center with 34.5 MW IT capacity in Phoenix, Arizona, in 2019 \cite{Water_DataCenterEnergy_Tradeoff_Arizona_Real_Measurement_WUE_Monthly_2022_KARIMI2022106194}; (b) Monthly PUE and WUE comparisons of two data centers in Phoenix, Arizona, in 2019
    (one with 54 MW IT capacity using dry coolers, and another one with 34.5 MW IT capacity using cooling towers) \cite{Water_DataCenterEnergy_Tradeoff_Arizona_Real_Measurement_WUE_Monthly_2022_KARIMI2022106194}; (c) Consumptive ratios of selected companies and data center sites in Table~\ref{tab:dc_metrics_2024}. The site-level consumptive ratios are for a large technology company's U.S. data center fleet, excluding air-cooled sites.}\label{fig:datacenter_pue_wue_consumptiveratio_arizona} 
    \end{figure}

More concretely, industrial disclosures from major technology companies
indicate that water-cooled data centers can use 25 to 35\% less
electricity than air-cooled data centers 
\cite{Amazon_AI_Water_EnergyWater_Tradeoff_Video,Water_AI_Nexus_10_65_Percent_Energy_WhitePaper_2025}, especially during the summer 
``when the grid experiences peak summer loads and regional power demand is at its highest'' \cite{Amazon_Water_400m_25_35per_PeakPowerReduction_AI_Cloud_Louisiana_PressRelease_2026}. The annual energy savings from evaporative cooling are also substantial \cite{Google_Water_10percent_Less_Energy_CliamteCounscious_Blog}.
For example, in Nevada, a data center that primarily relies on non-evaporative cooling exhibits a 6.4 to 10.2\% higher \emph{annual} average power usage effectiveness (PUE, defined as the ratio of total data center energy to IT energy) than another facility operated by the same company that relies heavily on evaporative cooling \cite{Google_SustainabilityReport_2025}. More energy savings from evaporative cooling are also observed in the monthly PUE and WUE measurements from real-world data centers, which are not often publicly available (Figure~\ref{fig:datacenter_pue_consumption_arizona}) \cite{Water_DataCenterEnergy_Tradeoff_Arizona_Real_Measurement_WUE_Monthly_2022_KARIMI2022106194}. These benefits of water evaporative cooling can further translate into reductions in carbon emissions~\cite{Carbon_SustainbleAI_CaroleWu_MLSys_2022_wu2022sustainable}, public health risks~\cite{Shaolei_Health_Impact_2030_arXiv_2024}, and indirect water use associated with electricity generation~\cite{Water_StressWeighted_SustainableComputing_DingYi_2025_10.1145/3757892.3757904,Shaolei_Water_AI_Thirsty_CACM}, as well as, critically, reductions in peak power demand to mitigate the stress on the grid during the summer months. 

In general, for server-level cooling, liquid-cooled high-density servers (e.g., AI servers) can typically operate at higher allowable inlet temperatures than air-cooled servers (e.g., general-purpose compute, storage, or networking equipment). This higher thermal tolerance expands the range of ambient conditions under which facility-level ``free'' air cooling can be utilized, thereby reducing reliance on power-intensive mechanical chillers or evaporative cooling systems for a greater portion of the year. 
Nonetheless, even for a leading technology company's state-of-the-art data center under construction which uses a ``water-efficient closed-loop, liquid-cooled system that recirculates the same water'' \cite{Meta_Lebanon_Water_AI_NoWaterMajority_100million_2026}, a substantial water capacity of 8~MGD is still secured to support facility-level (evaporative) cooling during peak heat conditions \cite{Water_Lebanon_LEAP_DataCenter_Meta_Final_2025}.

A portion of the water withdrawn by data centers is discharged as wastewater, most commonly into municipal wastewater treatment facilities. For example, a large data center currently under construction has been allocated  a water capacity of 8~MGD and a wastewater capacity of 4~MGD \cite{Water_Lebanon_LEAP_DataCenter_Meta_Final_2025},
while the first phase of another planned data center
receives an initial capacity of 2~MGD of water and 0.57~MGD of wastewater \cite{Google_Water_8MGD_Vrginia_WaterUtilityAgreement_2026}.  Depending on the quality of the supplied water and operational settings, the required wastewater capacity generally scales linearly with the water withdrawal. Therefore, we use data center water withdrawal as a primary metric for both water and wastewater capacity requirements.

In addition to the use of water for cooling, some data centers may directly manage fire water systems to comply with safety codes and ensure sufficient flow (typically measured in gallons per minute, GPM) during fire emergencies. For instance, one data center states 
that it has requested a peak capacity of 1.2 million gallons per day for fire suppression \cite{OpenAI_Water_1point2_MGD_Peak_forFire_ResponseToPublicRecords_WI_Dept_NaturalResources_Jan_2026}
in response to inquiries regarding its large peak need,
which is reportedly utilized (for cooling) during the hottest days of the year \cite{OpenAI_Water_1point2_MGD_Peak_News_2025}.  
Domestic water use includes needs such as restrooms and kitchens, and is typically minimal. 
Unless explicitly stated, we exclude both fire and domestic water from our quantitative analysis.

\paragraph{Scopes~2 and~3.} The indirect water use of data centers
is associated with electricity generation (Scope~2) as well as upstream supply chain activities (Scope~3), including rare earth mining and semiconductor manufacturing. While the Scope~2 indirect water use for electricity generation is typically larger than direct water use, it is non-potable in most cases and therefore not directly relevant to public water supply \cite{Shaolei_Water_AI_Thirsty_CACM}. 
The U.S. Energy Information Administration (EIA) routinely reports water withdrawal for the electric power sector, which is one of the largest freshwater-withdrawing sectors in the United States and comparable to agriculture in total withdrawal volumes (42.5\% for thermoelectric power vs. 43\% for crop irrigation relative to the total water withdrawal in the contiguous United States) \cite{Water_WaterWithdrawal_US_Electricity_EIA_2023,Water_US_Public_WaterAnalysis_USGS_Report_2025_Medalie2025}. It has also recently begun publishing monthly data on water withdrawal and consumption for individual power plants \cite{ElectricityDataBrowser_WaterWithdrawalConsumption_EIA_Website}.
Additionally, as water withdrawals can affect immediate water availability and downstream water quality, water permits are issued based on total withdrawal volumes and applicable consumptive use limits \cite{Water_RightsLaw_ConsumptiveIncluded_California_2026,Water_WaterRights_ConsumptiveConstraintMitigation_SRBC_website}. 
Therefore, it is important to quantify Scope~2 water withdrawal and consumption in order to assess the \emph{operational} impacts of data centers on regional watersheds.
A recent study indicates that a large technology company's Scope~3 water  accounts for over 99\% of its total corporate water footprint~\cite{Apple_WaterStrategy_Report_2025}. In practice, however, information on Scope~3 water use is often incomplete and limited~\cite{Water_FinanceInitiative_Benchmark_Website_2025}, and is therefore typically excluded from full-scope analyses of data center water use in most studies.

\vspace{.05in}
\noindent\emph{In this paper, we focus on the impact of data centers on public water systems. Therefore, while Scope 2 and 3 water remains crucial, we consider only direct Scope~1  water withdrawal and refer to it as water use or demand, unless otherwise noted.}

\subsection{U.S. Public Water Systems}\label{sec:us_public_water}

In the United States, a public water system is defined as any infrastructure, publicly or privately owned, that supplies water for human drinking through pipes or other constructed conveyances to at least 15 service connections or to an average of at least 25 people for at least 60 days per year \cite{EPA_PublicWaterSystem_website}. 
Nationwide, there are approximately 150,000 public water systems \cite{EPA_PublicWaterSystem_website} and 
16,000  municipal wastewater
treatment facilities \cite{Water_EPA_Wastewater_16000systems_2024}.

A public water system typically draws water from surface sources such as lakes, rivers, or reservoirs, though some rely on groundwater from aquifers. The water is treated to meet the EPA safety standards under the Safe Drinking Water Act \cite{EPA_Water_SafeDrinkingWaterAct_Summary_Website}, using methods like filtration and disinfection. After treatment, water may be stored in tanks and is then distributed to communities through a network of mains and smaller service lines. In addition to potable water, some water systems also provide reclaimed or recycled water for non‑potable purposes, such as irrigation or industrial use. While wastewater is collected and processed separately subject to the Clean Water Act \cite{EPA_CleanWaterAct_Summary_Website}, drinking water and wastewater systems are increasingly considered as an integrated network, allowing coordinated planning and management of water resources. 

For the convenience of presentation,
in this paper, we simply use the term ``Public Water System'' to refer to the entire system cycle, including sourcing, treatment, distribution, wastewater processing, and water recycling.

\paragraph{Community water systems.} A subset of public water systems that supply water to the same population year-round are known as \emph{community water systems}. There are roughly 50,000 community water systems in the United States, of which approximately 40,000 are small systems each serving no more than 3,300 people, about 9,000 are medium systems each serving between 3,301 and 100,000 people, and only 708 are large systems serving more than 100,000 people \cite{EPA_WaterDrinking_625billion_2021_7th_Report_Congress_2023}.
Nearly all hyperscale and colocation data centers in the United States are supplied by community water systems (mostly from potable water sources), with only a few exceptions drawing water from private groundwater sources (Table~\ref{tab:dc_metrics_2024}) that often require comprehensive environmental reviews and can impose additional availability risks 
\cite{Water_DataCenter_FAQ_NoSurplus_SRBC_2025}.

Compared to roughly 3,000 power utilities nationwide that are interconnected through regional transmission networks \cite{EIA_UtilityNumber_3000_2017_website}, the 50,000 community water systems are significantly more fragmented, with many lacking sufficient financial and technical capacity. According to the most recent EPA surveys conducted prior to the AI boom, total funding needs for U.S. water infrastructure are projected to reach approximately \$1.3 trillion over a 20-year period.\footnote{All monetary values are in U.S. dollars. According to the U.S. Bureau of Labor Statistics \cite{Inflation_Calculator_US_BLS_Website}, \$1 in January 2021 has the same purchasing power as \$1.24 in December 2025. For cost estimates using values between 2021 and 2025, inflation adjustments are not applied unless noted.} This includes \$625 billion (in 2021 dollars) for drinking water systems between 2021 and 2040 \cite{EPA_WaterDrinking_625billion_2021_7th_Report_Congress_2023} and \$630 billion (in 2022 dollars) for wastewater treatment and stormwater control between 2022 and 2042 \cite{EPA_WaterWaste_CleanNeeds_630billion_2022_Report_Congress_2024}. More recent third-party analyses that incorporate updated stormwater requirements and PFAS compliance costs estimate the total funding needs at \$3.4 trillion (in 2025 dollars) over the period from 2025 to 2044 \cite{Water_ValueCompaign_3point4_trillion_Report_2025}.

As highlighted in the EPA's 2024 report to Congress \cite{EPA_WaterAffordability_NeedsAssessment_Report_Congress_2024}, water affordability (defined as water service costs not exceeding 3.0\% to 4.5\% of household income) has emerged as a significant challenge nationwide, with an estimated 12.1 to 19.2 million U.S. households lacking access to affordable water services. Combined with increasing climate risks \cite{Water_ClimateRiskIndex_Municipal_US_Drinking_Water_CMU_2026_lyle2026climate},
the affordability challenge has further intensified the financial pressures faced by many community water systems.  Mitigating these risks requires
improvements in capacity, planning, and investment.

\paragraph{Capacity expansion.}
Public water systems are designed to safely and reliably meet maximum demand at all times and across all pressure zones, with additional margins and dedicated fire flow capacity to address extreme conditions such as prolonged heatwaves and droughts. Accordingly, drinking water treatment plants, pump stations, storage tanks and reservoirs, transmission mains, distribution networks, and wastewater treatment facilities must all be appropriately sized and carefully planned to accommodate peak scenarios.  As a result, capacity expansion and upgrades to water infrastructure often require multiple years to complete, particularly for complex projects such as treatment plant expansions. These projects typically involve environmental review, water rights permitting and allocation, multi-agency approvals, funding authorization, detailed engineering design, and construction, all of which can substantially extend implementation timelines \cite{Water_DataCenter_FAQ_NoSurplus_SRBC_2025,Water_WaterRights_ConsumptiveConstraintMitigation_SRBC_website}.
For example, the process of identifying, developing, and bringing a new water source into service can take roughly 20 years or longer~\cite{Google_Water_8MGD_2MGDinitial_CountyWebsite_2026}.
 
Moreover, many communities rely on reservoirs naturally replenished by rainfall or snowpack, which are highly sensitive to drought conditions and further constrain the development of additional reliable water supplies \cite{Water_DataCenter_FAQ_NoSurplus_SRBC_2025}. In addition, unlike cross-state electricity transmission, transporting water over long distances is considerably more challenging and generally not a viable option. Therefore, compared with power grid expansion, public water system upgrades are often equally challenging and time-consuming, and in some cases even more so due to strong local hydrology constraints.

\subsection{Water Capacity Bottleneck for Data Centers}\label{sec:us_public_water}

A large data center that relies on evaporative cooling can consume millions of gallons of water per day during the hottest periods of the year. For example, a planned data center in Virginia is  projected to use up to 8~MGD at full build-out \cite{Google_Water_8MGD_Vrginia_WaterUtilityAgreement_2026}, or reportedly as much as 30 times the town's current largest water user (a beverage bottling company) \cite{Google_Water_8MGD_300million_30x_BottlingCompany_Virginia_News_2025}.

On the other hand, a public water system's short-term ability to accommodate new service connection requests largely depends on its available capacity, 
as decided by the more restrictive of the remaining physical capacity and the uncommitted capacity (i.e., total physical capacity minus already allocated capacity). 
The system's full design capacity is therefore not the primary determinant. 
A public water system that serves, or is already committed to serving, existing users may still deny a new service connection if its available capacity is limited, even when the requested capacity is substantially smaller than that of some existing users.
 For example, a large technology company recently stated that its new data centers in Louisiana will draw water only from the host community's ``verified surplus water'' capacity, ensuring ``no strain on local water supplies'' \cite{Amazon_Water_400m_25_35per_PeakPowerReduction_AI_Cloud_Louisiana_PressRelease_2026}.
More details of capacity planning and management can be found in
Appendix~\ref{appandix:capacity_expansion}.

\begin{table}[!t]
\centering
\caption{U.S. water treatment surplus capacity (in MGD) by system service population category for all water sources and systems based on the EPA's most recent nationwide community water system survey (2006) \cite{EPA_Water_CommunitySurvey_Website}. The average surplus is defined as the design capacity minus the peak daily flow (Table~19 in \cite{EPA_Water_CommunitySurvey_2006_Vol_2_2009_EPA2009CWSSVol2}). *As the median surplus is not reported, we use the median design capacity minus the median peak daily treatment production (Table~9 in \cite{EPA_Water_CommunitySurvey_2006_Report_2011_EPA2011NationalChars},
which does not further disaggregate systems serving ``over 10,000'' people by size).
}\label{table:surplus_capacity_epa}
\resizebox{\textwidth}{!}{%
\tiny
\begin{tabular}{l|c|c|c|c|c|c|c|c|c}
\hline
\textbf{Category} & \textbf{\begin{tabular}[c]{@{}c@{}}100 or \\ Less\end{tabular}} & \textbf{\begin{tabular}[c]{@{}c@{}}101 - \\ 500\end{tabular}} & \textbf{\begin{tabular}[c]{@{}c@{}}501 - \\ 3,300\end{tabular}} & \textbf{\begin{tabular}[c]{@{}c@{}}3,301 - \\ 10,000\end{tabular}} & \textbf{\begin{tabular}[c]{@{}c@{}}10,001 - \\ 50,000\end{tabular}} & \textbf{\begin{tabular}[c]{@{}c@{}}50,001 - \\ 100,000\end{tabular}} & \textbf{\begin{tabular}[c]{@{}c@{}}100,001- \\ 500,000\end{tabular}} & \textbf{\begin{tabular}[c]{@{}c@{}}Over \\ 500,000\end{tabular}} & \textbf{All Sizes} \\ \hline
\textbf{Average Surplus} & 0.08 & 0.14 & 0.34 & 0.68 & 0.84 & 2.48 & 3.86 & 9.41 & 0.72 \\ \hline
\textbf{Median Surplus*} & 0.03 & 0.10 & 0.23 & 0.89 &  \multicolumn{4}{c|}{3.00} & 0.20 \\ \hline
\end{tabular}%
}
\end{table}

Unfortunately, many U.S. public water systems lack sufficient available capacity to accommodate new, large-scale industrial peak demands \cite{Water_DataCenter_FAQ_NoSurplus_SRBC_2025,EPA_WaterDrinking_625billion_2021_7th_Report_Congress_2023}. 
Table~\ref{table:surplus_capacity_epa} summarizes the surplus water treatment capacity of U.S. community water systems based on the EPA's most recent nationwide survey conducted in 2006 \cite{EPA_Water_CommunitySurvey_Website}. The results indicate that the available surplus capacity is limited across all system sizes when compared with the water capacity demand of large data centers. 
 Systems relying on surface water generally have larger total and surplus capacities than those relying on groundwater sources \cite{EPA_Water_CommunitySurvey_2006_Vol_2_2009_EPA2009CWSSVol2}.
Although this survey has not been updated, the current situation is likely similar or potentially more constrained, as reflected by the rapidly growing water infrastructure funding needs (after inflation adjustment) shown in the EPA's 2023 Report to Congress \cite{EPA_WaterDrinking_625billion_2021_7th_Report_Congress_2023}. 
Meanwhile, water capacity expansions are typically implemented incrementally in line with population growth, making it unlikely that the current nationwide statistics of water treatment surplus capacity differs substantially from the values reported in Table~\ref{table:surplus_capacity_epa}. For example, under California's Drinking Water State Revolving Fund (DWSRF) policies \cite{ca_dwsrf_capacity}, funding for capacity expansions is generally limited to 10\% above the existing maximum daily demand. This rule is intended to ensure engineering reliability, maintain service to existing customers, and reduce financial and operational risk.

Moreover, according to California's 2025 report on water resources \cite{California_WaterResources_Report_2025_DWR2025_Bulletin161}, more than 50\% of urban water suppliers in California report zero annual surplus capacity, while an additional 4.5\% report annual shortages that require supplier response actions. Among suppliers reporting monthly data, 20\% project shortages during certain months of the coming year that require shortage response measures. 
Although the data on daily supply and demand is not available, the fraction of suppliers experiencing shortages of daily capacity for some days of the year is likely substantially higher than 20\%. This suggests that accommodating new large water users, such as data centers that rely on evaporative cooling during the summer, can be challenging for many urban water suppliers under existing capacity constraints.

Table~\ref{table:surplus_capacity_epa} reports only the treatment plant capacity, while other critical components such as distribution pipes, wastewater processing, storage, and water source availability can also be bottlenecks. In addition, after accounting for operational safety margins, reliability headroom, fire suppression requirements, usage categories, and distribution losses, the practically \emph{allocable} surplus capacity becomes even smaller. The allocable capacity is the most relevant factor for connecting large water users, even when the system's total capacity is large.
For example, in Norwalk, Iowa, a city with approximately 15,000 residents, only 0.65~MGD of water capacity is designated as industrial reserve for new large water users as of March~2026, even though other existing water uses may be more substantial \cite{DataCenter_Water_point4_MGD_point65_MGD_Industrial_Norwalk_Iowa_Website_2026}.

Consequently, depending on local climate conditions and cooling system design, supporting a 100~MW IT load using evaporative cooling---which is modest compared with emerging gigawatt-scale AI facilities and generally requires a peak water capacity allocation of approximately 0.5--2.5~MGD (Appendix~\ref{appendix:capacity_need_100MW_IT_load})---can frequently exceed the available surplus capacity of public water systems. This estimate does not include deployments in hot climates, where elevated ambient temperatures can substantially increase water demand, potentially pushing the required peak withdrawal capacity beyond 5~MGD. 
In fact, many data center projects have already required substantial upgrades to local water infrastructure, even when their  water capacity demand is as low as approximately 0.1~MGD 
\cite{Water_Microsoft_point64MGD_Expansion_Leesburg_FinalOfficial_2024,Water_Leesburg_point11MGD_5million_StackDataCenter_Approval_LoudounNow_News_November_2024,Water_RateIncrease_7point3_otherwise23_27_withoutGoogle_TheDalles_Website_2025,Google_Water_8MGD_Vrginia_WaterUtilityAgreement_2026,Water_Lebanon_LEAP_DataCenter_Meta_Final_2025,Amazon_Water_400m_25_35per_PeakPowerReduction_AI_Cloud_Louisiana_PressRelease_2026,OpenAI_Community_175million_Investment_VantageDC_News_2025}.

Importantly, this constraint is not limited to small or medium systems. For example, without substantial capacity expansion, even a large water utility such as Loudoun Water, 
which serves more than 300,000 residents and many data centers in Northern Virginia \cite{Water_LoudounWater_Future90MGD_Current40MGD_Website}, 
could face difficulty accommodating the total water capacity 
requests of data centers within its service territory 
if all the data centers were to rely on evaporative cooling 
(Appendix~\ref{appendix:water_loudoun_hypothetical}).

To balance grid stress and water use, some data centers have increasingly adopted hybrid cooling strategies, such as applying evaporative cooling to a portion of IT loads and/or supplementing dry coolers with direct evaporative assistance only during summer periods. Nonetheless, even if only 10\% of IT loads are cooled with direct evaporative assistance during the summer in a cold climate, the growing scale of AI data centers can still lead to substantial peak water demand. For example, when fully built, an AI-dedicated data center is planned for approximately 0.7~MGD of peak water demand, which can be substantial for the available surplus capacity of a small or medium public water system in the host community in Wisconsin \cite{AI_Water_Microsoft_PeakingFactor_30_MountPleasant_Wisconsin_News_WPR_2025}. 
In another case of a host town with less than 2 MGD available remaining physical capacity (according to the water utility's 2024 annual report) \cite{Water_PortWashington_WaterUtility_4MGD_LessThan2MGD_Available_Website}, a peak water capacity request of 1.2~MGD reportedly triggers a major water infrastructure upgrade exceeding \$100 million \cite{OpenAI_Community_175million_Investment_VantageDC_News_2025}.
More recently, a leading technology company's state-of-the-art data center under construction employs a ``closed-loop, liquid-cooled system that recirculates the same water'' \cite{Meta_Lebanon_Water_AI_NoWaterMajority_100million_2026} and relies on water evaporation for facility-level cooling only during the hottest days of the year. However, the data center may not be able to operate at full capacity until its allocated water capacity of 8~MGD becomes available, as anticipated in 2031~\cite{Water_Lebanon_LEAP_DataCenter_Meta_Final_2025,Meta_Lebanon_Water_AI_NoWaterMajority_100million_2026}.
In another instance involving a new data center project in Virginia, the county record~\cite{Google_Water_8MGD_Vrginia_New_WaterSource_Agreement_2026} indicates that the future expansion of water capacity from 2 MGD to 8 MGD to serve the data center ``will limit the water supply available [to the water authority's service area] and will impair economic and other development opportunities in those localities that are dependent on water supply.''
As a result, future expansion of the data center project may face substantial delays while new water supply sources are developed, a process which itself is a ``lengthy and costly project''~\cite{Google_Water_8MGD_Vrginia_New_WaterSource_Agreement_2026}.

Therefore, choosing between air-cooled and evaporative 
systems for facility-level cooling to manage the tradeoff between water and electricity use may no longer be solely a design decision.
Instead, the availability of public water capacity is increasingly emerging as a binding, yet often underestimated, constraint and projected to surpass power access as a limiting factor for data center siting within the next five years~\cite{Water_CommunityOpposition_Rising_E3_WhitePaper_2025}.

\section{Understanding the Characteristics of Data Center Water Use}

This section characterizes data center water usage, emphasizing its highly variable and spiky nature, which can pose significant challenges for public water systems.
We then examine the impact of data centers on public water systems and the limitations of existing volumetric-based water footprint studies, which often overlook temporal variability and peak water demand that are critical constraints for data centers.

\subsection{How Data Centers Use Water Differently from Other Users}

To understand the characteristics of data center water use, we review public reporting sources,
including data center sustainability reports and water utility data, and identify two key features: a high consumptive ratio and a high peaking factor.

\subsubsection{High consumptive ratio}

The consumptive ratio quantifies the fraction of withdrawn water that is actually consumed, providing a measure of the impact on downstream water availability. We see from Table~\ref{tab:dc_metrics_2024} that data centers have a much higher consumptive ratio than typical industrial or residential users served by public water systems, because a large fraction of water withdrawal is lost to evaporation in facility-level cooling systems. 
Specifically, data centers typically exhibit high consumptive water ratios, ranging from 70 to 90\%, in contrast to urban residential homes that consume only 5--15\% of their total water withdrawals~\cite{Water_ConsumptionToWithdrawal_Ratio_owidwaterusestress_2017}. Manufacturing facilities generally have a consumptive rate of around 20\%~\cite{Apple_WaterStrategy_Report_2025}, and office buildings are often estimated to consume about 10\% of their water use~\cite{Google_SustainabilityReport_2025}. For comparison, the average consumptive rate for public water supply in the contiguous United States is approximately 12\% \cite{Water_US_Public_WaterAnalysis_USGS_Report_2025_Medalie2025}.

\begin{table*}[!t]
\centering
\caption{2024 U.S. data center water and energy use of selected companies. An asterisk ``$^*$'' denotes  values estimated or assigned based on third-party sources. A dagger ``$^\dagger$'' denotes ratios derived from reported water consumption values when withdrawal-based ratios are not explicitly disclosed. A dash ``--'' indicates that the corresponding value is not reported and cannot be plausibly estimated.
Approximately 94\% of Hyperscale-1's water use is potable, although the source is not explicitly identified as ``municipal water.''
The details of data sources
    are available in Appendix~\ref{appendixsubsec:wue_for_our_scenarios}.
}
\label{tab:dc_metrics_2024}
\resizebox{\textwidth}{!}{
\begin{tabular}{lcccccccc}
\toprule
\multirow{2}{*}{\textbf{Company}} & \textbf{IT Energy} & \textbf{Total Energy} & \multirow{2}{*}{\textbf{PUE}} & \textbf{Water Consumption} & \textbf{Water Withdrawal} & \textbf{WUE} & \textbf{Consumptive}  &\textbf{Municipal}\\
& \textbf{(MWh)} & \textbf{(MWh)} & & \textbf{(ML)} & \textbf{(ML)} & \textbf{(L/kWh)} & \textbf{Ratio} & \textbf{Ratio}  \\
\midrule
Hyperscale-1 & 20,417,908 & 22,255,520 & 1.09 & 23,120 & 29,510 &1.13 & 0.78 & -- \\
Hyperscale-2 &  7,764,474 &  9,006,789 & 1.16 & 2,951 & 3,934 & 0.38 & 0.75$^*$ & 99.45\%$^{\dagger}$\\
Hyperscale-3 &  11,828,810 &  12,775,115 & 1.08 & 1,698 & 2,365 & 0.14 & 0.72 & 99.79\%\\
Hyperscale-4 &  1,564,220 &  1,705,000 & 1.09$^*$ & 2,132 & 2,842 & 1.36 & 0.75$^*$ & 93.54\% \\
Hyperscale-5 & 16,004,477$^*$ & 18,245,104 & 1.14 & 1,560 & 2,081 & 0.10 & 0.75$^*$ & -- \\
\midrule
Colocation-1 &  4,533,626 &  5,667,033 & 1.25$^*$ & 2,793 & 3,724 & 0.62 & 0.75$^*$ & 99.79\%$^{\dagger}$  \\
Colocation-2 &  1,741,150 &  2,420,198 & 1.39 & 1,654 & 2,153 & 0.95 & 0.77 & 94.49\% \\
Colocation-3 &  2,584,146 &  3,617,805 & 1.40 & 2,119 & 2,323 & 0.82 & 0.91 & 100.00\%\\
Colocation-4 &  3,520,345 &  5,139,703 & 1.46 & 1,021 & 1,312 & 0.29 & 0.78 & -- \\
Colocation-5 &  1,237,664 &  1,522,327 & 1.23 & 1,584 & 2,112 & 1.28 & 0.75$^*$ & --  \\
Colocation-6 &    598,551 &    826,000 & 1.38 & 431 & 575 & 0.72 & 0.75$^*$ & --  \\
Colocation-7 &    623,706 &    829,529 & 1.33 & 36 & 48 & 0.06 & 0.75$^*$ & -- \\
\midrule
\textbf{Hyperscale} & \textbf{57,579,889} & \textbf{63,987,528} & \textbf{1.11} &\textbf{31,460} & \textbf{40,732} & \textbf{0.55} & \textbf{0.77} & -- \\
\textbf{Colocation} & \textbf{14,839,188} & \textbf{20,022,595} & \textbf{1.35} & \textbf{9,638} & \textbf{12,246} & \textbf{0.65} & \textbf{0.79} & -- \\
\bottomrule 
\end{tabular}
}
\end{table*}

Unlike other industrial processes where water can often be recycled or returned to the system, evaporative cooling converts water into vapor to remove heat, making these losses generally unavoidable.
In fact, some data centers may add chemicals to treat public water on-site, allowing for more cycles of concentration in cooling towers. This reduces blow-down water and further increases the consumptive ratio.
Even with liquid-cooled servers that allow higher server-level temperature setpoints, water evaporation are still preferred for facility-level cooling during summer peak periods in many places \cite{Amazon_Water_400m_25_35per_PeakPowerReduction_AI_Cloud_Louisiana_PressRelease_2026}, keeping consumptive ratios high. 

For the same amount of water withdrawal, a high consumptive ratio can pose additional water  challenges for public water systems, particularly in drought-prone regions with limited supply. 
These challenges are further amplified by the complex water rights and applicable consumptive constraints legally imposed on the host community's public water systems \cite{Water_WaterRights_ConsumptiveConstraintMitigation_SRBC_website}. 
As a result, data centers' higher water consumptive ratio may have a greater impact on water availability, a potential concern discussed in a recent study on the Washington Metropolitan Area water supply \cite{Water_Planning_WashingtonMetro_2050Prediction_DataCenter_PeakingFactor_3point62_Report_2025}.

\subsubsection{High peaking factor}\label{sec:high_peaking}
Similar to power grids, public water systems are engineered to reliably meet maximum demand at all times and across all pressure zones, with additional safety margins \cite{Water_DrinkingWaterRegulation_California_Website}. Accordingly, a user's peak water demand is a critical factor for water infrastructure planning, system resilience, and overall operational reliability.

Nonetheless, most data center operators disclose only annual total water use, often aggregated across an entire corporate fleet, providing limited visibility into facility-level peak demand \cite{Google_SustainabilityReport_2025,Facebook_SustainabilityReport_2025,Microsoft_SustainabilityReport_2024,Water_DrinkingWaterRegulation_California_Website}. To address this limitation, we review publicly available records to characterize data centers' peak water demand. In particular, we focus on the daily peaking factor whenever applicable, defined as the ratio of maximum daily water use to average daily water use. This metric is widely used in water infrastructure planning because it directly informs capacity sizing for treatment, storage, and distribution systems \cite{Water_DrinkingWaterRegulation_California_Website,Water_Planning_TheDallesOregon_2024,Water_Planning_WashingtonMetro_2050Prediction_DataCenter_PeakingFactor_3point62_Report_2025,Water_DrinkingWaterRegulation_California_File_2025}. Other key design parameters, such as peak hourly flow rate and instantaneous flow expressed in gallons per minute, are typically derived from the daily peaking factor using additional multiplicative adjustment factors \cite{Water_Planning_TheDallesOregon_2024,Water_DrinkingWaterRegulation_California_File_2025}.

\paragraph{Evaporative cooling towers.}
We first examine the monthly water usage effectiveness (WUE, defined as the ratio of total water consumption to IT energy use) for a large colocation data center in Phoenix, Arizona, serving a nearly constant IT load in 2019 \cite{Water_DataCenterEnergy_Tradeoff_Arizona_Real_Measurement_WUE_Monthly_2022_KARIMI2022106194}. Cooling towers are a widely-used conventional cooling system, offering stable performance and broad applicability across diverse climates \cite{CoolingTower_SPX}.
The ratio of the peak monthly WUE to the average monthly WUE is approximately 1.46, indicating seasonal variation in water intensity despite relatively stable computing demand.
Assuming a constant consumptive ratio (i.e., the fraction of withdrawn water that is consumed), seasonal variation in WUE can serve as a proxy for variation in withdrawal. In the absence of the facility-specific daily peaking factor, regulatory guidance commonly applies a minimum adjustment factor of 1.5 \cite{Water_DrinkingWaterRegulation_California_File_2025},
yielding an estimated daily peaking factor of at least 2.19.

In a more recent case of The Dalles, Oregon, the water pressure zone (Zone 310) in which a hyperscale data center using cooling towers is located and accounts for the majority of zonal water demand, exhibits a measured peaking factor of 2.21 \cite{Water_Planning_TheDallesOregon_2024,Google_SustainabilityReport_2025,Water_Google_TheDalles_ReserviorCapacity_Forest_OregonPublicBroadcasting_2026}. Consequently, given that non-data center water demands are typically less ``spiky,'' the data center's daily peaking factor may exceed 2.21.

\paragraph{Dry cooling with evaporative assistance.} 
An increasingly adopted configuration is air-based dry cooling supplemented with evaporative assistance during periods of high ambient temperature. 
Under typical server temperature setpoints and local climatic conditions, evaporative assistance may operate for only 5 to 15\% of the year, and possibly up to 40\% in hotter climates \cite{Microsoft_DataCenter_40percentEvaporative_Arizona_2025}.
As a result, water use is highly concentrated in a limited number of hot days, leading to a relatively high peaking factor even when total annual water consumption is substantially lower compared to conventional cooling towers.
As evaporative operation is triggered primarily on the hottest days, the high water demand is temporally aligned with periods of peak thermal stress, further amplifying the maximum hourly and daily water use. Consequently, although annual water use may be substantially reduced, the infrastructure implications associated with peak demand of evaporative assistance systems can remain significant.

\begin{figure}[!t]
    \centering
    \subfloat[Data Center Water Withdrawal]{
    \includegraphics[width=0.315\textwidth,valign=b]{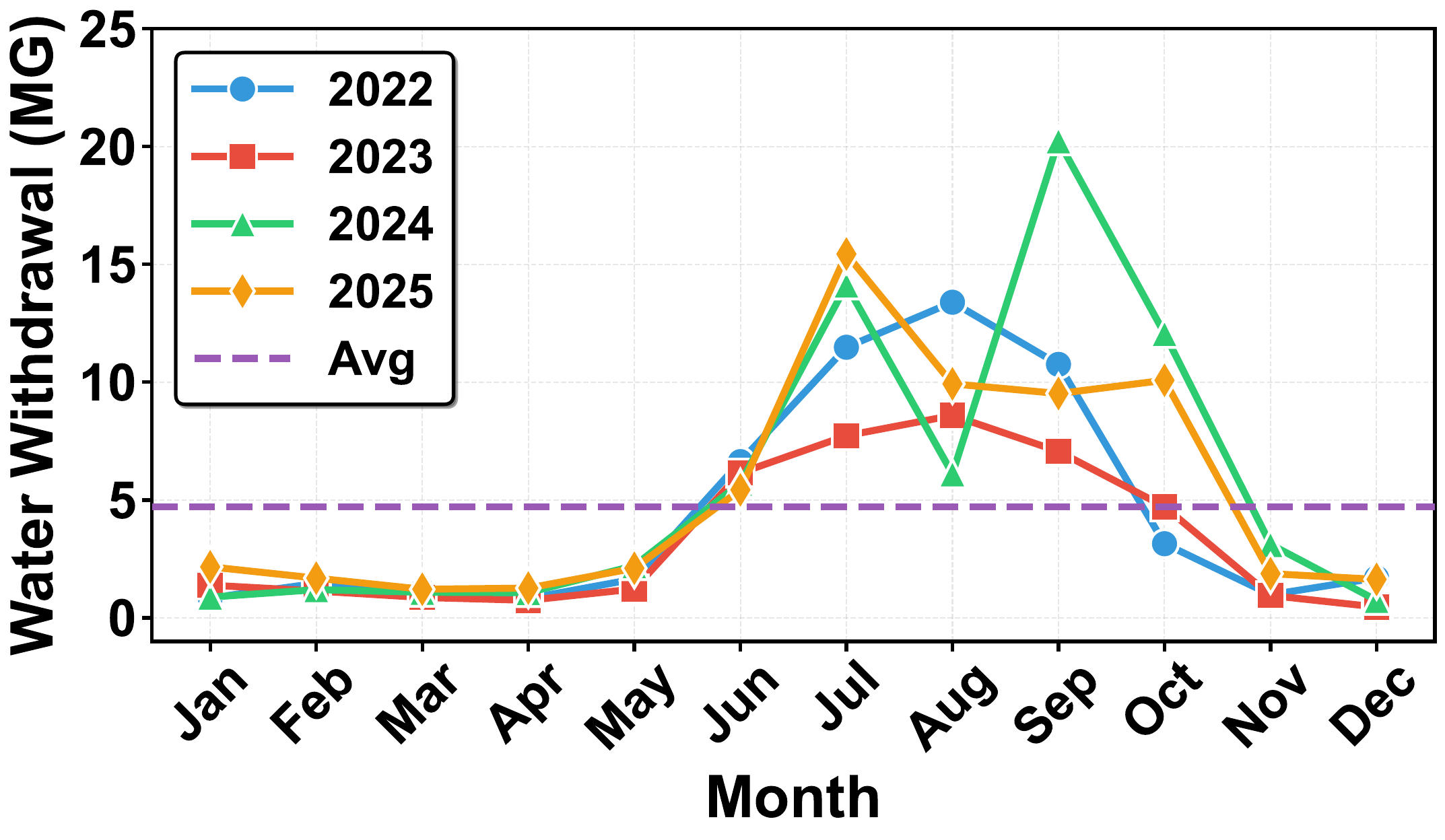}\label{fig:microsoft_monthly_water} 
    }
    \subfloat[Ratio of Data Center Water Withdrawal]{
    \includegraphics[width=0.315\textwidth,valign=b]{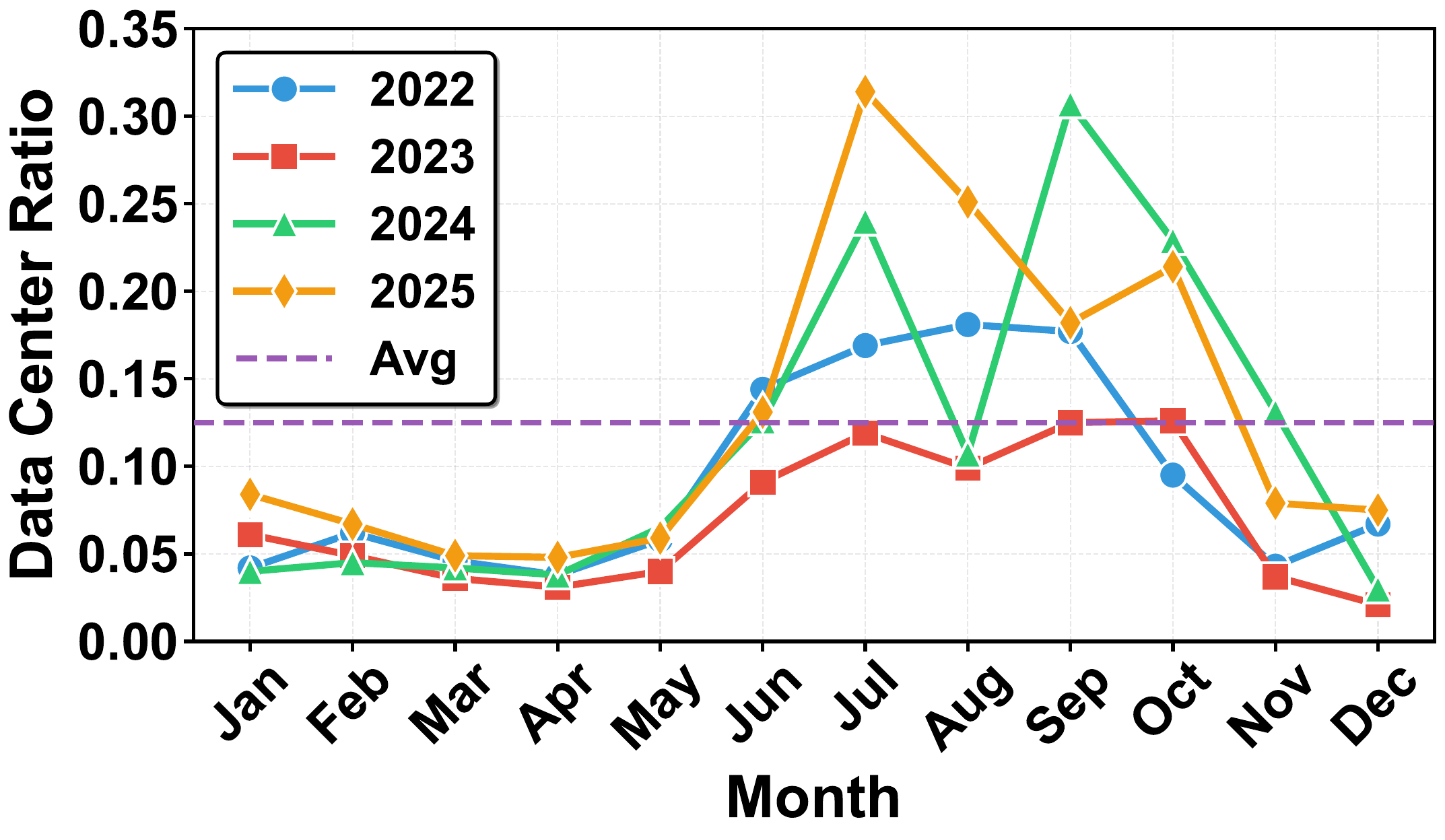}\label{fig:microsoft_monthly_water_ratio} 
    }
    \subfloat[Monthly Peaking Factor]{
    \includegraphics[width=0.315\textwidth,valign=b]{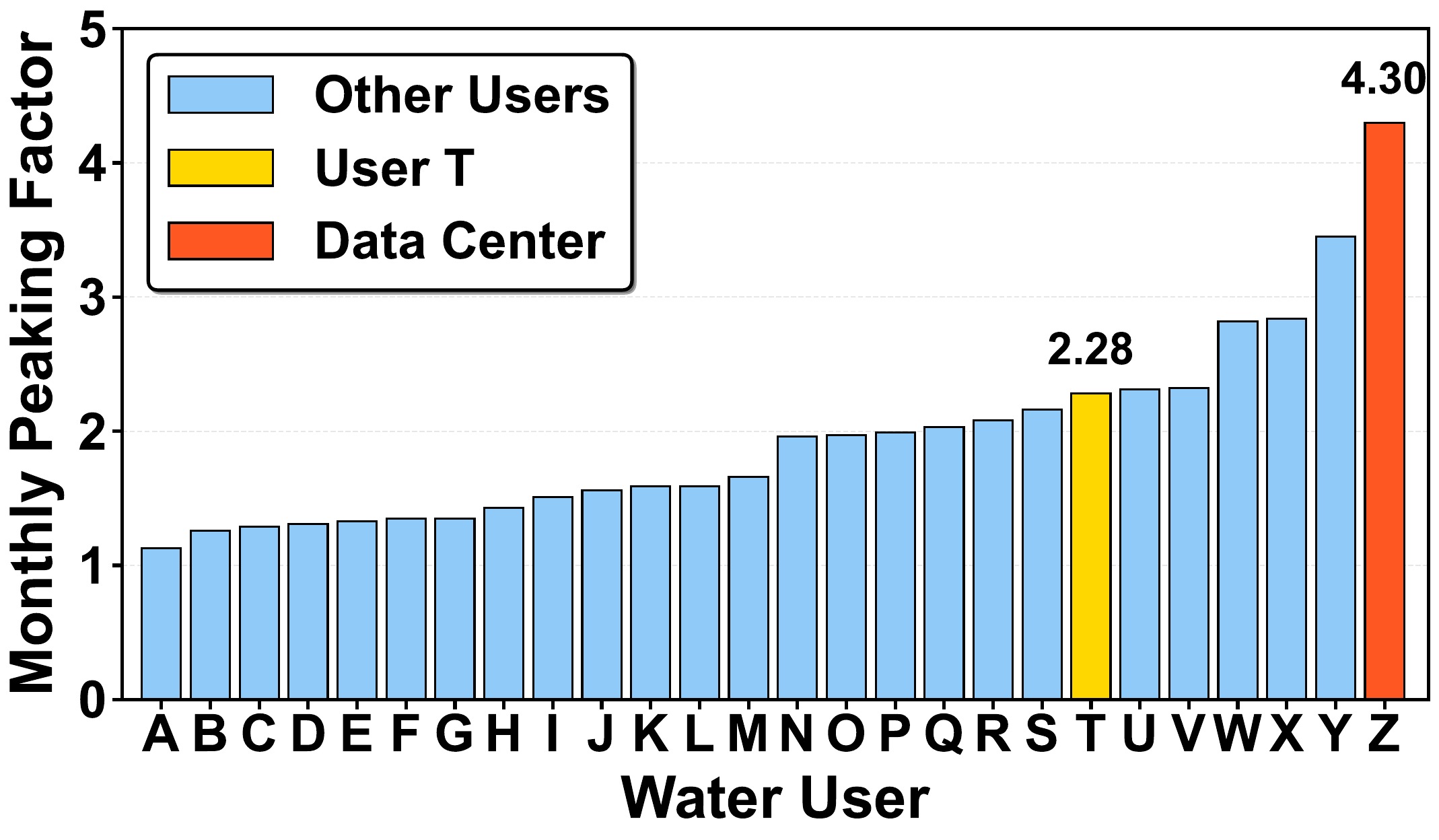}\label{fig:microsoft_monthly_peakingfactor} 
    }
    \caption{Water use data collected from the monthly financial reports available at \cite{Water_WestDesMoines_IA_Website}. (a) Monthly water use of a hyperscale data center in Iowa; (b) The ratio of the data center's water use to the total of 20 largest water users  served by same water works; (c) Monthly peaking factors (i.e., ratio of the maximum monthly water use to the average use between 2022 and 2025) of all the large water users counted by the water works.} \label{fig:west_des_moines_water_datacenter}
\end{figure}

Figure~\ref{fig:microsoft_monthly_peakingfactor} presents the monthly water withdrawal of a hyperscale data center in Iowa, indicating a ratio of peak monthly withdrawal to average monthly withdrawal of 4.30.
It is important to note that the disclosed withdrawal volume includes a small portion of domestic water use, which typically exhibits a peaking factor in the range of 1.5 to 2.5. Consequently, the monthly peak-to-average water withdrawal ratio attributable solely to the data center cooling load is likely higher than 4.30.
Applying a minimum adjustment factor of 1.5 to translate average demand to maximum day demand \cite{Water_DrinkingWaterRegulation_California_File_2025} yields an estimated daily peaking factor of at least 6.45.

We further present in Figure~\ref{fig:microsoft_monthly_water_ratio} the percentage contribution of the data center's monthly water use relative to the total water use of the top 20 largest users served by the same water works. This analysis further highlights the pronounced seasonal variability of data centers that rely on evaporative assistance during the summer. 
Although the data center's water use is relatively low during cooler months, its elevated summer water demand is sufficiently large that it becomes the  largest annual water user in both 2024 and 2025 served by the water works~\cite{Water_WestDesMoines_IA_Website}.

By comparison, we also present in Figure~\ref{fig:microsoft_monthly_peakingfactor} the monthly peaking factors for other large water users reported in the financial disclosures  from 2022 to 2025 \cite{Water_WestDesMoines_IA_Website}. Some water users appear in the reports for fewer than four years; for these users, we compute the ratio of their maximum monthly water use to their average monthly use over the period in which they are reported.  We can see that all other large water users exhibit substantially lower peaking factors than the data center. 
Note that, water user T, a large corporate campus, exhibits an abnormally high water use of approximately 15 million gallons in August 2023, exceeding more than ten times its average monthly use between 2022 and 2025. This spike is likely associated with campus expansion and infrastructure upgrades during that period, as indicated by the city's building permit reports (May to July 2023) \cite{WestDesMoines_IA_City_BuildingPermitReports_Website},  resulting in a one time increase in water use.
It does not exhibit the recurring annual pattern observed for the data center and other users. We therefore exclude this anomalous month when calculating the monthly peaking factor for user T.

\paragraph{Peaking factor for planning.} The total water capacity allocated to a data center is often not publicly disclosed, frequently on the grounds of business confidentiality. In one rare case with disclosed water demand~\cite{AI_Water_Microsoft_PeakingFactor_30_MountPleasant_Wisconsin_News_WPR_2025}, a large AI-dedicated hyperscale data center in Wisconsin requests a total water capacity of 0.7~MGD to cool its non-AI servers, while its average daily water demand is only approximately 23,000 gallons, including relatively stable domestic use but excluding fire protection water. This implies a peaking factor exceeding 30. Due to the cold climate in Wisconsin, it is estimated that up to only 480 hours of cooling per year will require water \cite{AI_Water_Microsoft_PeakingFactor_30_MountPleasant_Wisconsin_News_WPR_2025}, which reduces the total annual water use while increasing the peaking factor. 
In addition, the corresponding peaking factor for wastewater discharge at the same facility is approximately 15, which is substantially higher than that of non-data center users.

In another case regarding a hyperscale data center in Leesburg, Virginia, government records on water allocation and diesel generator capacity indicate that the planned daily peak WUE is estimated at 1.15~L/kWh as of March 2025 (Appendix~\ref{appendix:water_loudoun_hypothetical}), whereas the reported annual WUE for the operator's data center fleet across Northern Virginia is 0.14~L/kWh as of 2023 (which may be lower due to efficiency improvements in 2025)~\cite{Shaolei_Water_AI_Thirsty_CACM}. This comparison suggests that the peaking factor for the Leesburg data center campus is approximately 8 or higher.

Similarly, in an innovation district under development in Indiana, a leading technology company's data center hosting AI and core products is estimated to have a planned daily peaking factor of 6.3 or higher, whereas other users in the district typically exhibit peaking factors below 2.0, except for one hospital at 3.0, based on water allocation agreements or pre-agreements (Appendix~\ref{appendix:water_allocation_leap_indiana}). 
When the full water capacity of 8~MGD is delivered in 2031, the data center's allocation would account for 32\% of the total 25~MGD of new water capacity, making it likely the largest single user 
 in the district based on allocated capacity.

The elevated peaking factor is also evident at an aggregated level when multiple data centers are geographically clustered, such as in Northern Virginia. For example, a recent report analyzing water utility data indicates that the measured daily peaking factor of data centers within the Prince William Water service area reaches 10 in 2024~\cite{Water_Planning_WashingtonMetro_2050Prediction_DataCenter_PeakingFactor_3point62_Report_2025}. However, when a large number of data centers with diverse cooling system designs and operational configurations are aggregated, the overall peaking factor is moderated by multiplexing effects. Nevertheless, the weighted estimate for the actual peaking factor (not for infrastructure planning purposes) in Northern Virginia remains in the range of 3.5 to 3.7 under different scenarios (Table 6-5 of~\cite{Water_Planning_WashingtonMetro_2050Prediction_DataCenter_PeakingFactor_3point62_Report_2025}), which is still substantially higher than the typical \emph{aggregated} peaking factor of 1.5 to 2.5 for non-data center users \cite{Water_Residential_109GPD_Summer_California_LAO_Website}.

\paragraph{Summary.}
The water use of data centers exhibits pronounced seasonal and even daily spikes, resulting in peaking factors estimated to range from 3 to 10, depending on ambient temperature and cooling system configuration. Cooling systems that rely on evaporative assistance during the summer generally exhibit substantially higher peaking factors than conventional evaporative cooling towers. Additionally, in cooler climates and/or when server-level cooling operates at higher temperature setpoints that only need evaporative assistance for a limited number of days, the peaking factor can be even higher. For planning purposes, similar to power capacity redundancy \cite{Water_Planning_WashingtonMetro_2050Prediction_DataCenter_PeakingFactor_3point62_Report_2025,UptimeInstitute_Tier_Certification_Update_2021}, data centers often request more water than their actual demand to hedge against extreme heat events and/or rapid future load increases. Finally, we note that while water storage tanks can buffer short-term fluctuations in hourly demand, they are often less effective for prolonged events such as extended or month-long heatwaves that data centers typically prepare for. As a result, data centers employing evaporative cooling generally reserve substantial daily peak water capacity allocations to hedge against these risks. Nonetheless, as water capacity is increasingly becoming a constrained societal resource (e.g., in some cases, it can take 20 years or longer to identify, develop, and bring new water supply sources into service \cite{Google_Water_8MGD_2MGDinitial_CountyWebsite_2026}), 
the combination of water tanks and other water-aware computing-based solutions (Appendix~\ref{appendix:recommendation_additional}) presents a promising approach to lowering the peak water demand and strengthening public water system resilience.

\subsection{Impacts of Data Centers' Peak Water Demand on Public Water Systems}

Similar to power systems that are designed to provide a reliable electricity supply with stable voltage maintenance, public water systems need to reliably meet maximum demand at all times and across all pressure zones, with additional safety margins and dedicated fire flow capacity to address extreme conditions such as prolonged heatwaves. 
Nonetheless, within the broader context of underfunded infrastructure and growing water affordability concerns \cite{EPA_WaterDrinking_625billion_2021_7th_Report_Congress_2023,EPA_WaterWaste_CleanNeeds_630billion_2022_Report_Congress_2024,EPA_WaterAffordability_NeedsAssessment_Report_Congress_2024}, many community water systems have limited \emph{surplus} capacity available for large data centers, and developing new water sources can require a long lead time \cite{Water_DataCenter_FAQ_NoSurplus_SRBC_2025}. 

\paragraph{Water utility responses to data center peak demands.}
Some water utilities have begun to express concerns regarding 
data centers' peak water demands. For example,  West Des Moines Water Works stated in 2022 that it ``will only consider future data center projects'' if such projects can ``demonstrate and implement technology to significantly reduce peak water usage from the current levels,'' in order to preserve water availability for other users \cite{AI_Water_ChatGPT_Iowa_PeakDemand_News_AP_2023}. 

When asked about the water capacity requested by new data centers that would exceed the county's available supply, a representative of the Newton County, Georgia, water authority stated that ``they [data centers] are taking up the community wealth'' and ``we just don't have the water''~\cite{AI_Water_Georgia_CountyDailyUse_News_NYTimes_2025}. In Newton County, upgrades and capacity expansions needed to meet the new industrial-scale water demands and future growth are reportedly projected to increase local water rates by 33\% over the next two years, compared with a typical annual rise of approximately 2\%, unless additional external funding is secured~\cite{AI_Water_Georgia_CountyDailyUse_News_NYTimes_2025}.

In the United Kingdom, Thames Water, the country's largest water utility, has reportedly discussed options for ``restricting or reducing or objecting to'' data centers' peak water use during the hottest periods of the year \cite{AI_Water_Land_Impact_LincolnInsitute_ThamesWater_UK_Restrict_DataCenter_Policy_2025}.

\paragraph{Corporate-funded infrastructure upgrades.}  In response to concerns such as those above, community-corporate partnerships have increasingly been established by data center projects, paralleling investments in power infrastructure \cite{Microsoft_CommunityFirst_AI_Infrastructure_Blog_2026}. For example, a planned data center in Virginia will receive up to 8~MGD of water capacity at full build-out \cite{Google_Water_8MGD_Vrginia_WaterUtilityAgreement_2026}, which would reportedly be approximately 30 times greater than the town's current largest water user (a beverage bottling facility) and require roughly \$300 million in water infrastructure investments, anticipated to be borne by the data center operator \cite{Google_Water_8MGD_300million_30x_BottlingCompany_Virginia_News_2025}. 
In another case, a data center operator has funded water storage infrastructure to support the host town and enhance its water security \cite{Google_Water_Project_Infrastructure_Security_TheDalles_Oregon_BenTownsend_Blog_October_2025}.
More recently, a major technology company pledged up to \$400 million to upgrade local public water infrastructure to support evaporative cooling at its data center campus, ``[reducing] electricity demand by 25–35\% during periods of peak summer loads when regional power demand is highest'' \cite{Amazon_Water_400m_25_35per_PeakPowerReduction_AI_Cloud_Louisiana_PressRelease_2026}.

In a public water system, even though the total system-wide capacity is available, any component, including drinking water treatment plants, pump stations, storage tanks and reservoirs, transmission mains, distribution networks, and wastewater treatment facilities, can become a bottleneck for meeting the water demand of data centers. For example, 
despite sufficient aggregate system capacity in Leesburg, Virginia, a technology company agreed to pay \$25 million to fund water and wastewater improvements to secure approximately 0.64 MGD of water capacity and
``ensure the cost of serving our facilities [i.e., the data center campus] does not fall on local ratepayers''
\cite{Water_Microsoft_point64MGD_Expansion_Leesburg_FinalOfficial_2024,Microsoft_CommunityFirst_AI_Infrastructure_Blog_2026}.
Similarly, another data center with a relatively modest water capacity need of 0.1 MGD and 
wastewater need of 0.08 MGD requries
utility upgrades, including the construction of a new pump station, 
reportedly costing \$5.4 million paid by the data center
\cite{Water_Leesburg_point11MGD_5million_StackDataCenter_Approval_LoudounNow_News_November_2024}.  
In another case, a small host community with less than 2 MGD available remaining capacity (according to the water utility's 2024 annual report) \cite{Water_PortWashington_WaterUtility_4MGD_LessThan2MGD_Available_Website}, faced a peak water capacity request of 1.2~MGD, which the data center indicates is for fire protection \cite{OpenAI_Water_1point2_MGD_Peak_forFire_ResponseToPublicRecords_WI_Dept_NaturalResources_Jan_2026}, reportedly triggering a major water infrastructure upgrade exceeding \$100 million \cite{OpenAI_Community_175million_Investment_VantageDC_News_2025}.

In a more complex scenario, a city's water reservoir situated on federal lands needs capacity expansion to support future growth, which would necessitate comprehensive environmental review and federal approval. To expedite the process, the city is considering alternative approaches, such as legislative measures to transfer federal land ownership \cite{Water_Google_TheDalles_ReserviorCapacity_Forest_OregonPublicBroadcasting_2026}. This example highlights the non-financial and regulatory constraints that can limit the effective capacity of public water systems.

\paragraph{``Unmet'' water demand.} Importantly, while power capacity shortages for data centers are widely recognized \cite{DoE_DataCenter_EnergyReport_US_2024} and have even been described as a ``crisis'' \cite{AI_Energy_Crisis_Nature_Article_bourzac2024fixing}, the increasingly binding constraint of Scope~1 water availability has received much less attention. Treating water constraints in isolation can undermine the viability of water evaporative cooling, which is recognized as ``the most efficient means of cooling in many places'' \cite{Google_SustainabilityReport_2025}. In the absence of available water capacity, data centers may
have to rely on dry cooling technologies, which can increase electricity demand, even for state-of-the-art AI facilities that operate with relatively high server-level cooling temperature setpoints \cite{Microsoft_Water_Zero_MoreEnergy_DataCenter_2024}. 
For example, in Newton County, Georgia, a recent data center project in 2025 reportedly requested approximately 6~MGD of water capacity \cite{AI_Water_Georgia_CountyDailyUse_News_NYTimes_2025},
which could not be accommodated under the county's existing water infrastructure \cite{Water_ResourcesAnalysis_NewtonGeorgia_December_2024}.
Moreover, although many data centers served by Loudoun Water in Northern Virginia use dry cooling as facilitated by higher server-level cooling temperature setpoints, this also partially reflects Loudoun Water's potential capacity constraints at the time those data centers were connected---if all the data centers had employed water-based evaporative cooling, Loudoun Water's available capacity could have been depleted or even exceeded on the hottest days (Appendix~\ref{appendix:water_loudoun_hypothetical}).

Consequently, the ``\emph{unmet} water demand'' of data centers---the water that would have been used if it were available---is often effectively viewed as ``zero water demand'' in public discourse.
This silent shift may further exacerbate existing power capacity shortages, in addition to contributing to other externalities, including increased public health risks \cite{Shaolei_Health_Impact_2030_arXiv_2024}.

\subsection{Limitations of Total Volumetric Water Footprints}\label{sec:limitations_volumetric}

We conclude this section by emphasizing the contrast between the discussion above and prior work on data centers' overall water footprint.  Existing studies predominantly focus on quantifying and reducing the total water consumption of data centers~\cite{DoE_DataCenter_EnergyReport_US_2024,Water_Carbon_AI_Projection_2030_FengqiYou_Cornell_NatureSustainability_2025_xiao2025environmental,AI_Water_Carbon_Transparancy_Alex_deVriesGao_Pattern_Journal_2025_DEVRIESGAO2026101430,Water_DataCenterFootprint_EnvironmentalResearcHLetters_VT_2021_siddik2021environmental,Shaolei_Water_AI_Thirsty_CACM,Shaolei_Water_SpatioTemporal_GLB_TCC_2018_7420641}, often including indirect water consumption for electricity generation~\cite{IEA_AI_Energy_Report_2025,Water_DataCenterFootprint_EnvironmentalResearcHLetters_VT_2021_siddik2021environmental} and sometimes incorporating regional water stress adjustments~\cite{Water_StressWeighted_SustainableComputing_DingYi_2025_10.1145/3757892.3757904,Water_Wise_Carbon_Water_YankaiJiang_DeveshTiwari_PPoPP_2025_10.1145/3710848.3710891}.
Moreover, amid broader global challenges of water imbalances described as ``water bankruptcy'' \cite{Water_Bankruptcy_GlobalWaterBankruptcy_UnitedNationsUniversity_2026}, 
several technology companies have pledged substantial financial resources to become ``Water Positive,'' aiming to manage their water consumption responsibly and mitigate associated ecosystem impacts through initiatives such as water replenishment and restoration projects \cite{Google_Water_Stewardship_ProjectPortfolio_2025,Meta_Water_Volumetric_Report_2025,Microsoft_Water_2023,Amazon_Water_2023}.
 Nonetheless, these studies and efforts exhibit the following limitations and may lead to unintended consequences.

\paragraph{Water withdrawal.} While the existing studies are valuable for understanding overall impacts of data centers on global and regional long-term water resources and ecosystems, most of them, except for a few cases~\cite{Shaolei_Water_AI_Thirsty_CACM,Water_Planning_WashingtonMetro_2050Prediction_DataCenter_PeakingFactor_3point62_Report_2025}, tend to overlook the effects of water withdrawal, which is critical for short-term water availability, water quality, and allocation~\cite{Water_WaterRights_ConsumptiveConstraintMitigation_SRBC_website,Water_Consumption_Withdrawal_WorldResourceInstitute}. For example, the U.S. Energy Information Administration (EIA) routinely reports water withdrawal for the electric power sector, 
which is one of the largest freshwater-withdrawing sectors in the United States and comparable to agriculture in total withdrawal volumes (42.5\% for thermoelectric power vs. 43\% for crop irrigation relative to the total water withdrawal in the contiguous United States) \cite{Water_WaterWithdrawal_US_Electricity_EIA_2023,Water_US_Public_WaterAnalysis_USGS_Report_2025_Medalie2025}.
It has also recently begun publishing monthly data on water withdrawal and consumption for individual power plants~\cite{ElectricityDataBrowser_WaterWithdrawalConsumption_EIA_Website}. Importantly, public water systems almost always use water withdrawal as the primary metric for planning and allocation, and technology companies include Scope 1 water withdrawals in their annual sustainability disclosures~\cite{Google_SustainabilityReport_2025,Facebook_SustainabilityReport_2025}.

\paragraph{Cross-sector comparisons.} The distinction between municipal (typically potable) water for Scope~1 and non-potable water for Scope~2 is often missing from aggregate water consumption numbers, while Scope~3 water is rarely included due to the lack of reliable data in the public domain.
However, the water footprint reported for other sectors often reflects the full lifecycle \cite{WaterFootprintNetwork_Glossary_Website}. This can unintentionally lead to misinformed cross-sector comparisons.
For example, the full-scope lifecycle water use for certain animal-derived products, such as hamburgers---including rainwater stored in the soil
(i.e., green water) used to grow feed crops for patty production, whereas only less than 2\% of the meat's overall water footprint may come from public water systems (Appendix~\ref{appendix:animal_water}) \cite{Water_Beef_1point1_0point8_Municipal_UNESCO_2010_Mekonnen2010}---is sometimes directly compared with the operational or Scope~1 water use of data centers or AI model inference.  
Similarly, water-intensive golf courses, whose potable water use is increasingly regulated and accounts for only around 10\% or less of total water applied (Appendix~\ref{appendix:golf_course_water}), are sometimes compared against a data center's Scope~1 municipal (potable) water use.
Fundamental inconsistencies in accounting scopes and/or water types render such comparisons uninformative, obscuring the challenges faced by public water systems and misinforming public discourse.

\paragraph{Peak Scope 1 water use.} Most importantly, while existing studies have examined the spatial impacts of data center water use \cite{Shaolei_Water_SpatioTemporal_GLB_TCC_2018_7420641,Water_Carbon_AI_Projection_2030_FengqiYou_Cornell_NatureSustainability_2025_xiao2025environmental,Water_StressWeighted_SustainableComputing_DingYi_2025_10.1145/3757892.3757904}, they do not account for the high peak demand of data centers---particularly daily peak Scope~1 water use---which is a key consideration for public water system management. This omission neglects the higher peaking factor associated with data centers
compared to typical public water users (Figure~\ref{fig:microsoft_monthly_peakingfactor}),
 an increasing challenge given that many U.S. public water systems are aging and under-resourced
~\cite{EPA_Water_InfrastructureResilience_website,Water_DataCenter_FAQ_NoSurplus_SRBC_2025}.

\section{Quantifying and Forecasting Data Centers' Peak Water Demand}\label{sec:quantify_impact}

This section 
presents a quantitative analysis
of U.S. data centers' total peak water demand, with projections through 2030. The results reveal that in the high-growth case and without substantial water usage reduction, U.S. data center expansion could require up to 1,451 million gallons per day (MGD) in new water capacity from 2024 to 2030, with infrastructure valuations reaching as high as \$15-58 billion.
If hypothetically pooled, this capacity would be sufficient to supply New York City for much of the year, except during a limited number of peak demand days.
When significant water reductions are realized at a compound annual rate of 10\%, 
the new water capacity need could decrease to 227–604 MGD, 
with a corresponding valuation range of \$2–9 billion and \$6–24 billion, 
depending on the IT load growth rates.
Importantly, these impacts are highly concentrated on individual communities
hosting data centers, highlighting the need of sustained corporate-community partnerships to address them effectively.

\subsection{Growth Rates, Scenarios, and Data Sources}

We first present a general methodology to quantify data centers' peak water demand. Analogous to using IT power load as the basis for data center design (e.g., cooling systems, battery sizing, and backup generators), we use peak daily water use as the primary metric. This metric is widely employed in water infrastructure management and directly informs capacity sizing for treatment, storage, distribution, and wastewater handling \cite{Water_Planning_WashingtonMetro_2050Prediction_DataCenter_PeakingFactor_3point62_Report_2025,Water_Planning_TheDallesOregon_2024,Water_ResourcesAnalysis_NewtonGeorgia_December_2024}.
While the actual daily peak water demand of a data center can be directly measured for post-hoc reporting and analysis, estimating it in advance is valuable for guiding infrastructure planning. For planning purposes, the peak daily water demand can therefore be calculated as:
\begin{equation}\label{eqn:peak_WaterWithdrawal_Total}
PeakWaterDemand = \beta \cdot \frac{TotalWaterConsumption}{T},
\end{equation}
where $TotalWaterConsumption$ is the total Scope 1 water use planned over $T$ days, and $\beta \in [1, \infty)$ is the estimated or planned peaking factor.

\subsubsection{Growth rates}
We consider three data center growth cases informed by the recent U.S. data center energy report published by researchers at Lawrence Berkeley National Laboratory (LBNL) \cite{DoE_DataCenter_EnergyReport_US_2024}, complemented by industry market analyses \cite{DataCenter_Demand_20Percent_CAGR_2025to2030_JLL_Report_2025,DataCenter_Energy_McKinsey_AI_20_30Growth_US_2030_Website_2025}. 
These scenarios capture a range of plausible medium-term expansion trajectories for U.S. data center capacity under accelerating AI-driven demand.

\begin{itemize}
\item \textbf{Low growth:} The total U.S. data center energy increases
at a compound annual growth rate (CAGR) of 13\% as considered in \cite{DoE_DataCenter_EnergyReport_US_2024}. For
 IT loads of hyperscale and colocation data centers,
 the CAGRs are
  21\% and 23\%, respectively.

\item \textbf{Mid growth:} This represents the algorithmic mean of the low- and high-growth cases for IT loads.

\item \textbf{High growth:} The total U.S. data center energy increases
at a CAGR of 27\% as considered in \cite{DoE_DataCenter_EnergyReport_US_2024}. For
 IT loads of hyperscale and colocation data centers,
 the CAGRs are
  29\% and 32\%, respectively.

\end{itemize}

While the LBNL report projects growth through 2028, we extend the same growth rates through 2030, to align with the study period used in a recent analysis of U.S. data centers' carbon and water footprints \cite{Water_Carbon_AI_Projection_2030_FengqiYou_Cornell_NatureSustainability_2025_xiao2025environmental} and with projected growth rates from other market analyses (e.g., CAGR of approximately 20\% through 2030~\cite{DataCenter_Demand_20Percent_CAGR_2025to2030_JLL_Report_2025} and approximately 22.4\% over 2023--2030 under a medium-growth scenario~\cite{DataCenter_Energy_McKinsey_AI_20_30Growth_US_2030_Website_2025}).

\subsubsection{Scenarios}

Previous studies \cite{DoE_DataCenter_EnergyReport_US_2024,Water_Planning_WashingtonMetro_2050Prediction_DataCenter_PeakingFactor_3point62_Report_2025} consider constant or increasing average water usage effectiveness (WUE, defined as the ratio of total water consumption to total IT energy). In our analysis, we incorporate annual industry-wide WUE reductions 
by assuming compound annual rates of 5\% and 10\% 
under moderate and optimistic scenarios, respectively. 
These adjustment rates reflect continued technological and operational 
improvements in data center water efficiency, 
as well as tightening capacity constraints in many U.S. public water systems, 
which may render evaporative cooling infeasible or restrict its deployment 
in certain regions.
Specifically,
we consider the following three scenarios.
\begin{itemize}
\item \textbf{Baseline:} Industry-wide WUEs for hyperscale and colocation data centers remain constant at 2024 levels through 2030.

\item \textbf{Moderate:} Industry-wide WUEs for hyperscale and colocation data centers decrease at a compound annual rate of 5\% between 2024 and 2030.

\item \textbf{Optimistic:} Industry-wide WUEs for hyperscale and colocation data centers decrease at a compound annual rate of 10\% between 2024 and 2030.

\end{itemize}

While individual companies may achieve greater WUE reductions, the industry-wide adjustments applied in our study are consistent with previous observations and align with common reduction targets reported by data center operators \cite{QTS_SustainabilityReport_2024,Microsoft_SustainabilityReport_2025}.
For example, a major technology company's goal of reducing water intensity by 40\% by 2030 \cite{Microsoft_SustainabilityReport_2025}, relative to its 2020 level, lies between our Moderate and Optimistic scenarios. 
Importantly, under the Moderate and Optimistic scenarios, 
the projected U.S.-wide average WUE in 2030 
(Table~\ref{tab:wue_scenarios}) 
is approximately 0.43~L/kWh and 0.31~L/kWh, respectively. 
These values are lower than the modeled WUE projections reported in 
\cite{DoE_DataCenter_EnergyReport_US_2024} and are consistent with prevailing industry trends,  thereby representing 
plausible forward-looking estimates. 
In addition to the Baseline, Moderate, and Optimistic scenarios, we include two references---Reference (LBNL) and Reference (NS)---which apply WUE values based on the modeled results in \cite{DoE_DataCenter_EnergyReport_US_2024} and \cite{Water_Carbon_AI_Projection_2030_FengqiYou_Cornell_NatureSustainability_2025_xiao2025environmental}, respectively.\footnote{For our analysis, we re-calculate the total water \emph{consumption} for Reference (LBNL) rather than directly using the 2024–2028 values reported in \cite{DoE_DataCenter_EnergyReport_US_2024}. Specifically, we follow the standard annual on-site WUE definition, resulting in lower water consumption estimates than those originally reported. See Table~\ref{tab:annual_water_consumption} in the appendix for details.}

Currently, the majority of data center water use is supplied by public potable sources (e.g., more than 94\% for a large technology company \cite{Google_SustainabilityReport_2025}), despite the adoption of reclaimed or recycled water in some cases, which is often also provided by municipal utilities \cite{Water_Planning_WashingtonMetro_2050Prediction_DataCenter_PeakingFactor_3point62_Report_2025}. In the following analysis, we do not further differentiate between potable and non-potable water. Instead, we discuss the role of non-potable water in Section~\ref{sec:recommendation_discussion} as a potential approach to mitigating the impacts of data centers on public water systems while preserving the peak power reduction benefits of evaporative cooling.

\subsubsection{Data Sources}

We briefly describe the data sources and settings for our analysis, with details available in Appendix~\ref{appendixsec:water_demand_method}. For each IT load growth case (low, mid, and high growth), we quantify the annual water consumption,
withdrawal, and peak water withdrawal  for hyperscale and colocation data centers separately, while excluding potential water use from the smaller and declining segment of other data center types (Section~\ref{sec:datacenter_type}). 

Rather than relying on simulated WUE models \cite{DoE_DataCenter_EnergyReport_US_2024,Water_Carbon_AI_Projection_2030_FengqiYou_Cornell_NatureSustainability_2025_xiao2025environmental} that may exhibit substantial discrepancies from actual water use, we base our water efficiency calculations on industrial disclosures. Specifically, we review the 2024 sustainability reports from five leading hyperscale data center operators and seven large colocation data center providers (Table~\ref{tab:dc_metrics_2024}), which collectively represent about 86\% and 22\% of the hyperscale and colocation IT loads 
(relative to the 2024
 mid values in Table~\ref{tab:it_energy})
in the United States, respectively. Although these industrial reports have varying levels of reporting quality, they are among the most reliable and comprehensive data currently available in the public domain.

Additionally, we review multiple public sources, including government records, water utility data, and planning documents, to conservatively determine peaking factors to estimate the peak water demand. Finally, public reports are used to estimate the capacity valuation of U.S. public water system infrastructure needed to support the new data centers' peak water demand from 2024 to 2030.

While we have collected public data sources to the best of our ability and aimed to cover most plausible scenarios, projection uncertainties inherently remain, especially given the rapid growth of the data center industry and the challenge of obtaining detailed industrial data. Our parameter choices are generally conservative to avoid overestimation. As more comprehensive and preferably audited data in standardized formats become available, refinements to our analysis may be necessary.

\subsection{Results}

We now present the analysis of U.S. data centers' Scope 1 water use from 2024 to 2030, including the annual water withdrawal, consumption, total water capacity need, and valuation of the capacity need.

\subsubsection{Annual water withdrawal}

\begin{figure}[!t]
    \centering
    \subfloat[Low Growth]{
    \includegraphics[width=0.315\textwidth,valign=b]{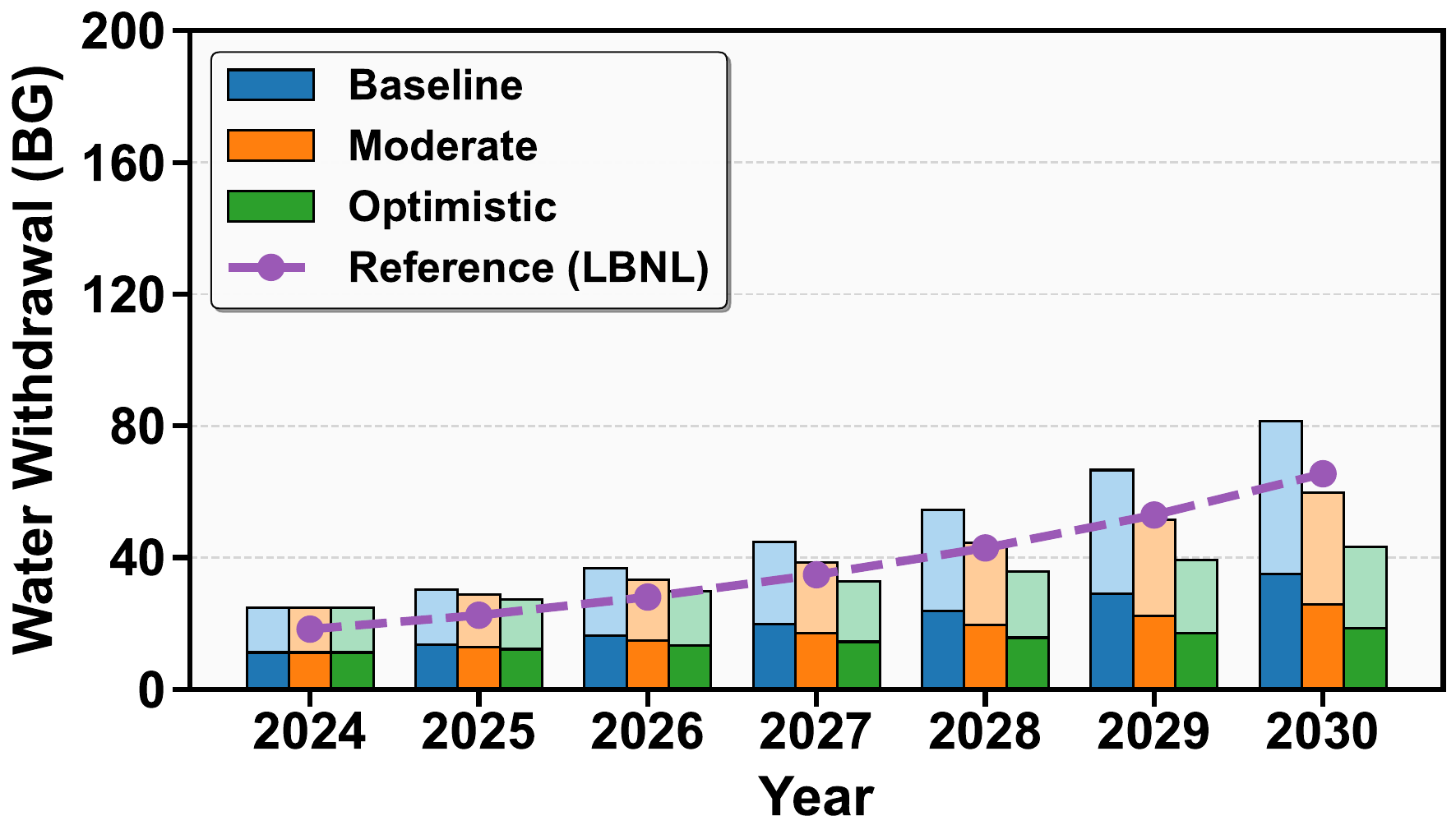} 
    }
    \subfloat[Mid Growth]{
    \includegraphics[width=0.315\textwidth,valign=b]{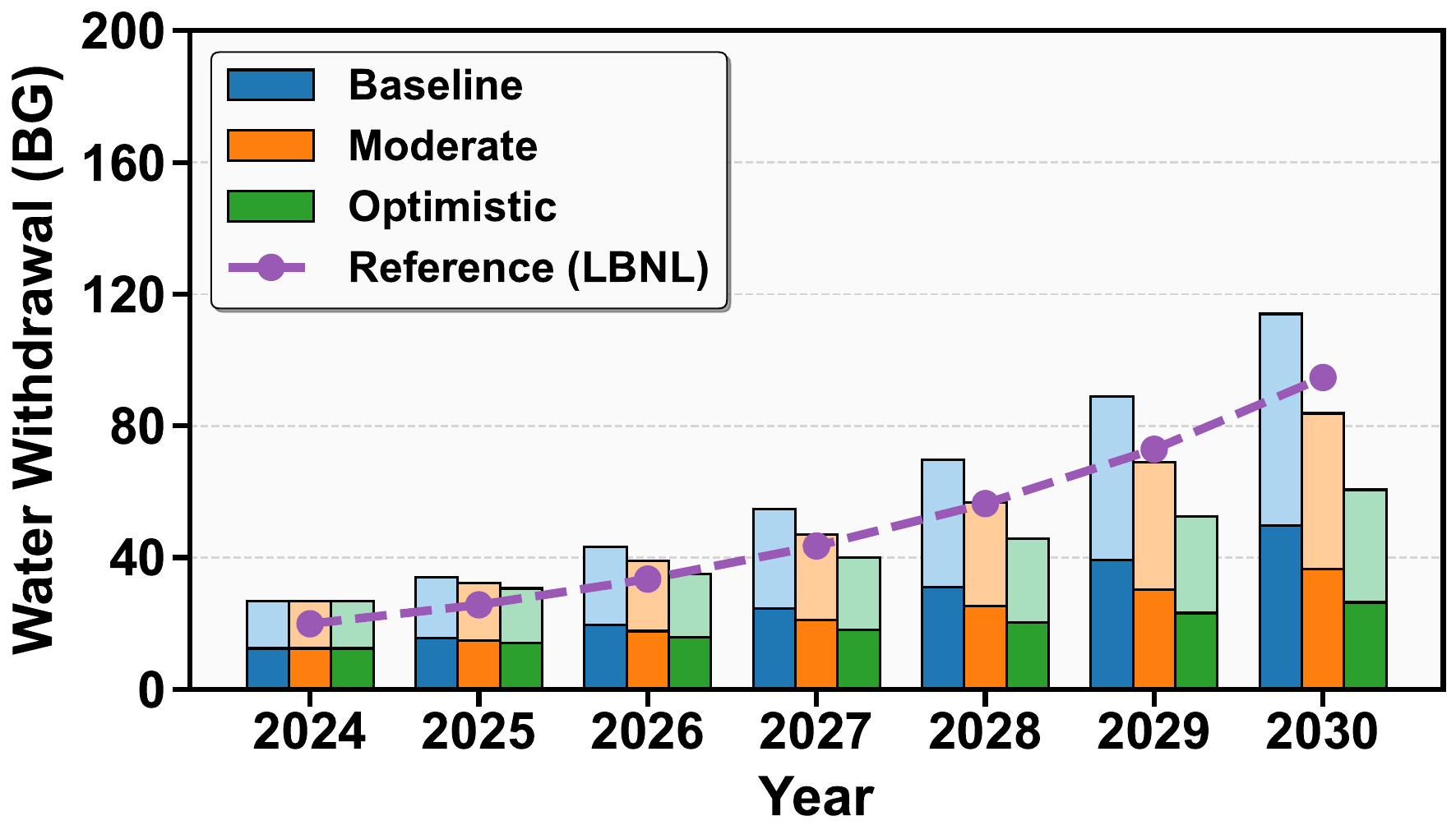} 
    }
    \subfloat[High Growth]{
    \includegraphics[width=0.315\textwidth,valign=b]{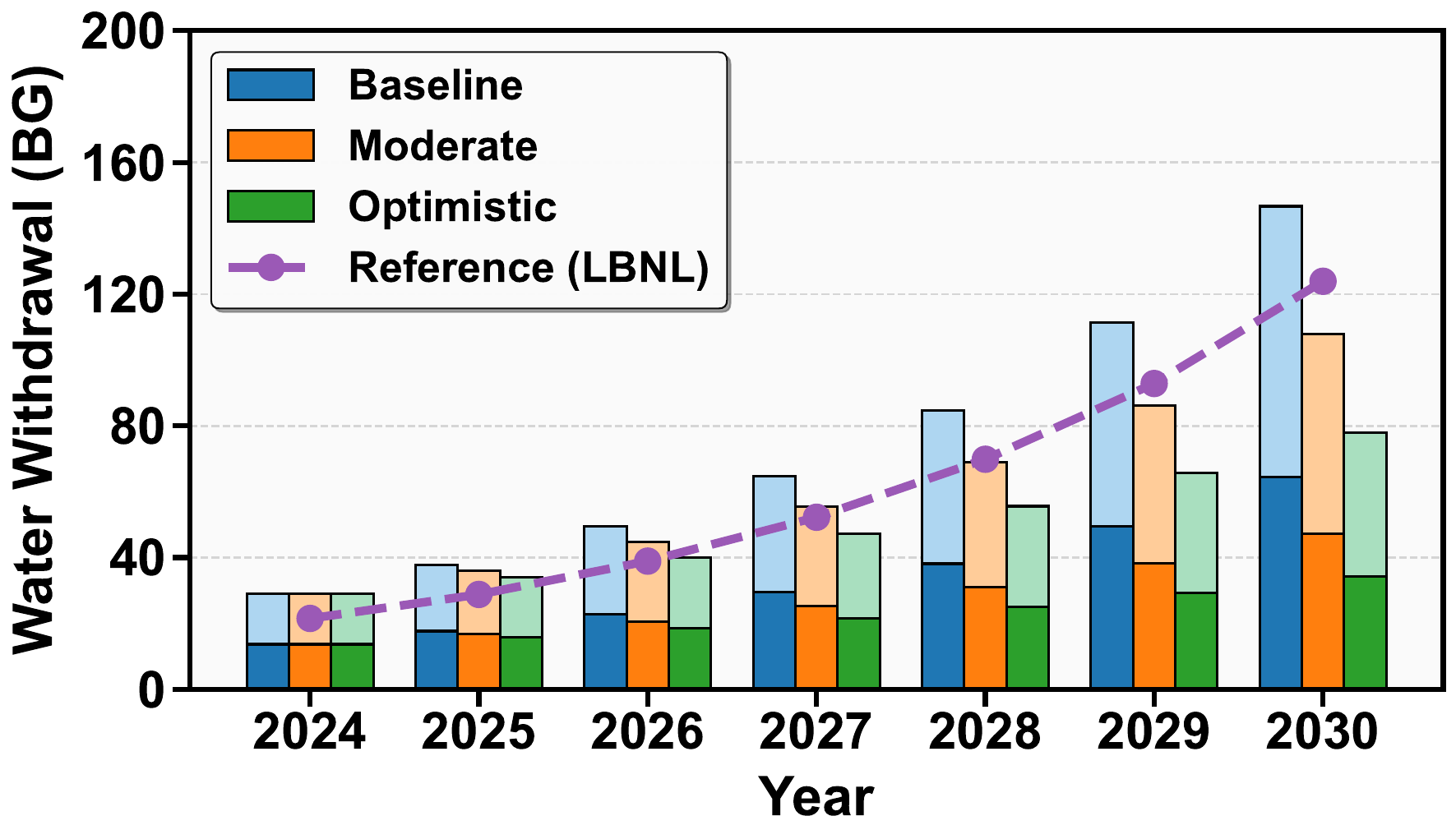} 
    }
    \caption{Annual water withdrawal in billion gallons (BG) under Reference, Baseline, Moderate, and Optimistic scenarios, 2024–2030. For Baseline, Moderate, and Optimistic scenarios, dark and light shades represent hyperscale and colocation data centers, respectively. The dashed line indicates total water withdrawal for Reference (LBNL).}\label{fig:total_water_withdrawal}
\end{figure}

We first present the total annual water withdrawal by U.S. data centers from 2024 to 2030, which serves as the basis for our water capacity analysis. The results are shown in Figure~\ref{fig:total_water_withdrawal}, with detailed values provided in Table~\ref{tab:annual_water_withdrawal}. 

We observe that our Baseline scenario, in which water efficiency remains constant through 2030, generally yields a higher water withdrawal number than Reference (LBNL). In contrast, the Moderate and Optimistic scenarios, reflecting industry-wide WUE reductions, yield lower projected withdrawals than Reference (LBNL), highlighting the potential benefits of adopting more water-efficient cooling technologies for U.S. data centers in the future.

Specifically, Figure~\ref{fig:total_water_withdrawal} illustrates an upward trend in annual water withdrawals by data centers from 2024 to 2030 across all growth scenarios. In the low-growth case, total withdrawals generally remain below 80 billion gallons even under the Baseline scenario, whereas the high-growth case projects a substantial increase that could exceed 140 billion gallons by 2030 under the Baseline scenario. Comparing the Baseline and Optimistic scenarios highlights that substantial improvements in cooling efficiency and operational practices could potentially reduce the projected total water demand by approximately half. Our results also distinguish between hyperscale and colocation data centers, showing that colocation data centers account for the majority of usage, although both segments scale substantially as overall demand rises. The growing trajectory in the mid and high growth cases underscores that, without a shift toward moderate or optimistic water efficiency improvements, the continued data center expansion can exert significant and accelerating pressure on public water resources. This finding is consistent with a recent internal projection by a leading technology company, which estimates that its Scope 1 water usage could increase by approximately 75\% relative to the 2024 level \cite{AI_Water_Microsoft_173percent_2030_compared_2024_2025estimate_News_NYTimes_2026}, despite substantial improvements in water efficiency.

Based on the most recent study \cite{Water_US_Public_WaterAnalysis_USGS_Report_2025_Medalie2025}, public water supply systems withdraw an average of 35,400~million gallons per day (MGD), including 4,219~MGD of consumptive use. This represents approximately 14.5\% of the total water withdrawn annually, on average, within the contiguous United States during water years 2010--2020 for major categories including crop irrigation, public supply, and thermoelectric power. Although public supply accounts for a smaller share of total withdrawals, it provides drinking water and directly supports essential human uses, making it a distinct category of water use that is comparatively more critical and constrained.

When benchmarking against total public water supply, we use the  baseline values in the most recent study \cite{Water_US_Public_WaterAnalysis_USGS_Report_2025_Medalie2025} without adjusting for future system-wide changes. Nationally, under the Baseline scenario, total U.S. data center water withdrawals in 2030 are projected to represent approximately 0.6\% to 1.1\% of total public water withdrawals in the low- and high-growth cases, respectively. 
Specifically, without significant improvements in water efficiency and in the high-growth case, total U.S. data center water withdrawals in 2030 could rival the current annual water supply of Los Angeles (from 2019 to 2024), which serves approximately 4 million residents \cite{Water_LosAngeles_DWP_146billion_gallon_per_year_2019_2024}. 

Under the Optimistic scenario, which assumes substantial water efficiency gains, the national share of U.S. data center water withdrawals decreases further to approximately 0.3\% to 0.6\%, depending on the growth rates.
This highlights the potential benefits of implementing water-efficient technologies.

Although the national share of data center withdrawals relative to total public supply remains modest, public water is fundamentally a local and seasonal resource. It is considerably more difficult to transfer across regions than electricity and is constrained by local water infrastructure, hydraulic conditions, water rights, and permitting requirements.
Thus, the impacts of data centers' Scope 1 water use should be interpreted in the context of local public water infrastructure, particularly the \emph{available} surplus capacity of public water systems, which is often constrained due to longstanding underinvestment (Section~\ref{sec:us_public_water}).

\subsubsection{Annual water consumption}

\begin{figure}[!t]
    \centering
    \subfloat[Low Growth]{
    \includegraphics[width=0.315\textwidth,valign=b]{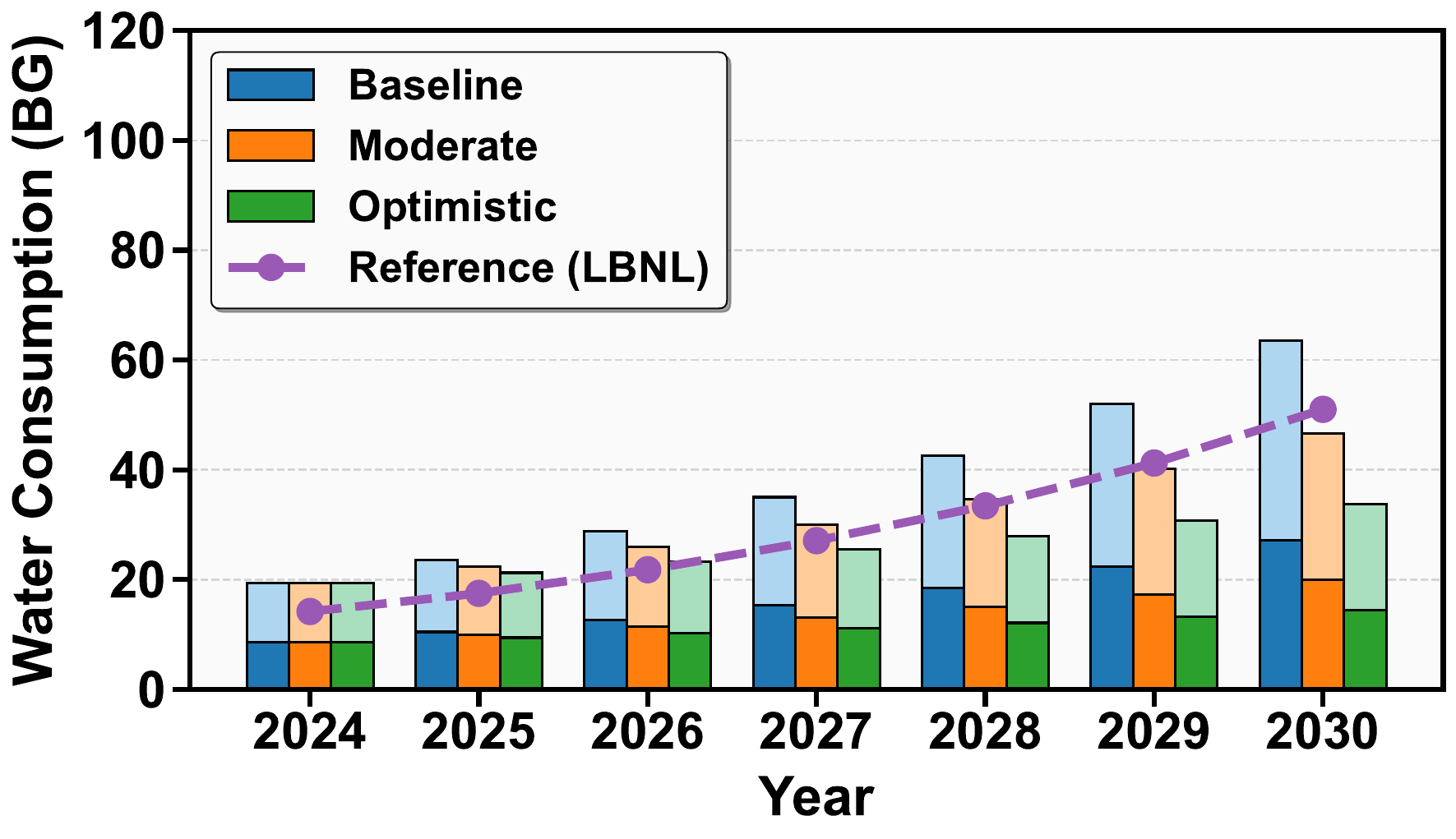} 
    }
    \subfloat[Mid Growth]{
    \includegraphics[width=0.315\textwidth,valign=b]{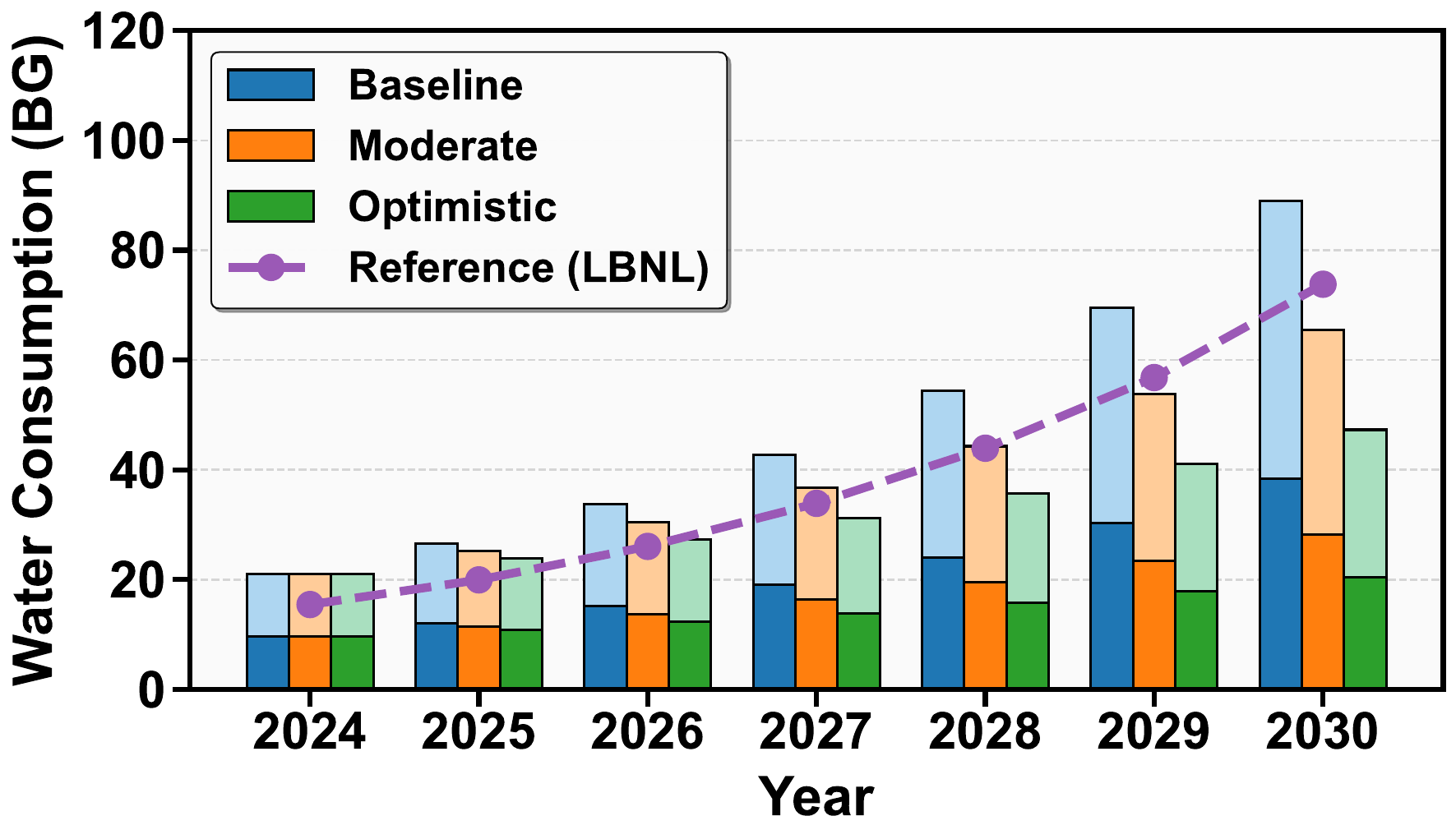} 
    }
    \subfloat[High Growth]{
    \includegraphics[width=0.315\textwidth,valign=b]{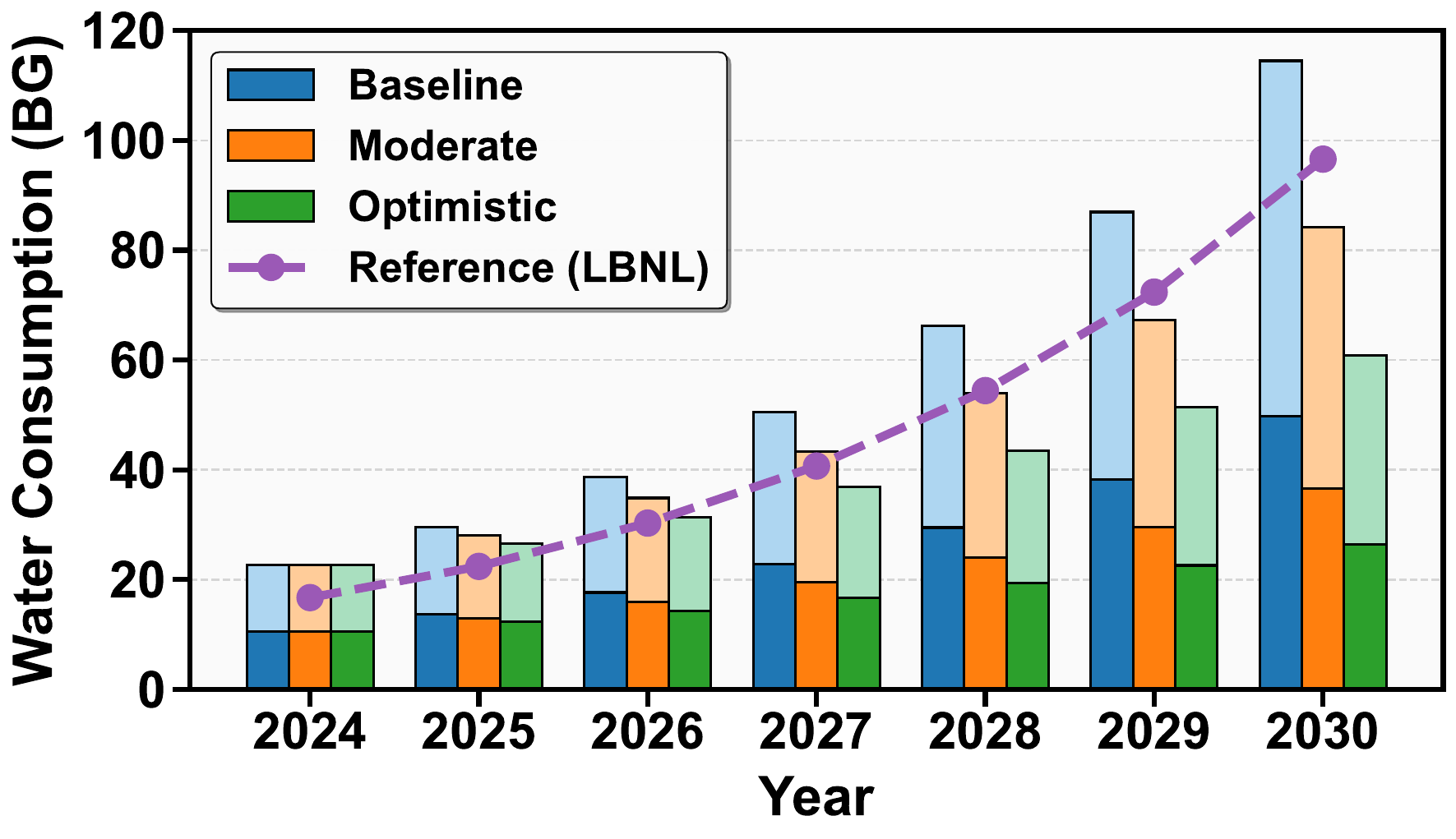} 
    }
    \caption{Annual water consumption in billion gallons (BG) under the Reference, Baseline, Moderate, and Optimistic scenarios from 2024 to 2030. For the Baseline, Moderate, and Optimistic scenarios, dark and light shades represent hyperscale and colocation data centers, respectively. The dashed line indicates the result for Reference~(LBNL).
}\label{fig:annual_water_consumption}
\end{figure}

Water consumption is an important, and sometimes regulated \cite{Water_RightsLaw_ConsumptiveIncluded_California_2026,Water_WaterRights_ConsumptiveConstraintMitigation_SRBC_website}, performance metric for public water systems, as it directly affects regional water availability, particularly during drought conditions \cite{Water_Planning_WashingtonMetro_2050Prediction_DataCenter_PeakingFactor_3point62_Report_2025}.
Thus, while our primary focus is on data centers' Scope~1 water withdrawals, we also present Scope~1 water consumption in Figure~\ref{fig:annual_water_consumption} for completeness, with detailed values provided in Table~\ref{tab:annual_water_consumption}. As shown, total U.S. data center water consumption follows a trend similar to total water withdrawals in Figure~\ref{fig:total_water_withdrawal}.

Compared with the contiguous U.S. public water supply's total consumptive use of 4,219~MGD \cite{Water_US_Public_WaterAnalysis_USGS_Report_2025_Medalie2025}, data centers' water consumption in 2030 is projected to represent approximately 4\% to 7\% under the Baseline scenario and 2\% to 4\% under the Optimistic scenario, depending on the IT load growth rates. These shares are notably higher than the corresponding withdrawal shares, reflecting the higher consumptive ratio of data center water use compared with other users served by public water systems. 

While these figures represent national aggregates and already indicate a non-negligible impact, the implications of data centers' consumptive water use are more appropriately assessed at local and seasonal scales, where certain public water systems may experience disproportionately greater impacts.

\begin{table*}[!t]
\centering
\caption{Average daily demand and water capacity need (million gallons per day), and capacity valuation from 2024 to 2030 under Reference, Baseline, Moderate, and Optimistic scenarios. The valuations are expressed
as ``(low, high)'' cost estimates.}
\label{tab:water_mdd_valuation_scenarios}
{\fontsize{6}{7}\selectfont 
\renewcommand{\arraystretch}{1.1}
\begin{tabular}{c|c|c|cc|ccc|ccc|ccc}
\toprule
\multirow{2}{*}{\textbf{Year}} & \multirow{2}{*}{\textbf{Metric}} & \multirow{2}{*}{\textbf{Growth}}
& \multicolumn{2}{c|}{\textbf{Reference}}
& \multicolumn{3}{c|}{\textbf{Baseline}}
& \multicolumn{3}{c|}{\textbf{Moderate}}
& \multicolumn{3}{c}{\textbf{Optimistic}} \\
\cline{4-14}
& & & \textbf{LBNL} & \textbf{NS}
& \textbf{Total} & \textbf{Hyper} & \textbf{Colo}
& \textbf{Total} & \textbf{Hyper} & \textbf{Colo}
& \textbf{Total} & \textbf{Hyper} & \textbf{Colo} \\
\hline
\multirow{6}{*}{2024} & \multirow{3}{*}{ADD} 
  & Low  & 50 & 206 & 68 & 31 & 37 & 68 & 31 & 37 & 68 & 31 & 37 \\
& & \textbf{Mid}  & \textbf{55} & \textbf{221} & \textbf{74} & \textbf{34} & \textbf{40} & \textbf{74} & \textbf{34} & \textbf{40} & \textbf{74} & \textbf{34} & \textbf{40} \\
& & High & 59 & 236 & 79 & 38 & 42 & 79 & 38 & 42 & 79 & 38 & 42 \\
\cline{2-14}
& \multirow{3}{*}{Capacity} 
  & Low  & 225 & 927 & 306 & 138 & 168 & 306 & 138 & 168 & 306 & 138 & 168 \\
& & \textbf{Mid}  & \textbf{245} & \textbf{995} & \textbf{332} & \textbf{153} & \textbf{178} & \textbf{332} & \textbf{153} & \textbf{178} & \textbf{332} & \textbf{153} & \textbf{178} \\
& & High & 265 & 1064 & 357 & 169 & 189 & 357 & 169 & 189 & 357 & 169 & 189 \\
\hline
\multirow{6}{*}{2030} & \multirow{3}{*}{ADD} 
  & Low  & 179 & 599 & 223 & 96 & 127 & 164 & 71 & 93 & 118 & 51 & 67 \\
& & \textbf{Mid}  & \textbf{259} & \textbf{832} & \textbf{312} & \textbf{136} & \textbf{176} & \textbf{230} & \textbf{100} & \textbf{129} & \textbf{166} & \textbf{72} & \textbf{94} \\
& & High & 339 & 1065 & 402 & 176 & 226 & 295 & 130 & 166 & 214 & 94 & 120 \\
\cline{2-14}
& \multirow{3}{*}{Capacity} 
  & Low  & 807 & 2694 & 1003 & 433 & 570 & 737 & 318 & 419 & 533 & 230 & 303 \\
& & \textbf{Mid}  & \textbf{1167} & \textbf{3744} & \textbf{1406} & \textbf{613} & \textbf{793} & \textbf{1033} & \textbf{451} & \textbf{583} & \textbf{747} & \textbf{326} & \textbf{421} \\
& & High & 1528 & 4793 & 1809 & 794 & 1015 & 1330 & 583 & 746 & 961 & 422 & 539 \\
\hline
\multirow{6}{*}{\begin{tabular}[c]{@{}c@{}}\textbf{2024$\rightarrow$2030}\end{tabular}} & \multirow{3}{*}
{\begin{tabular}[c]{@{}c@{}}\textbf{Capacity}\\ \textbf{Increase}\end{tabular}}
  & Low  & 582 & 1767 & 697 & 295 & 402 & 431 & 180 & 251 & 227 & 92 & 135 \\
& & \textbf{Mid}  & \textbf{922} & \textbf{2749} & \textbf{1074} & \textbf{460} & \textbf{614} & \textbf{702} & \textbf{297} & \textbf{404} & \textbf{415} & \textbf{172} & \textbf{243} \\
& & High & 1262 & 3730 & 1451 & 625 & 826 & 972 & 415 & 558 & 604 & 253 & 351 \\
\cline{2-14}
& \multirow{3}{*}{\begin{tabular}[c]{@{}c@{}}\textbf{Capacity}\\ \textbf{Valuation}\\(\$ Billion)\end{tabular}} 
  & Low  & (6, 23) & (18, 71) & (7, 28) & (3, 12) & (4, 16) & (4, 17) & (2, 7) & (3, 10) & (2, 9) & (1, 4) & (1, 5) \\
& & \textbf{Mid}  & \textbf{(9, 37)} & \textbf{(27, 110)} & \textbf{(11, 43)} & \textbf{(5, 18)} & \textbf{(6, 25)} & \textbf{(7, 28)} & \textbf{(3, 12)} & \textbf{(4, 16)} & \textbf{(4, 17)} & \textbf{(2, 7)} & \textbf{(2, 10)} \\
& & High & (13, 50) & (37, 149) & (15, 58) & (6, 25) & (8, 33) & (10, 39) & (4, 17) & (6, 22) & (6, 24) & (3, 10) & (4, 14) \\
\bottomrule
\end{tabular}
}
\end{table*}

\subsubsection{New water capacity demand}

A data center's total water capacity requirement, in terms of maximum daily water demand, is a critical planning consideration and may not be fully supported by existing public water systems. To quantify this water capacity requirement, the average daily demand (ADD) needs to be adjusted using a peaking factor. In this study, we review multiple public records and \emph{conservatively} adopt an average peaking factor of 4.5, reflecting the practice that data centers often provision water capacity with worst-case considerations and redundancy, analogous to approaches used in power infrastructure planning.  As evaporative-assisted cooling becomes more common, it can improve the annual water efficiency while simultaneously increasing the peaking factor, since such systems typically operate only 5 to 15\% of the year. For example, it is common for leading technology companies' planned peaking factors to range from 6 to 8 or even higher for their state-of-the-art data centers under construction. Additional details regarding
the peaking factor are available in Appendix~\ref{appendix:peaking_factor}.

The results are summarized in Table~\ref{tab:water_mdd_valuation_scenarios}. While we aggregate U.S. data centers' water capacity needs, it is important to note that an individual data center's actual peak daily water use can be lower than its total capacity requirement (which typically accounts for redundancy) and may not align nationally on the same day, as the hottest days driving the peak water demand generally occur asynchronously across regions.

From 2024 to 2030, the estimated new water capacity needed to support U.S. data centers' peak water demand ranges from 697 to 1,451~MGD under the Baseline scenario. For comparison, New York City delivers approximately 1~billion gallons of water per day to nearly 9~million residents in 2024 \cite{Water_NYC_1512MGDin1979_1billionNow_2025}. Hypothetically, if all the new water capacity needed by U.S. data centers from 2024 to 2030 in the high-growth case were pooled, it could supply New York City for most of the year, except for the few peak days in the summer. While supplying water at the scale of New York City is far more complex than simple aggregation, this comparison highlights the magnitude of U.S. data centers' total water capacity needs due to their high peaking factors.

Under the Optimistic scenario, which assumes substantial water reductions, the new total water capacity need decreases to between 227 and 604~MGD, depending on IT load growth rates between 2024 and 2030. Even with continuous efficiency gains corresponding to a 10\% compound annual reduction in WUE, high-growth IT loads can still drive the total water capacity need to a level that, if aggregated hypothetically, could supply roughly half of New York City's daily public water demand (2024 level).

\subsubsection{Valuation of new water capacity}

We now quantify the valuation of the total new water capacity needed 
for U.S. data centers from 2024 to 2030. To establish a plausible range of cost estimates, we review public reports
on 17 water infrastructure projects of varying complexity, type, and geography, 
including 6 projects specifically related to data centers.
Table~\ref{tab:cost_water_infra_project_survey} indicates that a reasonable valuation (in 2021–2025 dollars, without inflation adjustment) 
for water infrastructure is \$10–\$40 million per MGD, depending on factors such as project complexity.

\begin{figure}[!t]
    \centering
    \subfloat[Low Growth]{
    \includegraphics[width=0.315\textwidth,valign=b]{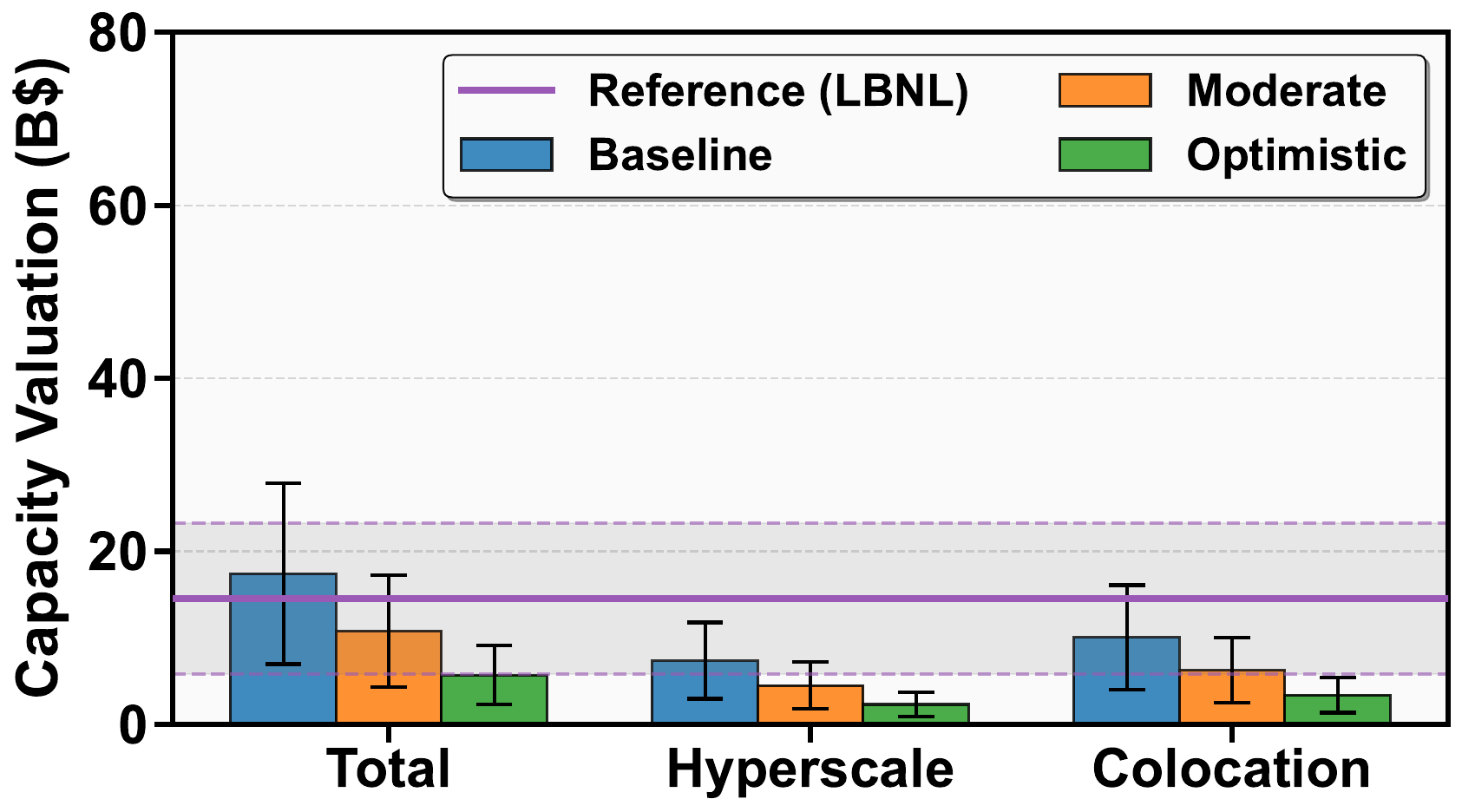} 
    }
    \subfloat[Mid Growth]{
   \includegraphics[width=0.315\textwidth,valign=b]{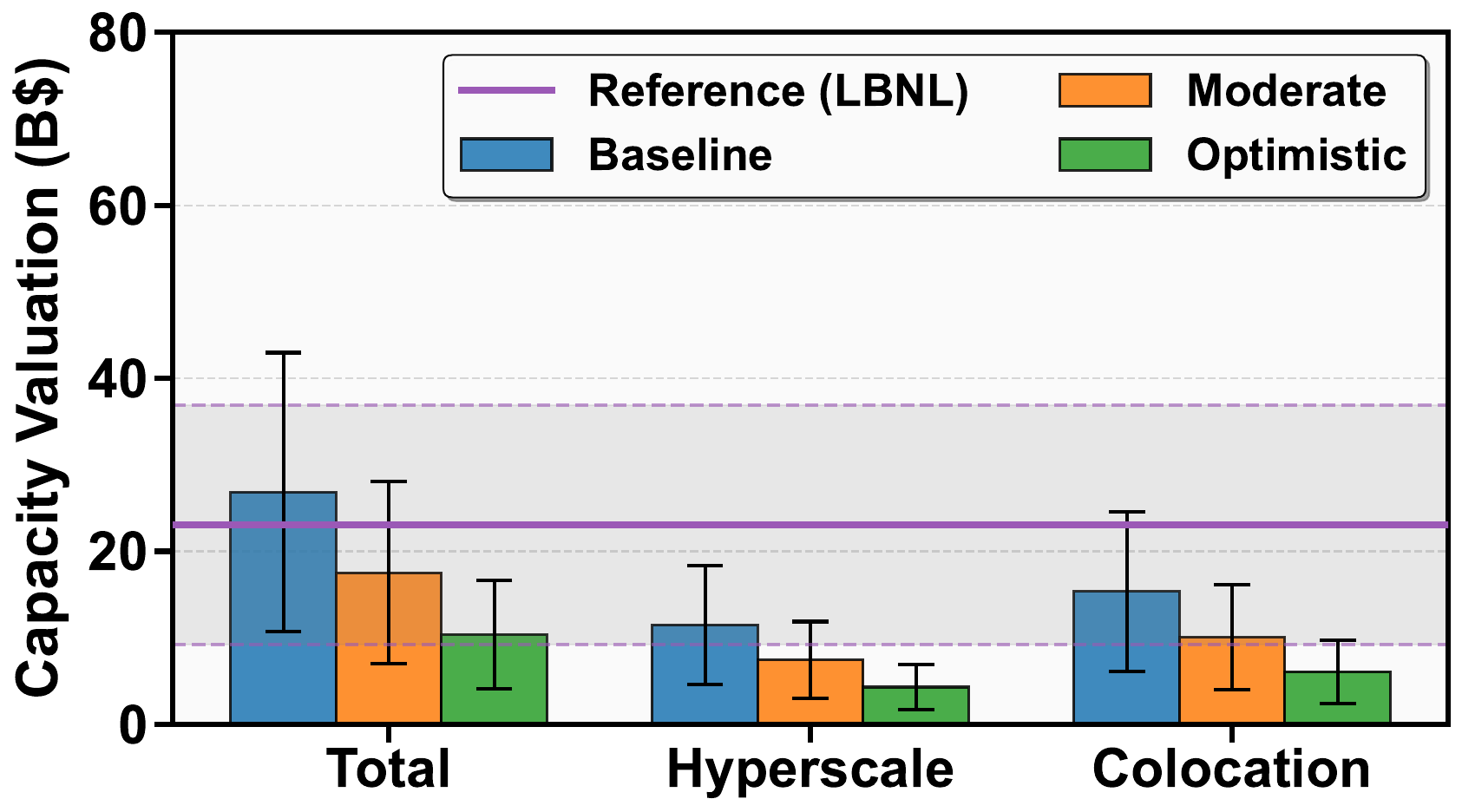} 
    }
    \subfloat[High Growth]{
   \includegraphics[width=0.315\textwidth,valign=b]{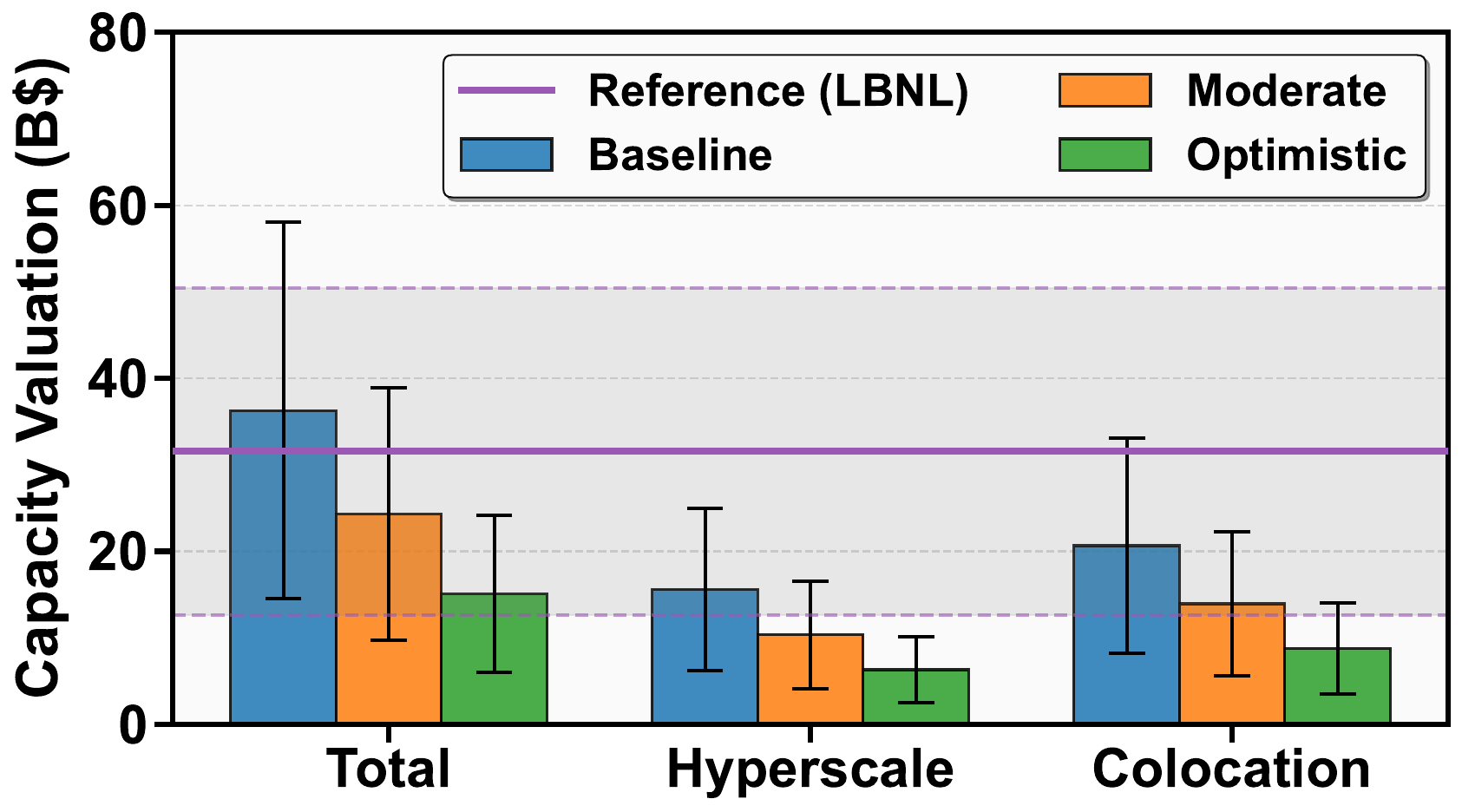} 
    }
    \caption{Valuation of the new water infrastructure capacity to meet
    U.S. data centers' new peak water demand from 2024 to 2030 under different scenarios for low, mid, and high growth projections.}\label{fig:delta_mdd_scenarios}
\end{figure}

Using this cost range, Figure~\ref{fig:delta_mdd_scenarios} and Table~\ref{tab:water_mdd_valuation_scenarios} present 
the total valuation of the new water capacity needed to support U.S. data centers through 2030. 
The overall valuation is generally on the order of \$10 billion. 
Specifically, under the Baseline scenario, the valuation ranges from \$7–28 billion in the low-growth case
and \$15–58 billion in the high-growth case. 
Under the Optimistic scenario, the valuations decrease to \$2–9 billion and \$6–24 billion 
in the low-growth and high-growth cases, respectively.

Although these billion-dollar-level investments are modest compared to the overall AI and data center industry, 
they may not be affordable for many public water systems that are already grappling with financial challenges 
and large funding needs \cite{EPA_WaterDrinking_625billion_2021_7th_Report_Congress_2023,EPA_WaterWaste_CleanNeeds_630billion_2022_Report_Congress_2024}. Thus, corporate–community partnerships to expand water infrastructure are particularly important, 
meeting data centers' water needs while avoiding increased water rates for local ratepayers amid rising water affordability challenges \cite{EPA_WaterAffordability_NeedsAssessment_Report_Congress_2024}.

Importantly, 
our valuation may differ from the actual new funding requirement, 
as some of this capacity may already be available
and (partially) funded by existing sources such as state–federal support, bonds, or current ratepayer contributions. 
Nonetheless, similar to ``Water Positive'' initiatives that replenish water to offset Scope~1 consumption \cite{Meta_Water_Volumetric_Report_2025,Google_Water_Stewardship_ProjectPortfolio_2025}, 
even if the water capacity is already available, data centers can add new capacity to proactively offset their capacity needs, 
allowing the host community to retain limited available infrastructure capacity for future growth. 
Further details are provided in Section~\ref{sec:recommendation_discussion}.

\section{Recommendations}\label{sec:recommendation_discussion}

In this section, we provide a set of key recommendations to address the growing water capacity demand of U.S. data centers 
and its impact on public water systems, while additional recommendations to address the overall water efficiency
are available in Appendix~\ref{appendix:recommendation_additional}.

\paragraph{Recommendation 1: Report peak water usage.} 
While total annual water withdrawal and consumption are important metrics, data centers' peak water demand is critical for public water system planning and management. However, data center operators rarely disclose their peak water demand. 
We therefore recommend that data center annual reports include per-site peak water metrics, including withdrawal, consumption, and discharge, to provide a comprehensive view of operational water impacts on public water systems.

Additionally, to assess peak water efficiency, we recommend extending the standard annualized water usage effectiveness (WUE) metric 
to define the \emph{peak} WUE (\pwue) as follows.

{\definition[\pwue]\label{def:peakWUE}{
The peak water usage effectiveness, \pwue, has a unit of L/kWh and is defined as
\begin{equation}\label{eqn:peakWUE}
\pwue = \max_{t \in \mathcal{T}} \frac{WaterConsumption(t)}{ITEnergy(t)},
\end{equation}
where $WaterConsumption(t)$ and $ITEnergy(t)$ denote the total Scope~1 water consumption 
and IT energy consumption on day $t$, respectively, and $\mathcal{T}$ represents the period considered, 
such as a calendar or fiscal year.
}}

The \pwue metric can be determined from either actual measurements for post-hoc reporting and analysis, 
or from projected values for planning purposes. Additionally, when
the IT load remains relatively flat, 
\pwue can be simply expressed as $\pwue = \beta \cdot WUE$,
where $\beta \in [1,\infty)$ is the daily peaking factor (either measured or planned), 
and $WUE \geq 0$ is the annual WUE commonly reported in the literature \cite{DoE_DataCenter_EnergyReport_US_2024} 
and in industry disclosures \cite{Facebook_SustainabilityReport_2025}.

\paragraph{Recommendation 2: Develop corporate-community partnerships.}
The lack of technical and financial  capacity remains a critical challenge for many U.S. public water systems~\cite{EPA_WaterDrinking_625billion_2021_7th_Report_Congress_2023}. 
While some state laws require large water users to make ``pro rata'' fair contributions to fund water infrastructure expansion \cite{Google_Water_8MGD_2MGDinitial_CountyWebsite_2026,Google_Water_8MGD_Vrginia_New_WaterSource_Agreement_2026}, additional corporate-community partnerships that go beyond these legal requirements may further strengthen the resilience of water infrastructure in host communities.
 By providing funding or technical assistance for local water system upgrades, data centers can help mitigate the added strain their operations place on public water supplies. Additionally,
sourcing non-potable water for industrial use, such as recycled or reclaimed wastewater, can help reduce demand on municipal drinking water supplies for both data center operations and community needs~\cite{xAI_WaterRecylingPlant_80million_Investment_Memphis_News_2025}. We therefore recommend that data centers actively participate in these partnerships, aligning corporate responsibility with community well-being and long-term water resilience.

\paragraph{Recommendation 3: ``Pipe Neutral'' 
commitments.} 
Technology companies have increasingly committed to ``Water Positive'' initiatives by replenishing more water than their Scope~1 consumption \cite{Meta_Water_Volumetric_Report_2025,Google_Water_Stewardship_ProjectPortfolio_2025}. 
However, these replenishment efforts focus on total volumetric benefits and may not address the impacts of data centers' peak water use on public water systems. 
As a complement to the recent community-first AI approaches \cite{Microsoft_CommunityFirst_AI_Infrastructure_Blog_2026}, 
we recommend that data centers add new water capacity to meet their needs, 
avoiding water rate increases for local ratepayers while allowing the host community to retain limited water infrastructure capacity for future growth.  
We refer to this approach as ``Water Capacity Neutral,'' or colloquially, ``Pipe Neutral.'' 

To quantify the net contribution of a data center to public water systems, 
we formally define a metric called the Water Capacity Impact (\ouralg) score as follows.

{\definition[\ouralg]\label{def:ouralg}{
The Water Capacity Impact (\ouralg) score is defined as
\begin{equation}\label{eqn:ouralg}
\ouralg = \frac{C_{\text{Added}} - C_{\text{Allocated}}}{C_{\text{Available}}},
\end{equation}
where $C_{\text{Added}}$, $C_{\text{Allocated}}$, and $C_{\text{Available}}$ denote the water capacity added by the data center, 
allocated to the data center, and currently available for allocation in the host community, respectively.
}}

A \ouralg\ score less than $-1$ indicates that the available water capacity is insufficient to meet the data center's needs; 
a score between $-1$ and $0$ indicates a net resource usage of the water capacity; 
and a score greater than $0$ indicates that the data center contributes more water capacity to the public water system than it needs, 
representing a positive contribution. 
A concrete example is a large technology company’s recent commitment of up to \$400 million to local public water infrastructure to support its new Louisiana data centers \cite{Amazon_Water_400m_25_35per_PeakPowerReduction_AI_Cloud_Louisiana_PressRelease_2026}, effectively reversing the otherwise negative \ouralg score.

\paragraph{Recommendation 4: Coordinated water-power planning.} 
Water-based evaporative cooling uses water, but can reduce peak power demand by 10--35\% compared to waterless dry cooling
\cite{Amazon_AI_Water_EnergyWater_Tradeoff_Video,Google_Water_10percent_Less_Energy_CliamteCounscious_Blog}. 
Thus, 
water used for evaporative assistance---even if applied only on the hottest days---functions like a ``water battery,'' alleviating peak-hour demand on the electrical grid during the summer.  
Indeed, while regional water availability remains important, evaporative cooling, if used responsibly, can be an economically competitive approach to shaving peak power demand compared to traditional generation methods (Appendix~\ref{appendix:water_power_cost_comparison}). This aligns with a leading technology company's statement
that ``water is the most efficient means of cooling in many places''
  \cite{Google_SustainabilityReport_2025}.
In addition, some large data centers have recently secured up to 8~MGD of peak water capacity to reduce the electricity demand during the summer \cite{Google_Water_8MGD_Vrginia_WaterUtilityAgreement_2026,Water_Lebanon_LEAP_DataCenter_Meta_Final_2025}, with associated water infrastructure upgrades costing hundreds of millions of dollars \cite{Google_Water_8MGD_300million_30x_BottlingCompany_Virginia_News_2025}.

On the other hand, opportunistically utilizing available surplus power, such as solar power during summer midday when water evaporative rates are highest, can reduce peak water demand without overloading the grid. This strategy is particularly beneficial when public water systems are stressed during heatwaves.

We therefore recommend a coordinated water-power strategy to optimally manage the tradeoff: (1) responsibly leveraging water 
 to alleviate grid stresses while minimizing impacts on local water resources; and (2) opportunistically utilizing power to reduce the burden on public water systems without overloading the grid.

\section{Conclusion}

In this paper, we examine data centers' direct water withdrawal in the United States and their impacts on public water systems. We first show that, compared to typical water users such as residential buildings, data centers have a higher consumptive ratio and a higher peaking factor, even when their annual water use remains modest. Using public sources including industrial disclosures and government records, we quantify the U.S. data center water use from 2024 to 2030. Our analysis shows that, under the Baseline scenario, data centers' total water withdrawal in 2030 accounts for 0.6\%--1.1\% of total public water withdrawals in the contiguous United States, while their water consumption reaches 4\%--7\% of total public water consumption due to the high consumptive ratio.  
Most importantly, under the Baseline scenario,
U.S. data centers are projected to require 697--1,451~MGD of new water capacity, valued at up to \$58~billion and comparable to New York City's average daily supply of roughly 1,000~MGD. Under an Optimistic scenario with a 10\% compound annual reduction in water use intensity, these figures decrease to approximately 30--50\% of the Baseline scenario values. While these numbers are national aggregates, local and seasonal impacts on specific public water systems may be disproportionately larger.  

Finally, we provide recommendations to address the growing water demand of U.S. data centers, including reporting peak water use, 
developing corporate-community partnerships,
adopting a Water Capacity Neutral approach (colloquially ``Pipe Neutral'') to allow host communities to retain limited water capacity, and implementing coordinated water-power planning to optimally manage the tradeoff.
Together, these strategies enhance operational efficiency, protect local water systems, and ensure long-term community resilience while supporting technological advancement and data center growth.

\section*{Acknowledgement}

The authors would like to thank Lisa McFadden (Water Environment Federation) and Gregory Pierce (UCLA)
for their feedback and comments on the initial draft of this paper.

\vspace{0.3in}

{
       \bibliographystyle{unsrt}

}

\newpage
\appendix

\section*{Appendix}

\section{Detailed Methodology for Quantifying and Forecasting Data Centers' Water Demand}\label{appendixsec:water_demand_method}

This section presents the detailed methodology
of quantifying and forecasting data centers' water demand
It describes the data sources and calculation procedures used to estimate IT equipment energy consumption (hereafter referred to as IT energy) and water use effectiveness (WUE) under different scenarios from 2024 to 2030, which are subsequently used to quantify the average daily demand (\add) and water capacity demand
measured in terms of the maximum daily demand (\mdd).

\subsection{Key Metrics and Definitions}
We first define the key metrics used in this study. Unless otherwise specified, energy, water consumption, and water withdrawal refer to annual values and include only data center operations, excluding office and other non–data-center activities. When converting annual quantities to daily averages,
we apply 365 days per year, without distinguishing leap and non-leap years. All energy- and water-related quantities presented in the subsequent analysis are computed based on these definitions.

\begin{gather}
\mathrm{WUE} = \frac{\mathrm{On\text{-}site\ Water\ Consumption}}{\mathrm{IT\ Energy}} \label{eq:wue} \\
\mathrm{PUE} = \frac{\mathrm{Total\ Energy}}{\mathrm{IT\ Energy}} \label{eq:pue} \\
\mathrm{Water\ Consumptive\ Ratio} = \frac{\mathrm{Water\ Consumption}}{\mathrm{Water\ Withdrawal}} \label{eq:consumptive_ratio} \\
\mathrm{ADD} = \frac{\mathrm{Water\ Withdrawal}}{365} \label{eq:add} \\
\mathrm{MDD} = \mathrm{ADD} \times \mathrm{Peaking\ Factor} \label{eq:mdd}
\end{gather}

\subsection{IT Energy}
\label{appendixsubsec:it_energy_est}
The Lawrence Berkeley National Laboratory (LBNL) report
\cite{DoE_DataCenter_EnergyReport_US_2024}
provides a detailed classification of data center types,
which are further aggregated into three groups in our study: hyperscale, colocation, and others.
As WUEs can differ across data center types, we estimate IT energy  at the data center type level to ensure consistent WUE calculation and aggregation.

LBNL reports total the U.S. data center electricity consumption, PUE, and WUE for 2024--2028 under low and high growth rates (Table~\ref{tab:lbnl_energy_pue_it}). In principle, aggregate IT energy can be derived from total energy and PUE using Eq.~\eqref{eq:pue}. Over 2024--2028, the reported total data center energy exhibits compound annual growth rates (CAGRs) of approximately 13\% and 27\% under the low and high rates, respectively. These growth rates are broadly consistent with other projections, which report CAGRs of approximately 20\% through 2030~\cite{DataCenter_Demand_20Percent_CAGR_2025to2030_JLL_Report_2025} and approximately 22.4\% over 2023--2030 under a medium-growth scenario~\cite{DataCenter_Energy_McKinsey_AI_20_30Growth_US_2030_Website_2025}. Given this consistency, the total data center energy for 2029 and 2030 is extrapolated by applying the corresponding CAGRs derived from the 2024--2028 period.

\begin{table}[!ht]
\centering
\caption{LBNL-reported U.S. data center total energy, PUE, WUE, and  aggregate IT energy under low and high assumptions (2024--2030).
The values for 2029 and 2030 are extended based on the LBNL values for 2024 to 2028.
}
\label{tab:lbnl_energy_pue_it}
\small
\begin{tabular}{c|cc|cc|cc|cc}
\toprule
\multirow{2}{*}{\textbf{Year}} 
& \multicolumn{2}{c|}{\textbf{WUE (L/kWh)}} 
& \multicolumn{2}{c|}{\textbf{PUE}} 
& \multicolumn{2}{c|}{\textbf{Total Energy (TWh)}} 
& \multicolumn{2}{c}{\textbf{IT Energy (TWh)}} \\
\cline{2-9}
 & Low & High & Low & High & Low & High & Low & High \\
\hline
2024&0.39 &0.40& 1.33 & 1.45 & 185.00 & 232.00 & 139.10 & 160.00 \\
2025&0.40 &0.42& 1.23 & 1.45 & 203.00 & 303.00 & 165.04 & 208.97 \\
2026&0.42 &0.44& 1.19 & 1.40 & 238.00& 388.00 & 200.00 & 277.14 \\
2027&0.44 &0.46& 1.17 & 1.37 & 279.00& 481.00 & 238.46 & 351.09 \\
2028&0.45 &0.48& 1.15 & 1.35 & 325.00 & 578.00 & 282.61 & 428.15 \\
2029&0.46 &0.49& 1.13 & 1.33 & 367.25 & 734.06 & 325.00 & 551.92 \\
2030&0.47 &0.50& 1.11 & 1.31 & 414.99 & 932.26 & 373.87 & 711.65 \\
\bottomrule
\end{tabular}
\end{table}

For PUE, LBNL reports values through 2028. PUE values for 2029 and 2030 are estimated by extending the observed trend, assuming an annual reduction of 0.02. Details regarding WUE are discussed in Section~\ref{appendixsubsec:reference}. Using the estimated total energy and PUE values, aggregate IT energy for 2024--2030 is computed accordingly. Details are shown in Table~\ref{tab:lbnl_energy_pue_it}.

However, LBNL does not report the annual distribution of total energy across data center types for 2025 to 2027. As a result, IT energy derived directly from total energy and PUE would only yield aggregate values and would not support type-specific estimation. 
To address this limitation, we construct type-specific IT energy estimates using a bottom-up approach based on the disaggregated IT energy in 2024 and 2028 reported by LBNL. Therefore, the aggregate IT energy derived from total energy and PUE is not adopted in our type-specific analysis. Instead, it only serves as a reference to evaluate the reasonableness of our resulting bottom-up IT energy estimates.

Specifically, in the LBNL report, IT energy is decomposed into server, network, and storage components. Server energy is reported for 2024 and 2028 under low and high rate assumptions for each data center type. In contrast, network and storage energy are reported for 2024--2028 only at the national level, without low-high distinctions and without differentiation by data center type.

Based on the reported server energy values for 2024 and 2028, the CAGRs of server energy over 2024--2028 are approximately 22\% and 31\% for hyperscale data centers under the low and high rate assumptions, respectively, and approximately 24\% and 34\% for colocation data centers. In contrast, server energy in the ``Others'' category decreases over this period, yielding negative CAGRs of approximately $-7\%$ and $-8\%$ under the low and high assumptions, respectively. These type-specific growth rates are then applied to construct annual server energy trajectories for 2024--2030 by interpolating intermediate years (2025--2027) and extrapolating later years (2029--2030), in line with the extension applied to the aggregate LBNL total energy projections.
Although the ``Others'' category shows higher values under the low assumption than under the high assumption, the terminology of ``low'' and ``high'' is consistently defined according to the total server energy assumptions reported by LBNL to maintain consistency across data center types. A mid estimate is further constructed as the arithmetic average of the corresponding low and high estimates. 

The server energy estimates by data center type are summarized in Table~\ref{tab:server_energy}.
\begin{table*}[!ht]
\centering
\caption{Type-specific U.S. data center server energy used in our IT energy estimation (2024--2030). Mid values are defined as the arithmetic average of the corresponding low and high estimates.}
\label{tab:server_energy}
\small
\begin{tabular}{c|ccc|ccc|ccc|ccc}
\toprule
\multirow{2}{*}{\textbf{Year}} & \multicolumn{3}{c|}{\textbf{Hyperscale (TWh)}} & \multicolumn{3}{c|}{\textbf{Colocation (TWh)}} & \multicolumn{3}{c|}{\textbf{Others (TWh)}} & \multicolumn{3}{c}{\textbf{Total (TWh)}} \\
\cline{2-13} 
 & Low & \textbf{Mid} & High & Low & \textbf{Mid} & High & Low & \textbf{Mid} & High & Low & \textbf{Mid} & High \\
\hline
2024 & 48.66 & \textbf{55.00} & 61.33 & 50.67 & \textbf{54.67} & 58.67 & 14.00 & \textbf{14.00} & 14.00 & 113.33 & \textbf{123.67} & 134.00 \\
2025 & 59.30 & \textbf{69.94} & 80.57 & 62.59 & \textbf{70.72} & 78.84 & 13.08 & \textbf{12.98} & 12.87 & 134.98 & \textbf{153.63} & 172.28 \\
2026 & 72.27 & \textbf{89.06} & 105.84 & 77.32 & \textbf{91.64} & 105.95 & 12.22 & \textbf{12.03} & 11.83 & 161.81 & \textbf{192.72} & 223.63 \\
2027 & 88.07 & \textbf{113.56} & 139.05 & 95.52 & \textbf{118.95} & 142.38 & 11.42 & \textbf{11.15} & 10.88 & 195.01 & \textbf{243.66} & 292.30 \\
2028 & 107.33 & \textbf{145.00} & 182.67 & 118.00 & \textbf{154.67} & 191.33 & 10.67 & \textbf{10.34} & 10.00 & 236.00 & \textbf{310.00} & 384.00 \\
2029 & 130.80 & \textbf{185.39} & 239.97 & 145.77 & \textbf{201.44} & 257.11 & 9.97 & \textbf{9.58} & 9.19 & 286.54 & \textbf{396.41} & 506.28 \\
2030 & 159.40 & \textbf{237.33} & 315.26 & 180.07 & \textbf{262.79} & 345.51 & 9.31 & \textbf{8.88} & 8.45 & 348.79 & \textbf{509.01} & 669.22 \\
\bottomrule
\end{tabular}
\end{table*}

For network and storage energy, the LBNL report provides aggregate U.S. totals for 2024--2028. Using the reported values in 2024 and 2028, we compute separate CAGRs for network energy (approximately 26\%) and storage energy (approximately 7\%). These growth rates are applied to extrapolate total network and storage energy for 2029 and 2030.
The resulting annual network and storage energy is then allocated across hyperscale, colocation, and others according to the year-specific server energy shares of the three data center types. This allocation is performed separately under the low and high assumptions. Consistent with the definition adopted above, the terms ``low'' and ``high'' refer to the corresponding LBNL total server energy assumptions rather than to alternative allocation rules or proportion levels. 
The corresponding estimates are summarized in Table~\ref{tab:stg_net_energy}.

\begin{table}[!ht]
\centering
\caption{Estimated U.S. data center network and storage energy used in our IT energy estimation, 2024--2030. Shares represent each component’s proportion relative to total server energy under the low and high assumptions.
}
\label{tab:stg_net_energy}
\begin{tabular}{c|cc|cc|cc}
\toprule
\multirow{2}{*}{\textbf{Year}} & \multicolumn{2}{c|}{\textbf{Energy (TWh)}} & \multicolumn{2}{c|}{\textbf{Storage Ratio}} & \multicolumn{2}{c}{\textbf{Network Ratio}} \\
\cline{2-7}
 & Storage & Network & Low & High & Low & High \\
\hline
2024 & 17.10 & 9.08 & 0.15 & 0.13 & 0.08 & 0.07 \\
2025 & 17.75 & 12.35 & 0.13 & 0.10 & 0.09 & 0.07 \\
2026 & 18.44 & 16.18 & 0.11 & 0.08 & 0.10 & 0.07 \\
2027 & 19.70 & 19.92 & 0.10 & 0.07 & 0.10 & 0.07 \\
2028 & 22.02 & 23.19 & 0.09 & 0.06 & 0.10 & 0.06 \\
2029 & 23.46 & 29.32 & 0.08 & 0.05 & 0.10 & 0.06 \\
2030 & 24.99 & 37.06 & 0.07 & 0.04 & 0.11 & 0.06 \\
\bottomrule
\end{tabular}
\end{table}
By summing server, network, and storage energy within each data center type, we obtain IT energy for three data center types in our analysis: Hyperscale, Colocation, and Others. Aggregating across the three types yields total U.S. IT energy. The resulting estimates are summarized in Table~\ref{tab:it_energy}. 

\begin{table*}[!h]
\centering
\caption{Estimated U.S. data center IT energy by data center type in our analysis (2024--2030). IT energy equals the sum of server, network, and storage energy, with network and storage energy allocated across types according to annual server energy shares. Mid values are defined as the arithmetic average of the corresponding low and high estimates.
}
\label{tab:it_energy}
\small
\begin{tabular}{c|ccc|ccc|ccc|ccc}
\toprule
\multirow{2}{*}{\textbf{Year}} & \multicolumn{3}{c|}{\textbf{Hyperscale (TWh)}} & \multicolumn{3}{c|}{\textbf{Colocation (TWh)}} & \multicolumn{3}{c|}{\textbf{Others (TWh)}} & \multicolumn{3}{c}{\textbf{Total IT (TWh)}} \\
\cline{2-13}
 & Low & \textbf{Mid} & High & Low & \textbf{Mid} & High & Low & \textbf{Mid} & High & Low & \textbf{Mid} & High \\
\hline
2024 & 59.90 & \textbf{66.81} & 73.31 & 62.38 & \textbf{66.25} & 70.13 & 17.23 & \textbf{16.98} & 16.74 & 139.51 & \textbf{149.85} & 160.18 \\
2025 & 72.52 & \textbf{83.59} & 94.65 & 76.55 & \textbf{84.58} & 92.62 & 16.00 & \textbf{15.56} & 15.12 & 165.08 & \textbf{183.73} & 202.38 \\
2026 & 87.73 & \textbf{104.98} & 122.23 & 93.87 & \textbf{108.11} & 122.35 & 14.84 & \textbf{14.25} & 13.66 & 196.43 & \textbf{227.34} & 258.25 \\
2027 & 105.96 & \textbf{131.93} & 157.90 & 114.93 & \textbf{138.30} & 161.68 & 13.74 & \textbf{13.05} & 12.35 & 234.63 & \textbf{283.28} & 331.92 \\
2028 & 127.89 & \textbf{166.03} & 204.18 & 140.61 & \textbf{177.23} & 213.86 & 12.71 & \textbf{11.95} & 11.18 & 281.21 & \textbf{355.21} & 429.21 \\
2029 & 154.89 & \textbf{209.94} & 264.99 & 172.62 & \textbf{228.26} & 283.91 & 11.81 & \textbf{10.98} & 10.15 & 339.31 & \textbf{449.18} & 559.05 \\
2030 & 187.76 & \textbf{266.12} & 344.49 & 212.11 & \textbf{294.83} & 377.55 & 10.97 & \textbf{10.10} & 9.24 & 410.84 & \textbf{571.05} & 731.27 \\
\bottomrule
\end{tabular}
\end{table*}

Comparing the IT energy values in Table~\ref{tab:it_energy} (our analysis) and Table~\ref{tab:lbnl_energy_pue_it} (LBNL), we find that the differences are reasonably small. Specifically, from 2024 to 2028, the percentage deviation remains constantly small, ranging from -1.8\% to +0.3\% under the low assumption and from -6.8\% to +0.2\% under the high assumption, indicating that our estimates are generally lower than or close to the LBNL values. In 2029–2030, the deviation becomes positive, reaching +4.4\% and +9.9\% for the low case and +1.3\% and +2.8\% for the high case, respectively. Although our estimates exceed the LBNL values in these later years, the differences remain moderate and within an acceptable range and  align well with other projections through 2030 (e.g., 
\cite{Water_Carbon_AI_Projection_2030_FengqiYou_Cornell_NatureSustainability_2025_xiao2025environmental,DataCenter_Demand_20Percent_CAGR_2025to2030_JLL_Report_2025,DataCenter_Energy_McKinsey_AI_20_30Growth_US_2030_Website_2025}).

Overall, the results suggest that our estimation remains close to the LBNL values. Unless otherwise specified, the IT energy values used in this study refer to those reported in Table~\ref{tab:it_energy}.

\subsection{Reference Scenarios}\label{appendixsubsec:reference}

This subsection describes the calculation procedures for the Reference (LBNL) and Reference (NS) scenarios. For both scenarios, the respective U.S.-wide average WUE is applied and multiplied by
the total IT energy (for all the data center types) to obtain
the U.S. data centers' total water.

\textbf{Reference (LBNL)~\cite{DoE_DataCenter_EnergyReport_US_2024}.}  
The LBNL report provides national-level WUE estimates for 2024--2028 under low and high assumptions only. For 2029--2030, WUE values are estimated by extending the observed trend, assuming an annual increase of 0.01. Water consumption under the LBNL reference is calculated by combining low IT energy with low WUE and high IT energy with high WUE.

\textbf{Reference (NS)~\cite{Water_Carbon_AI_Projection_2030_FengqiYou_Cornell_NatureSustainability_2025_xiao2025environmental}.}  
The recent study classifies data centers based on the adoption of advanced liquid cooling (ALC) and reports state-level PUE and WUE values with and without ALC for 2024. The Reference (NS) scenario in  our study adopts the baseline case defined in \cite{Water_Carbon_AI_Projection_2030_FengqiYou_Cornell_NatureSustainability_2025_xiao2025environmental}, which assumes an initial ALC adoption rate of 5\% in 2024, increasing at a CAGR of 20\%, together with the corresponding PUE and WUE values reported for that baseline case. Nationwide weighted averages are computed using each state's total data center energy consumption from EPRI~\cite{DataCenter_Energy_EPRI_AI_9Percent_US_2030_WhitePaper_2024}. Following the assumed ALC adoption trajectory, the weighted-average WUE in each year is calculated as:
\begin{align}
    \text{WUE}_{\text{year}} = (1 - \alpha_{\text{year}})\,\text{WUE}_{\text{base}} + \alpha_{\text{year}}\,\text{WUE}\_{\text{ALC}}_{\text{base}},
\end{align}
where $\alpha_{\text{year}}$ denotes the ALC adoption rate in a given year. While the study \cite{Water_Carbon_AI_Projection_2030_FengqiYou_Cornell_NatureSustainability_2025_xiao2025environmental} considers
various WUE scenarios and focuses on AI data centers which generally have low WUEs, we apply its (baseline) WUE reference to all data center types solely for orders-of-magnitude cross-checking,
helping to prevent overestimation;
instead, we rely primarily on the LBNL results as our reference.

The resulting WUE values for the Reference (LBNL) and Reference (NS) scenarios are summarized in Table~\ref{tab:wue_scenarios}.  Unless otherwise specified, the mid values of water consumption, water withdrawal, \add and \mdd are defined as the arithmetic average of the corresponding low and high quantities.

\begin{table}[!ht]
\centering
\caption{WUE (L/kWh) under Reference, Baseline, Moderate, Optimistic scenarios, 2024--2030.}
\label{tab:wue_scenarios}
\scriptsize
\renewcommand{\arraystretch}{1.1} 
\begin{tabular}{c|cc|c|cc|cc|cc|cc|cc|cc}
\toprule
\multirow{3}{*}{\textbf{Year}} & \multicolumn{3}{c|}{\textbf{Reference}} & \multicolumn{4}{c|}{\textbf{Baseline}} & \multicolumn{4}{c|}{\textbf{Moderate}} & \multicolumn{4}{c}{\textbf{Optimistic}} \\
\cline{2-16}
& \multicolumn{2}{c|}{LBNL}&  \multirow{2}{*}{NS} & \multicolumn{2}{c|}{\textbf{U.S.-wide}}&\multirow{2}{*}{Hyper} &  \multirow{2}{*}{Colo} & \multicolumn{2}{c|}{\textbf{U.S.-wide}}&\multirow{2}{*}{Hyper} &\multirow{2}{*}{Colo}& \multicolumn{2}{c|}{\textbf{U.S.-wide}}  &\multirow{2}{*}{Hyper} &  \multirow{2}{*}{Colo} \\
\cline{2-3}
\cline{5-6} 
\cline{9-10} 
\cline{13-14} 
&Low &High & &Low&High& & &Low&High& & &Low&High& &\\
\hline
2024& 0.385&0.395 & 1.582 & 0.525&0.534 & 0.546 & 0.650 & 0.525&0.534 & 0.546 & 0.650 & 0.525&0.534 & 0.546 & 0.650 \\
2025& 0.401&0.419 & 1.581 & 0.541&0.553 & 0.546 & 0.650 & 0.514&0.525 & 0.519 & 0.617 & 0.487&0.497 & 0.492 & 0.585 \\
2026& 0.420&0.444 & 1.579 & 0.554&0.566 & 0.546 & 0.650 & 0.500&0.511 & 0.493 & 0.586 & 0.449&0.459 & 0.443 & 0.526 \\
2027& 0.437&0.464 & 1.577 & 0.565&0.576 & 0.546 & 0.650 & 0.484&0.494 & 0.468 & 0.557 & 0.412&0.420 & 0.398 & 0.473 \\
2028& 0.450&0.480 & 1.575 & 0.573&0.584 & 0.546 & 0.650 & 0.467&0.475 & 0.445 & 0.529 & 0.376&0.383 & 0.359 & 0.426 \\
2029& 0.460&0.490 & 1.572 & 0.580&0.589 & 0.546 & 0.650 & 0.449&0.456 & 0.423 & 0.503 & 0.342&0.348 & 0.323 & 0.384 \\
2030& 0.470&0.500 & 1.569 & 0.585&0.593 & 0.546 & 0.650 & 0.430&0.436 & 0.402 & 0.477 & 0.311&0.315 & 0.290 & 0.345 \\
\bottomrule
\end{tabular}
\end{table}

\subsection{Baseline, Moderate, and Optimistic Scenarios}
\label{appendixsubsec:wue_for_our_scenarios}
This subsection describes the details for the Baseline, Moderate, and Optimistic WUE scenarios. All the three scenarios share the same 2024 WUE values and differ only in their assumed compound annual changes: 0\% per year for the Baseline scenario, $-5\%$ per year for the Moderate scenario, and $-10\%$ per year for the Optimistic scenario. Accordingly, the key task in scenario construction is to estimate type-specific WUE values for the base year 2024, from which WUE trajectories for 2025--2030 are derived.

\subsubsection{The Year of 2024}

To estimate the 2024 U.S.-level WUE for different data center types, we select a set of large data center operators to estimate WUE separately for hyperscale and colocation data centers, while WUE for the ``Others'' data center type is assumed to be zero for a conservative estimate. The estimation relies primarily on publicly available company-wide sustainability reports released for the calendar (or fiscal) year of 2024 released by 12 large operators, including five hyperscale operators and seven colocation operators~\cite{Google_SustainabilityReport_2025, microsoft_datacenter_sustainability_efficiency, microsoft_environmental_data_fact_sheet, meta_environmental_data_index, apple_environmental_report, amazon_aws_sustainability,digital_realty_impact_report, Equinix_SustainabilityReport_2024, QTS_SustainabilityReport_2024, cyrusone_sustainability_report, switch_sustainability,switch_esg_report, ntt_sustainability_report, edgeconnex_sustainability_report}.  We do not link these specific references to individual data center companies in order to avoid direct comparisons and maintain their anonymity while preserving our transparency to the extent possible.

The selected hyperscale and colocation data center operators collectively represent approximately 86\% and 22\% of U.S. hyperscale and colocation IT loads in 2024, respectively, based on the mid IT estimates reported in Table~\ref{tab:it_energy}.  Therefore, the WUE values aggregated across these companies provide a reasonable estimate of industry practices.

Although reporting formats and data granularity vary across companies, the overall data quality is generally sufficient for our analysis. Reported values are followed as closely as possible, and explicit assumptions are introduced only when necessary to ensure internal consistency. Based on these company-level data, U.S.-level type-specific WUE values for 2024 are derived and subsequently used to construct scenario-specific trajectories.

For each selected operator, we collect (and estimate if needed) its U.S. data center–level IT energy (MWh), total energy (MWh), PUE, water consumption (ML), water withdrawal (ML), WUE (L/kWh), the water consumptive ratio, municipal water ratio, and potable water ratio. The resulting values are summarized in Table~\ref{tab:dc_metrics_2024}. 
Note that, based on publicly disclosed data, we can reliably calculate the U.S. potable water ratio for only one data center operator (94.40\% for Hyperscale-1) and the global potable water ratios for two other operators. Therefore, we exclude potable ratios in Table~\ref{tab:dc_metrics_2024} and instead present them only in the operator-specific results below.

The reported values in Table~\ref{tab:dc_metrics_2024} are adopted directly whenever explicitly disclosed. When specific metrics are unavailable, values estimated or assigned based on clearly stated assumptions or publicly available third-party information are introduced and identified with an asterisk ``$^*$'' in Table~\ref{tab:dc_metrics_2024}.

Unless otherwise specified, the municipal and potable water ratios refer to shares of water directly supplied by the municipal/public water systems and potable water relative to
the total water withdrawal. If withdrawal-based ratios are explicitly reported, they are used directly; if only consumption-based ratios are disclosed, the corresponding ratios are derived from reported water consumption values and marked with a dagger symbol ($\dagger$). If neither withdrawal- nor consumption-based ratios are available, the corresponding entry is denoted by ``--'', indicating that the value is unreported and cannot be reliably estimated, although public potable water is likely used as a general industry practice.
Energy and water quantities are rounded to the nearest integer, while all other numerical values are rounded to two decimal places.

For hyperscale operators, total water consumption generally includes both data center water use and office-related water use. When separate reporting is unavailable, office water \emph{consumption} is estimated using an industry-standard consumptive ratio of 10\% \cite{Google_SustainabilityReport_2025}, while data center water consumption is estimated using a default consumptive ratio of 75\% \cite{Water_Planning_WashingtonMetro_2050Prediction_DataCenter_PeakingFactor_3point62_Report_2025}.
For colocation operators, office water use is assumed to be negligible, as they typically employ significantly fewer staff than hyperscalers, which maintain large teams of software developers and corporate employees; accordingly, reported water consumption is attributed primarily to data center operations.
Additional default settings and estimation details are provided in the operator-specific calculation procedures presented below.

\textbf{Hyperscale-1.} 
This operator reports data center water withdrawal and water consumption at the location/site level. We aggregate all U.S. data center sites to obtain total U.S. data center water withdrawal (approximately 29{,}510~ML) and water consumption (approximately 23{,}120~ML), yielding a water consumptive ratio of approximately 0.78. 
The operator reports North America–level electricity consumption (treated as U.S.) but does not separately disclose electricity use attributable to data center operations. To estimate U.S. data center electricity consumption, we apply the ratio of global data center electricity consumption to the combined global electricity consumption for data center and office operations (approximately 96\%). This yields an estimated U.S. data center electricity consumption of approximately 22~TWh. 
A global PUE value of 1.09 is then applied to derive U.S. data center IT energy, resulting in approximately 20~TWh. Finally, WUE is calculated as the ratio of U.S. data center water consumption to U.S. data center IT energy, yielding approximately 1.13~L/kWh.
In addition, the operator reports location-level potable water use. Based on the aggregated U.S. data center locations, the U.S.-level potable water ratio is approximately 94.40\%.

\textbf{Hyperscale-2.}
This operator reports Americas-level (North and South America combined) PUE and WUE values of 1.16 and 0.38~L/kWh for 2024, respectively, which we adopt as proxies for U.S. data center operations. Total water withdrawal and water consumption are reported at the North America level and are treated as U.S.-level totals in this study (5{,}789~ML and 3{,}136~ML, respectively).
U.S. data center water withdrawal and water consumption are estimated by applying a data center water consumptive ratio of 0.75 and an office water consumptive ratio of 0.10. 
Combined with the water withdrawal and consumption numbers,
this yields approximately 3{,}934~ML of U.S. data center water withdrawal and 2{,}951~ML of U.S. data center water consumption.
Using the estimated U.S. data center water consumption and the reported WUE, U.S. data center IT energy is calculated as approximately 7.8~TWh. Total electricity consumption is then derived by applying the reported PUE, yielding approximately 9.0~TWh. 
Although a North America–level electricity consumption of approximately 16.6~TWh is reported, it is not disaggregated for data center operations. We therefore base our estimation on the disclosed water quantities and estimate electricity use accordingly.
Note that by applying 3{,}136~ML (including office use but assumed here to be entirely for data centers) with a WUE of 0.38~L/kWh and a PUE of 1.16, we obtain an alternative estimate of 9.6~TWh for the North America–level total data center electricity consumption, which still differs from the reported total electricity consumption of 16.6~TWh in North America (including non-data center electricity consumption). This discrepancy is likely due to reporting or accounting inconsistencies by the operator. 
The operator reports global water consumption by source. Third-party water accounts for 5{,}775~ML out of a total water consumption of 5{,}807~ML worldwide. We use this proportion as a proxy for the U.S. data center municipal water ratio, yielding approximately 99.45\%.

\textbf{Hyperscale-3.} 
This operator reports data center water withdrawal at the location/site level, which allows us to aggregate all facilities explicitly identified as located in the United States and obtain total U.S. data center water withdrawal. Based on the disclosed values, U.S. water withdrawal is approximately 2{,}365~ML out of a global data center water withdrawal of approximately 4{,}143~ML, corresponding to a share of about 0.57. We directly sum the water withdrawals of all disclosed data center locations, and the resulting total is slightly different from the reported total water withdrawal 4{,}145~ML due to rounding errors. The U.S. data center water consumption is then estimated by scaling the reported global data center water consumption (2974~ML) according to the observed U.S. withdrawal share, yielding approximately 1{,}698~ML. This procedure implicitly assumes a uniform data center water consumptive ratio of approximately 0.72 across all facilities. The operator also reports global third-party water withdrawal of 5{,}625~ML out of a total withdrawal of 5{,}637~ML. We treat this proportion as a proxy for the U.S. municipal water ratio, yielding approximately 99.79\%.

The operator reports a global WUE of 0.19~L/kWh and a global PUE of 1.08. Because approximately 43\% of total data center water withdrawal/consumption occurs outside the United States, the reported global WUE may not accurately represent U.S.-level performance. We therefore do not directly adopt the global WUE for estimating U.S. operations. However, in the absence of U.S.-level PUE disclosure, we adopt the reported global PUE of 1.08 for U.S.-level estimation, since average PUE values are typically more consistent across U.S. and non-U.S. locations.

The operator also discloses location/site-based data center electricity consumption. By aggregating all facilities explicitly identified as located in the United States, we obtain total U.S. data center electricity consumption of approximately 12.8~TWh. Dividing this value by the global PUE yields U.S. IT energy of approximately 11.8~TWh. Using the estimated U.S. water consumption, we derive a U.S.-level WUE of approximately 0.14~L/kWh, which is lower than the reported global WUE.

\textbf{Hyperscale-4.}
The operator reports location-based data center electricity consumption by region, from which we aggregate approximately 1.7~TWh for U.S. hyperscale data centers (excluding colocation data centers). The reported global data center electricity consumption is approximately 2.5~TWh (including colocation data centers), implying that self-managed U.S. operations account for about 68\% of total data center electricity use. This share is subsequently used to scale global water-related quantities to the U.S. level. 

Specifically, global data center water withdrawal and discharge are reported as approximately 1{,}800~million gallons (MG) and 900~MG, respectively. We therefore compute water consumption as the difference between withdrawal and discharge, yielding approximately 900~MG. After separating data center and office water use using consumptive ratios of 0.75 for data centers and 0.10 for office operations, the resulting global data center water withdrawal is scaled using the 68\% U.S. energy share, yielding approximately 2{,}842~ML for U.S. data centers. Applying the default data center consumptive ratio of 0.75 results in an estimated U.S. data center water consumption of approximately 2{,}132~ML. To derive IT energy, we assume a PUE of 1.09, consistent with the value adopted for Hyperscale-1 due to the lack of disclosure by this operator, yielding an estimated U.S. data center IT energy of approximately 1.6~TWh. Dividing the estimated water consumption by IT energy results in a WUE of approximately 1.36~L/kWh.

The operator further reports total global water use of 1{,}756~MG (slightly different from 1,800 MG), consisting of 1{,}532~MG of freshwater, 197~MG of recycled water, and 27~MG of other alternative water sources. The report indicates that freshwater and recycled water are primarily supplied through municipal systems, with less than 5\% obtained from other providers. Accordingly, we attribute 95\% of the combined freshwater and recycled water volumes to municipal supply, yielding an estimated U.S. municipal water ratio of approximately 93.54\%.

\textbf{Hyperscale-5.}
For this operator, U.S. data center total electricity consumption and IT energy are not directly reported. The operator reports a North America–level WUE of 0.13~L/kWh and a PUE of 1.14, which are treated as values of its U.S. operations in this study.

Despite the lack of detailed electricity consumption data, Hyperscale-5 reportedly maintains one of the largest data center footprints in the United States and ranks among the top purchasers of renewable energy, surpassing many of its peers.
Thus, for the purpose of setting its contribution to average WUE within the hyperscale data center category,
we assume its U.S. IT energy to be 120\% of the average combined IT energy of the three largest hyperscale operators (Hyperscale-1, Hyperscale-3, and Hyperscale-2, in descending order), yielding approximately 16.0~TWh. Based on the reported PUE of 1.14, the corresponding total electricity consumption is estimated to be approximately 18.2~TWh.
A different electricity consumption value would only slightly affect the average WUE for the hyperscale data center type (e.g., by less than 5\% if the total energy consumption of Hyperscale-5 were adjusted to match that of the current largest Hyperscale-1).

Notably, the reported WUE for Hyperscale-5 is explicitly defined as water withdrawal per unit of IT energy. To maintain consistency, we convert the withdrawal-based WUE to a consumption-based WUE using a data center water consumptive ratio of 0.75, yielding an effective WUE of approximately 0.10~L/kWh. Applying this value to the estimated IT energy results in total water consumption of approximately 1{,}560~ML and total water withdrawal of approximately 2{,}081~ML.

\textbf{Colocation-1.}
The operator has significant non-U.S. presence and reports global-level water consumption and total electricity consumption. We assume a U.S. share of 50\% and apply this allocation factor to estimate U.S. data center water consumption of approximately 2{,}793~ML and U.S. data center electricity consumption of approximately 5.7~TWh.
As colocation data centers generally have higher PUE values
than hyperscalers,
a default PUE of 1.25 (assumed in this study) is then applied to derive U.S. data center IT energy, yielding approximately 4.5~TWh. WUE is computed as water consumption divided by IT energy, resulting in approximately 0.62~L/kWh. Finally, assuming a data center water consumptive ratio of 0.75, U.S. data center water withdrawal is estimated based on the calculated water consumption.

In addition, the operator reports global municipal water consumption of 1{,}475{,}293~kGal out of a total global water consumption of 1{,}475{,}762~kGal. We use this proportion as a proxy for the U.S. municipal water ratio, yielding approximately 99.97\%. Among the reported municipal water consumption, 849{,}319~kGal is identified as potable water, resulting in approximately 57.55\%. However, the low potable water ratio is possibly due to the operator's substantial non-U.S. presence, where stricter regulations on potable water use may apply; the U.S. potable water ratio is likely higher.

\textbf{Colocation-2.}
This operator reports global-level water withdrawal and water consumption, as well as energy consumption by region. We compute the ratio of reported energy consumption in the Americas to total global energy consumption and treat this ratio as the U.S. allocation factor in this study (approximately 40\%). 
Using this allocation factor, U.S. data center water withdrawal and water consumption are estimated by scaling down the corresponding global totals, yielding approximately 2{,}153~ML and 1{,}654~ML, respectively. The resulting water consumptive ratio is approximately 0.77. The operator reports a WUE of 0.95~L/kWh and a PUE of 1.39. Based on these values, IT energy is computed as water consumption divided by WUE, yielding approximately 1.7~TWh. 
The total electricity consumption is then derived by applying the reported PUE to the estimated IT energy.

In addition, the operator reports global municipal water withdrawal of 5{,}140~ML out of a total global withdrawal of 5{,}440~ML. We use this proportion as a proxy for the U.S. municipal water ratio, resulting in approximately 94.49\%. The operator further reports that non-potable sources account for 37\% of total water use at the global level. Accordingly,  the complementary share (63\%) is the potable water ratio. However, similar to Colocation-1, the low potable water ratio is likely due to the operator's substantial non-U.S. presence, where stricter regulations on potable water use may apply; the U.S. potable water ratio is likely higher.

\textbf{Colocation-3.}
For this operator, the vast majority of data center operations are located in the United States. The operator reports water consumption of 2{,}119~ML and water withdrawal of 2{,}323~ML, from which a water consumptive ratio of approximately 0.91 is computed. The operator also reports a WUE of 0.82~L/kWh and a PUE of 1.40. 
Based on these values, IT energy is calculated as water consumption divided by WUE, yielding approximately 2.6~TWh. Total data center electricity consumption (approximately 3.6~TWh) is then derived by applying the reported PUE to the estimated IT energy. 
In addition, the operator reports that all water withdrawal is sourced from third-party suppliers. Accordingly, we treat the municipal water ratio as 100\%.

\textbf{Colocation-4.}
For this operator, nearly all data center operations are located in the United States. The operator reports data center water consumption of approximately 1{,}021~ML and water withdrawal of approximately 1{,}312~ML, from which a water consumptive ratio of approximately 0.78 is calculated. The operator also reports a WUE of 0.29~L/kWh and a PUE of 1.46. 
Based on the reported WUE, U.S. data center IT energy is computed as water consumption divided by WUE, yielding approximately 3.5~TWh. Total data center energy is then derived by applying the reported PUE to the estimated IT energy.

\textbf{Colocation-5.}
For this operator, nearly all data center operations are located in the United States. The operator reports total electricity consumption and specifies a PUE of 1.23, calculated as a 12-month trailing average across seasoned sectors, which we adopt in this study. A WUE value of 1.28 L/kWh is also disclosed.
Using the reported electricity consumption (approximately 1.5 TWh) and the disclosed PUE, we derive the corresponding U.S. data center IT energy (approximately 1.2 TWh). Based on the reported WUE, total water consumption is then estimated at approximately 1,584 ML.
Finally, applying the default consumptive ratio of 0.75, we estimate total U.S. data center water withdrawal at approximately 2,112 ML.

\textbf{Colocation-6.}
For this non-U.S. operator, we assume that 20\% of its data center capacity is located in the United States. The operator reports global total energy consumption, which is treated as total global data center electricity consumption. Applying the 20\% allocation factor yields U.S. data center electricity consumption of approximately 0.8~TWh. 
Using the reported PUE of 1.38, IT energy is derived as total electricity divided by PUE, resulting in approximately 0.6~TWh. The reported WUE of 0.72~L/kWh is then applied to estimate U.S. data center water consumption based on the derived IT energy, yielding approximately 431~ML. Finally, by applying the water consumptive ratio of 0.75, U.S. data center water withdrawal is estimated from the calculated water consumption.

\textbf{Colocation-7.}
For this operator with significant non-U.S. presence, we assume that 50\% of its data center capacity is located in the United States. The operator reports global water withdrawal for evaporative cooling, which is treated as total data center water withdrawal in this study. Applying the 50\% allocation factor yields U.S. data center water withdrawal of approximately 48~ML. U.S. data center water consumption is then estimated by applying 
the default water consumptive ratio of 0.75, resulting in approximately 36~ML.
The operator also reports total purchased electricity at the global level, which is treated as total global data center electricity consumption, together with a reported PUE of 1.33. Applying the same 50\% allocation factor yields U.S. data center electricity consumption of approximately 0.8~TWh. IT energy is then derived as total electricity divided by PUE, resulting in approximately 0.6~TWh.
Based on the estimated U.S. data center water consumption and IT energy, WUE is computed as water consumption divided by IT energy, yielding approximately 0.06~L/kWh.

\textbf{Data center type-specific WUE.} 
We assign the 2024 U.S.-level hyperscale WUE as the IT energy–weighted average across the five hyperscale operators (approximately 0.55), and the 2024 U.S.-level colocation WUE as the IT energy–weighted average across the seven colocation operators (approximately 0.65); the WUE for others is set to 0. 
Note that while IT energy is estimated for some companies and may differ from the actual, undisclosed values, such differences affect only the weights in our aggregation, and our methodology of aggregating multiple companies provides a reasonably robust reflection of the industry's prevailing WUE.

\subsubsection{The Years of 2025 to 2030}

We obtain WUE trajectories under three different scenarios. For our Baseline, Moderate, and Optimistic scenarios, WUE is differentiated by data center type. 
Starting from the 2024 values, WUE varies with compound annual  rates of 0\%, -5\%, and -10\% under the Baseline, Moderate, and Optimistic scenarios, respectively. The data center type-specific IT energy values in our calculations are reported in Table~\ref{tab:it_energy}. Accordingly, water consumption is first computed separately for each data center type and then aggregated to obtain national totals. 

\begin{table*}[!t]
\centering
\caption{Annual water consumption (million gallons) under Reference, Baseline, Moderate, and Optimistic scenarios, 2024--2030. Mid values are defined as the arithmetic average of the corresponding low and high estimates.}
\label{tab:annual_water_consumption}
\scriptsize
\renewcommand{\arraystretch}{1.1}
\begin{tabular}{c |c|cc|ccc|ccc|ccc}
\toprule
\multirow{2}{*}{\textbf{Year}} & \multirow{2}{*}{\textbf{Growth}}
& \multicolumn{2}{c|}{\textbf{Reference}}
& \multicolumn{3}{c|}{\textbf{Baseline}}
& \multicolumn{3}{c|}{\textbf{Moderate}}
& \multicolumn{3}{c}{\textbf{Optimistic}} \\
\cline{3-13}
& & \textbf{LBNL} & \textbf{NS}
& \textbf{Total} & \textbf{Hyper} & \textbf{Colo}
& \textbf{Total} & \textbf{Hyper} & \textbf{Colo}
& \textbf{Total} & \textbf{Hyper} & \textbf{Colo} \\
\hline
\multirow{3}{*}{2024} & Low  & 14,191 & 58,323 & 19,351 & 8,647 & 10,704 & 19,351 & 8,647 & 10,704 & 19,351 & 8,647 & 10,704 \\
                      & \textbf{Mid}  & \textbf{15,453} & \textbf{62,643} & \textbf{20,984} & \textbf{9,615} & \textbf{11,369} & \textbf{20,984} & \textbf{9,615} & \textbf{11,369} & \textbf{20,984} & \textbf{9,615} & \textbf{11,369} \\
                      & High & 16,716 & 66,964 & 22,618 & 10,583 & 12,035 & 22,618 & 10,583 & 12,035 & 22,618 & 10,583 & 12,035 \\
\hline
\multirow{3}{*}{2025} & Low  & 17,489 & 68,951 & 23,606 & 10,469 & 13,137 & 22,426 & 9,946 & 12,480 & 21,245 & 9,422 & 11,823 \\
                      & \textbf{Mid}  & \textbf{19,946} & \textbf{76,742} & \textbf{26,581} & \textbf{12,066} & \textbf{14,515} & \textbf{25,252} & \textbf{11,463} & \textbf{13,789} & \textbf{23,923} & \textbf{10,859} & \textbf{13,063} \\
                      & High & 22,404 & 84,534 & 29,556 & 13,663 & 15,893 & 28,078 & 12,979 & 15,099 & 26,600 & 12,296 & 14,304 \\
\hline
\multirow{3}{*}{2026} & Low  & 21,797 & 81,964 & 28,772 & 12,664 & 16,108 & 25,967 & 11,429 & 14,537 & 23,305 & 10,258 & 13,047 \\
                      & \textbf{Mid}  & \textbf{26,045} & \textbf{94,860} & \textbf{33,706} & \textbf{15,154} & \textbf{18,552} & \textbf{30,420} & \textbf{13,677} & \textbf{16,743} & \textbf{27,302} & \textbf{12,275} & \textbf{15,027} \\
                      & High & 30,294 & 107,756 & 38,640 & 17,644 & 20,996 & 34,873 & 15,924 & 18,949 & 31,299 & 14,292 & 17,007 \\
\hline
\multirow{3}{*}{2027} & Low  & 27,090 & 97,781 & 35,018 & 15,296 & 19,722 & 30,024 & 13,115 & 16,909 & 25,528 & 11,151 & 14,377 \\
                      & \textbf{Mid}  & \textbf{33,890} & \textbf{118,053} & \textbf{42,777} & \textbf{19,045} & \textbf{23,733} & \textbf{36,676} & \textbf{16,328} & \textbf{20,348} & \textbf{31,185} & \textbf{13,884} & \textbf{17,301} \\
                      & High & 40,690 & 138,326 & 50,537 & 22,793 & 27,744 & 43,329 & 19,542 & 23,787 & 36,841 & 16,616 & 20,225 \\
\hline
\multirow{3}{*}{2028} & Low  & 33,433 & 117,016 & 42,590 & 18,462 & 24,128 & 34,689 & 15,037 & 19,652 & 27,943 & 12,113 & 15,830 \\
                      & \textbf{Mid}  & \textbf{43,932} & \textbf{147,809} & \textbf{54,381} & \textbf{23,968} & \textbf{30,413} & \textbf{44,293} & \textbf{19,522} & \textbf{24,772} & \textbf{35,679} & \textbf{15,725} & \textbf{19,954} \\
                      & High & 54,431 & 178,601 & 66,172 & 29,474 & 36,698 & 53,897 & 24,006 & 29,891 & 43,415 & 19,338 & 24,078 \\
\hline
\multirow{3}{*}{2029} & Low  & 41,237 & 140,939 & 51,980 & 22,359 & 29,621 & 40,221 & 17,301 & 22,920 & 30,694 & 13,203 & 17,491 \\
                      & \textbf{Mid}  & \textbf{56,806} & \textbf{186,576} & \textbf{69,476} & \textbf{30,306} & \textbf{39,171} & \textbf{53,759} & \textbf{23,450} & \textbf{30,309} & \textbf{41,025} & \textbf{17,895} & \textbf{23,130} \\
                      & High & 72,374 & 232,213 & 86,972 & 38,252 & 48,720 & 67,297 & 29,599 & 37,699 & 51,356 & 22,588 & 28,769 \\
\hline
\multirow{3}{*}{2030} & Low  & 51,016 & 170,280 & 63,502 & 27,104 & 36,398 & 46,679 & 19,924 & 26,756 & 33,747 & 14,404 & 19,343 \\
                      & \textbf{Mid}  & \textbf{73,808} & \textbf{236,685} & \textbf{89,009} & \textbf{38,416} & \textbf{50,593} & \textbf{65,430} & \textbf{28,239} & \textbf{37,190} & \textbf{47,303} & \textbf{20,416} & \textbf{26,887} \\
                      & High & 96,601 & 303,090 & 114,516 & 49,728 & 64,788 & 84,180 & 36,555 & 47,625 & 60,858 & 26,427 & 34,431 \\
\bottomrule
\end{tabular}
\end{table*}

The Reference (LBNL and NS) scenarios focus on WUE at the aggregate national level. To enable a consistent and direct comparison, we therefore also compute a total U.S.-average WUE for our three scenarios by aggregating across the data center types. For each year, the total water consumption is calculated as the sum of type-specific IT energy multiplied by the corresponding WUE values, and the U.S.-average WUE is then obtained by dividing this aggregate water consumption by total IT energy.
By construction, this total WUE is an IT energy–weighted average across data center types. Because IT energy differs under the low and high assumptions, total WUE is calculated separately for the two cases, yielding corresponding low and high total WUE values.

\begin{table*}[!t]
\centering
\caption{Annual water withdrawal (million gallons) under Reference, Baseline, Moderate, and Optimistic scenarios, 2024--2030. Mid values are defined as the arithmetic average of the corresponding low and high estimates.}
\label{tab:annual_water_withdrawal}
\scriptsize
\renewcommand{\arraystretch}{1.1}
\begin{tabular}{c |c|cc|ccc|ccc|ccc}
\toprule
\multirow{2}{*}{\textbf{Year}} & \multirow{2}{*}{\textbf{Growth}}
& \multicolumn{2}{c|}{\textbf{Reference}}
& \multicolumn{3}{c|}{\textbf{Baseline}}
& \multicolumn{3}{c|}{\textbf{Moderate}}
& \multicolumn{3}{c}{\textbf{Optimistic}} \\
\cline{3-13}
& & \textbf{LBNL} & \textbf{NS}
& \textbf{Total} & \textbf{Hyper} & \textbf{Colo}
& \textbf{Total} & \textbf{Hyper} & \textbf{Colo}
& \textbf{Total} & \textbf{Hyper} & \textbf{Colo} \\
\hline
\multirow{3}{*}{2024} & Low  & 18,287 & 75,160 & 24,795 & 11,195 & 13,600 & 24,795 & 11,195 & 13,600 & 24,795 & 11,195 & 13,600 \\
                      & \textbf{Mid}  & \textbf{19,910} & \textbf{80,711} & \textbf{26,894} & \textbf{12,448} & \textbf{14,446} & \textbf{26,894} & \textbf{12,448} & \textbf{14,446} & \textbf{26,894} & \textbf{12,448} & \textbf{14,446} \\
                      & High & 21,534 & 86,262 & 28,993 & 13,702 & 15,292 & 28,993 & 13,702 & 15,292 & 28,993 & 13,702 & 15,292 \\
\hline
\multirow{3}{*}{2025} & Low  & 22,513 & 88,758 & 30,246 & 13,554 & 16,691 & 28,734 & 12,877 & 15,857 & 27,221 & 12,199 & 15,022 \\
                      & \textbf{Mid}  & \textbf{25,668} & \textbf{98,758} & \textbf{34,064} & \textbf{15,622} & \textbf{18,443} & \textbf{32,361} & \textbf{14,841} & \textbf{17,521} & \textbf{30,658} & \textbf{14,059} & \textbf{16,598} \\
                      & High & 28,824 & 108,757 & 37,883 & 17,689 & 20,194 & 35,989 & 16,804 & 19,184 & 34,094 & 15,920 & 18,175 \\
\hline
\multirow{3}{*}{2026} & Low  & 28,033 & 105,414 & 36,863 & 16,396 & 20,467 & 33,269 & 14,798 & 18,471 & 29,859 & 13,281 & 16,578 \\
                      & \textbf{Mid}  & \textbf{33,485} & \textbf{121,958} & \textbf{43,192} & \textbf{19,620} & \textbf{23,572} & \textbf{38,981} & \textbf{17,707} & \textbf{21,274} & \textbf{34,986} & \textbf{15,892} & \textbf{19,093} \\
                      & High & 38,937 & 138,502 & 49,522 & 22,844 & 26,677 & 44,693 & 20,617 & 24,076 & 40,112 & 18,504 & 21,609 \\
\hline
\multirow{3}{*}{2027} & Low  & 34,814 & 125,664 & 44,863 & 19,804 & 25,059 & 38,464 & 16,980 & 21,485 & 32,705 & 14,437 & 18,268 \\
                      & \textbf{Mid}  & \textbf{43,539} & \textbf{151,665} & \textbf{54,812} & \textbf{24,657} & \textbf{30,155} & \textbf{46,994} & \textbf{21,140} & \textbf{25,854} & \textbf{39,958} & \textbf{17,975} & \textbf{21,983} \\
                      & High & 52,263 & 177,666 & 64,761 & 29,510 & 35,251 & 55,525 & 25,301 & 30,224 & 47,211 & 21,513 & 25,698 \\
\hline
\multirow{3}{*}{2028} & Low  & 42,941 & 150,295 & 54,559 & 23,902 & 30,657 & 44,439 & 19,468 & 24,970 & 35,796 & 15,682 & 20,114 \\
                      & \textbf{Mid}  & \textbf{56,407} & \textbf{189,783} & \textbf{69,674} & \textbf{31,031} & \textbf{38,643} & \textbf{56,750} & \textbf{25,275} & \textbf{31,475} & \textbf{45,713} & \textbf{20,359} & \textbf{25,354} \\
                      & High & 69,873 & 229,270 & 84,788 & 38,159 & 46,629 & 69,060 & 31,081 & 37,979 & 55,629 & 25,036 & 30,593 \\
\hline
\multirow{3}{*}{2029} & Low  & 52,940 & 180,936 & 66,585 & 28,948 & 37,637 & 51,522 & 22,399 & 29,123 & 39,318 & 17,094 & 22,224 \\
                      & \textbf{Mid}  & \textbf{72,904} & \textbf{239,452} & \textbf{89,007} & \textbf{39,236} & \textbf{49,770} & \textbf{68,872} & \textbf{30,360} & \textbf{38,511} & \textbf{52,558} & \textbf{23,169} & \textbf{29,389} \\
                      & High & 92,868 & 297,968 & 111,429 & 49,525 & 61,904 & 86,221 & 38,321 & 47,900 & 65,798 & 29,244 & 36,554 \\
\hline
\multirow{3}{*}{2030} & Low  & 65,468 & 218,519 & 81,338 & 35,091 & 46,247 & 59,791 & 25,795 & 33,996 & 43,227 & 18,649 & 24,578 \\
                      & \textbf{Mid}  & \textbf{94,691} & \textbf{303,653} & \textbf{114,020} & \textbf{49,737} & \textbf{64,284} & \textbf{83,815} & \textbf{36,561} & \textbf{47,254} & \textbf{60,595} & \textbf{26,432} & \textbf{34,163} \\
                      & High & 123,915 & 388,787 & 146,702 & 64,382 & 82,320 & 107,840 & 47,327 & 60,513 & 77,964 & 34,215 & 43,748 \\
\bottomrule
\end{tabular}
\end{table*}

The resulting WUE trajectories from 2024 to 2030 for all the scenarios are summarized in Table~\ref{tab:wue_scenarios}. The corresponding annual water consumption values under each scenario are reported in Table~\ref{tab:annual_water_consumption}. For our Baseline, Moderate, and Optimistic scenarios, both total and type-specific (hyperscale and colocation) values are provided. Annual water consumption values are reported without decimal places. As the differences between the low and high total WUE values under the Baseline, Moderate, and Optimistic scenarios are small, all WUE values in Table~\ref{tab:wue_scenarios} are reported to three decimal places to ensure clear numerical distinction.

\subsection{Consumptive Ratio}\label{appendix:consupmtive_ratio}
Type-specific consumptive ratios are applied for the three data center types in the calculation of \add\ and \mdd. Specifically, the U.S.-level consumptive ratio for hyperscale data centers is approximately 0.77, computed as the ratio of aggregated water consumption to aggregated water withdrawal across five hyperscale operators. The corresponding ratio for colocation data centers is approximately 0.79, calculated analogously across seven colocation operators. 

The consumptive ratio for ``Others'' is set to 0.75 by default~\cite{Water_Planning_WashingtonMetro_2050Prediction_DataCenter_PeakingFactor_3point62_Report_2025}. This assumption does not affect the Baseline, Moderate, or Optimistic scenarios, because the WUE for ``Others'' is set to zero in these scenarios and therefore does not contribute to either water consumption or water withdrawal.
In contrast, under the Reference (LBNL and NS) scenarios, all three data center types contribute to water consumption because a common U.S.-wide average WUE is applied to each type. 

After obtaining U.S.-level IT energy and WUE for each data center type, the type-specific annual water consumption is calculated as the product of IT energy and WUE. Water withdrawal is then derived by dividing water consumption by the corresponding type-specific consumptive ratio. The resulting annual withdrawal values are reported in Table~\ref{tab:annual_water_withdrawal}. Since \add is defined based on water withdrawal, it is computed as follows (without distinguishing leap and non-leap years):
\begin{align}
\text{ADD}_{t}
=
\frac{1}{365}
\sum_{i \in \mathcal{D}}
\frac{E^{\mathrm{IT}}_{i, t}\cdot \mathrm{WUE}_{i, t}}{\gamma_i},
\end{align}
where $t$ denotes the year, $i$ denotes the data center type ($\mathcal{D} = \{\mathrm{hyperscale, colocation, others}\}$), $E^{\mathrm{IT}}_{i,t}$ is the annual IT energy consumption of type $i$ in year $t$, $\mathrm{WUE}_{i,t}$ is the corresponding WUE, and $\gamma_i$ is the consumptive ratio associated with data center type $i$. 
Under the Reference (LBNL and NS) scenarios, a U.S.-wide average $\mathrm{WUE}_{i,t}$ is applied across hyperscale, colocation, and others. In contrast, under the Baseline, Moderate, and Optimistic scenarios, $\mathrm{WUE}_{i,t}$ differs by data center type and is set to zero for ``others''.

\subsection{Peaking Factor}\label{appendix:peaking_factor}

To estimate \mdd, a daily peaking factor is required to convert \add
into the water capacity demand measured in terms of the maximum daily demand. In Section~\ref{sec:high_peaking}, we observe that cooling towers exhibit a relatively lower peaking factor (measured or estimated at approximately 2.0–2.5, which will be higher for planning purposes to account for the worst case \cite{Google_Water_8MGD_2MGDinitial_CountyWebsite_2026}), whereas dry cooling with evaporative assistance shows a substantially higher peaking factor due to the concentration of water use on a limited number of days each year. For one leading technology company, the daily peaking factor is estimated at approximately 6.5 (based on the measured monthly peaking factor of 4.3) in Iowa, planned at 8.0+ in Leesburg, Virginia, and planned at 30+ in Wisconsin. Likewise, a leading technology company's state-of-the-art data center under construction to host AI and core products has secured an allocation of up to 8~MGD of water capacity, resulting in a peaking factor exceeding 6.3 based on the company's existing U.S.-wide average water withdrawal intensity (see Appendix~\ref{appendix:water_allocation_leap_indiana} for details).

A recent report analyzing water utility data indicates that the measured daily peaking factor of multiple data centers within the Prince William Water service area reached 10 in 2024~\cite{Water_Planning_WashingtonMetro_2050Prediction_DataCenter_PeakingFactor_3point62_Report_2025}. When hundreds of data centers with diverse cooling system designs and operational configurations are aggregated, the overall peaking factor is moderated by multiplexing effects. Nevertheless, the weighted estimate for the  actual peaking factor (not for infrastructure
planning purposes) in Northern Virginia remains in the range of 3.5 to 3.7 under different scenarios (Table 6-5 of~\cite{Water_Planning_WashingtonMetro_2050Prediction_DataCenter_PeakingFactor_3point62_Report_2025}). 

For infrastructure planning purposes, the peaking factor will be even higher than the measured value, as an additional safety margin is applied to account for worst-case scenarios and sharp rises in IT loads and cooling demand \cite{Google_Water_8MGD_2MGDinitial_CountyWebsite_2026}. This is analogous to additional backup generator capacity (commonly ``2N,'' or doubling the capacity relative to actual power needs~\cite{Water_Planning_WashingtonMetro_2050Prediction_DataCenter_PeakingFactor_3point62_Report_2025}) and power capacity reservation (e.g., a commonly assumed average power utilization factor of 50\%, i.e., reserving twice the actual power need~\cite{DoE_DataCenter_EnergyReport_US_2024}). Thus, even assuming a 25\% additional reservation, the effective peaking factor for hundreds of data centers in Northern Virginia~\cite{Water_Planning_WashingtonMetro_2050Prediction_DataCenter_PeakingFactor_3point62_Report_2025} would be approximately 4.5.

Therefore, to provide a conservative and reasonable estimate grounded in empirical values observed from available sources, we adopt an average industry-wide peaking factor of 4.5, without further increase through 2030. The water capacity demand  measured in terms of \mdd for planning purposes is subsequently calculated under the peaking factor of 4.5 using Eq.~\eqref{eq:mdd}. 

In the future, as evaporative-assisted cooling becomes more common, it can improve the annual water efficiency while simultaneously increasing the peaking factor, since such systems ``use zero water for a majority of the year'' \cite{Meta_Lebanon_Water_AI_NoWaterMajority_100million_2026}
and typically operate only 5 to 15\% of the year \cite{Amazon_Water_400m_25_35per_PeakPowerReduction_AI_Cloud_Louisiana_PressRelease_2026}.
Thus, the actual peaking factor is likely higher than 4.5, and this variation can be partially accommodated by the different growth rates used in our analysis. As peak water use or the peaking factor is rarely reported publicly, we recommend that data centers include site-level peaking factors in their annual reports to support improved water capacity management.

\subsection{Estimated Costs for Water Infrastructure Projects}\label{appendix:cost_water_10_40_mgd}

Water infrastructure project costs vary widely depending on factors such as project complexity, geographic location, labor costs, and inflation. For example, a recent public water infrastructure upgrade to support a leading technology company's data centers cost up to \$400 million for an undisclosed water capacity~\cite{Amazon_Water_400m_25_35per_PeakPowerReduction_AI_Cloud_Louisiana_PressRelease_2026}, whereas a smaller upgrade providing approximately 0.11~MGD of drinking water capacity and 0.08~MGD of wastewater capacity to a data center reportedly cost \$5.4 million~\cite{Water_Leesburg_point11MGD_5million_StackDataCenter_Approval_LoudounNow_News_November_2024}.

\begin{table}[!t]
\scriptsize
    \centering
    \caption{Estimated cost for water projects. Projects 12--17 are specifically or partly associated with data centers.}
    \label{tab:cost_water_infra_project_survey}
    \begin{tabularx}{0.92\textwidth}{@{} c X c @{}}
        \toprule
        \textbf{Project} & \textbf{Description} & \textbf{Unit Cost (\$M/MGD)} \\ 
        \midrule
        1 & Water treatment expansion only \cite{Water_ProjectCost_UnitCostStudy_Final_CityOfPhoenix_AZ_2024} & 7.6 \\
        2 & Water treatment expansion only \cite{Water_ProjectCost_UnitCostStudy_Final_CityOfPhoenix_AZ_2024} & 6.6 \\
        3 & Advanced water treatment, wastewater treatment, concentrate management, and solids handling \cite{Water_ProjectCost_UnitCostStudy_Final_CityOfPhoenix_AZ_2024} & 86 \\
        4 & Advanced water treatment, wastewater treatment, concentrate management, and solids handling \cite{Water_ProjectCost_UnitCostStudy_Final_CityOfPhoenix_AZ_2024} & 41 \\
        5 & Advanced water treatment, wastewater treatment, concentrate management, and solids handling \cite{Water_ProjectCost_UnitCostStudy_Final_CityOfPhoenix_AZ_2024} & 23--31 \\
        6  & Advanced water treatment, wastewater treatment, concentrate management, and solids handling \cite{Water_ProjectCost_UnitCostStudy_Final_CityOfPhoenix_AZ_2024} & 23--32 \\
        7  & New water reuse/recycling facility \cite{EPA_Water_WIFIA_SanDiegoProject_2024} & 46 \\
        8  & Water treatment expansion only \cite{Water_ResourcesAnalysis_NewtonGeorgia_December_2024} & 9 \\
        9  & Reservoir and new water treatment plant \cite{Water_ResourcesAnalysis_NewtonGeorgia_December_2024} & 12 \\
        10  & Wastewater treatment construction and expansion only \cite{Water_ResourcesAnalysis_NewtonGeorgia_December_2024} & 20 \\
        11 & Wastewater treatment facility improvement and expansion \cite{Water_ExpansionProject_20million_1.5MGD_Florida_2022} & 13\\
                  \midrule
        12 & Wastewater recycling facility only \cite{xAI_WaterRecylingPlant_80million_Investment_Memphis_News_2025} & 6 \\
        13 & Water (0.64 MGD) and sewer service upgrades \cite{Microsoft_CommunityFirst_AI_Infrastructure_Blog_2026,Water_Microsoft_point64MGD_Expansion_Leesburg_FinalOfficial_2024} & 40  \\
        14  & New water source and other necessary facilities (8 MGD) \cite{Google_Water_8MGD_300million_30x_BottlingCompany_Virginia_News_2025} & 37 \\
        15  & Water and wastewater services (0.11 MGD water and 0.08 MGD wastewater) \cite{Water_Leesburg_point11MGD_5million_StackDataCenter_Approval_LoudounNow_News_November_2024} & 50 \\
        16 & Supplying 25 MGD water to an innovation district \cite{Water_Lebanon_LEAP_WaterOverview_560million_2025} & 22\\
        17 & Wastewater system improvements and expansion \cite{Water_Lebanon_LEAP_DataCenter_Meta_Final_2025} & 23\\
        \bottomrule
    \end{tabularx}
\end{table}

To derive a reasonable cost range, we review multiple recent or planned water infrastructure projects documented in the public domain, including official cost estimates from local governments and water utilities as well as publicly reported project costs. Among these, Projects 12-17 are specifically or partly associated with data center water needs. The results are shown in Table~\ref{tab:cost_water_infra_project_survey}. 
All reported costs are in 2025 dollars or earlier without inflation adjustment; consequently, actual project costs may increase in future years due to inflation.

Notably, some reported cost estimates exclude transmission and distribution components, such as pipelines, which can represent a substantial share of total costs and often exceed the cost of water or wastewater treatment facilities themselves \cite{EPA_WaterDrinking_625billion_2021_7th_Report_Congress_2023}. Based on this review and to account for additional applicable infrastructure components including pipes, pumps, and storage, we adopt a conservative unit cost range of \$10–40 million per MGD of capacity for valuing data center water infrastructure requirements in our study. 

Importantly, our estimate of cost ranges is intended to represent the scale of infrastructure investment for each MGD capacity and should \emph{not} be interpreted as the actual financial obligation of a specific data center operator, 
which may vary depending on contractual arrangements, accounting rules, local rate structures, among others.
\section{Data Center Water Use in Loudoun County, Virginia }\label{appendix:water_loudoun_hypothetical}
We use Loudoun County, Virginia, as a case study, given its high concentration of data centers. As a large public/community water system according to the EPA's classification \cite{EPA_WaterDrinking_625billion_2021_7th_Report_Congress_2023},
Loudoun Water provides service to over 81,000 households as well as many data centers \cite{Water_LoudounWater_Future90MGD_Current40MGD_Website}. In 2024, the measured peak daily water \emph{withdrawal} by data centers in the Loudoun Water service area reached 10.9 MGD, resulting in 
a peak of 2,716
gallons per megawatt of IT load per day  (or 0.428 L/kWh) consolidated over all the served data centers \cite{Water_Planning_WashingtonMetro_2050Prediction_DataCenter_PeakingFactor_3point62_Report_2025}. The 10.9 MGD of water withdrawal includes both
potable water and non-potable water, corresponding to a total IT load of approximately
4,013 MW across the Loudoun Water service area.
The relatively low peak water withdrawal reflects the current practice
that many Northern Virginia data centers predominantly
rely on dry-cooling systems, despite the fact that dry coolers are more power-intensive than evaporative cooling, particularly during the summer months. Note that the same report \cite{Water_Planning_WashingtonMetro_2050Prediction_DataCenter_PeakingFactor_3point62_Report_2025} also notes a lower peak of 2,435 gallons per megawatt on Page 6-14, which would imply a higher corresponding IT load and, consequently, a higher peak water use in the following estimate. To remain conservative, we therefore adopt the larger value of 2,716 gallons per megawatt per day. 

We consider a hypothetical scenario in which all IT loads are cooled using the more power-efficient evaporative cooling method. To improve the result robustness, we consider the following methods to estimate the corresponding total peak daily water demand.

\textbf{The estimate in \cite{Water_Planning_WashingtonMetro_2050Prediction_DataCenter_PeakingFactor_3point62_Report_2025}.}
The report \cite{Water_Planning_WashingtonMetro_2050Prediction_DataCenter_PeakingFactor_3point62_Report_2025} assumes in Table~6-5 that, under a scenario in which 
evaporative systems account for 90\% of the share, peak water \emph{withdrawal} reaches 5,200 gallons per megawatt of IT load per day. 
This corresponds to a \pwue of approximately 0.615 L/kWh, which is
compared to (or even
substantially lower) than many of the average company-wide WUE values reported by leading colocation providers and technology companies (see Table~\ref{tab:dc_metrics_2024}). 
This discrepancy may reflect the particular operational configuration and local climatic conditions, as well as the possibility that some facilities classified as using evaporative cooling in fact serve only a limited portion of their IT load with evaporative systems \cite{Water_Planning_WashingtonMetro_2050Prediction_DataCenter_PeakingFactor_3point62_Report_2025}. Consequently, the estimate of 
5,200 gallons per megawatt of IT load per day (under the assumption
of 90\% evaporative cooling)
is specific to the selected set of data centers in the study \cite{Water_Planning_WashingtonMetro_2050Prediction_DataCenter_PeakingFactor_3point62_Report_2025} and may not be representative of the current average WUE or \pwue of U.S. data centers.
Nonetheless, under the assumed setting in \cite{Water_Planning_WashingtonMetro_2050Prediction_DataCenter_PeakingFactor_3point62_Report_2025},
if all data centers were to adopt evaporative cooling, peak water withdrawal would easily exceed 5,500 gallons per megawatt of IT load per day, corresponding to a total peak withdrawal of approximately 22.1 MGD. 
In practice, public water utilities also adopt more conservative planning assumptions by incorporating higher peaking factors than actually metered peaks and reserving additional capacity to manage worst-case scenarios, such as extreme heatwaves. This can effectively drive up the total water capacity need to 30 MGD or more.

\textbf{Industry disclosures.} To complement the estimates derived based on \cite{Water_Planning_WashingtonMetro_2050Prediction_DataCenter_PeakingFactor_3point62_Report_2025}, we consider industry disclosures. A large colocation provider reports an annual average WUE of 1.55~L/kWh for facilities using evaporative cooling in 2024 \cite{Equinix_SustainabilityReport_2024}. In comparison, a large technology company operating multiple data centers in the United States with a mix of water-intensive evaporative cooling and  non-evaporative systems reports an estimated annual WUE of 1.13~L/kWh \cite{Google_SustainabilityReport_2025}. Note that excluding the four air-cooled U.S. data center sites (whose energy consumption is not disclosed) from the 23 sites in total \cite{Google_SustainabilityReport_2025} would increase the estimated annual WUE for evaporatively cooled sites to above 1.13~L/kWh, bringing it closer to the 1.55~L/kWh reported for evaporative cooling by \cite{Equinix_SustainabilityReport_2024}. 
These company-wide WUE values provide empirical U.S./global benchmarks for water efficiency in large-scale data centers using evaporative cooling. 
 To account for the relatively cooler climate of Northern Virginia, we apply the reported company-wide average WUE values directly and do not incorporate additional peaking factors or redundancy-related capacity allocations. This approach yields estimates that are plausibly conservative.
Specifically, the implied peak daily water consumption in 2024 would range from 28.8 to 39.4 million gallons. Using a consumptive use ratio of 0.75 as assumed in \cite{Water_Planning_WashingtonMetro_2050Prediction_DataCenter_PeakingFactor_3point62_Report_2025}, the corresponding peak daily water withdrawal would range from 38.3 to 52.6 million gallons. 

\textbf{Water allocation example.} 
The total water capacity allocated to a data center is often not publicly disclosed. Nonetheless, public county records \cite{Water_Microsoft_point64MGD_Expansion_Leesburg_FinalOfficial_2024,Water_Microsoft_point59MGD_Intitial_Leesburg_Consulting_2023} indicate that, as of April 2024, the Town of Leesburg, Virginia, allocates 1.23 MGD of water capacity to a large data center campus operated by a leading technology company. This campus is also approved for approximately 314 MW of total diesel generator capacity by the Virginia Department of Air Quality as of March 2025 \cite{Virginia_AirPermitsDataCenter_Microsoft_Leesburg_Website}. 

Assuming the same ``2N'' redundancy configuration and an 80\% IT load utilization factor as in \cite{Water_Planning_WashingtonMetro_2050Prediction_DataCenter_PeakingFactor_3point62_Report_2025}, a 314 MW diesel generator capacity corresponds to an effective IT load of approximately 126 MW. When compared against the 1.23 MGD water capacity allocation, this implies roughly 9,762 gallons of water withdrawal per MW of IT load per day. While future load growth without additional water capacity allocation may be feasible and could reduce the effective \pwue, the current \pwue estimate more closely reflects the planned design value, which accounts for worst case operating conditions as well as anticipated future expansion.
Given that tens of large data centers with an estimated aggregate IT load of 4,013 MW are served by Loudoun Water, the total required water capacity could approach 40 MGD if all facilities were to rely on evaporative cooling.

In summary, if all data centers within the Loudoun Water service area were to rely on evaporative cooling during the summer months, 
the aggregate peak water capacity requirement would likely fall in the range of 20--50~MGD, 
even under conservative assumptions.
In contrast, Loudoun Water reports a total drinking water design capacity of 70 MGD, with a measured peak demand of 41 MGD and an available capacity of less than 30 MGD as of 2022 \cite{Water_LoudounWater_41MGDPeak_CountyDocuments_2022}. The capacity available for new allocation is likely even smaller after accounting for operational reserves and committed but not yet fully utilized allocations.
These figures suggest that the remaining infrastructure capacity, after meeting the needs of residential, commercial, and other existing users, would likely be insufficient to accommodate the aggregate peak water demand or allocation requests of data center customers as of the end of 2024, should all data centers rely on evaporative cooling. In addition, even if the total water treatment plant capacity may be sufficient, other components of the system, including pump stations and distribution pipelines, may constitute binding constraints in meeting the peak water demand associated with evaporative cooling across all the data centers.

Accordingly, in the absence of further expansion of water treatment and distribution infrastructure, accommodating the high peak demand associated with a full transition to evaporative cooling for all the data centers would likely be difficult and necessitate alternative strategies. These may include large-scale adoption of reclaimed water (which does not reduce the underlying consumptive water use of evaporative cooling)  as well as deployment of dry cooling systems (which shift peak resource burdens from regional water systems to the electricity grid, particularly during summer periods).

\section{Water Capacity Need for Cooling 100 MW IT Loads}\label{appendix:capacity_need_100MW_IT_load}

Assuming sufficient water capacity is available, we now estimate the water capacity need for cooling a 100 MW IT load using evaporative methods. We rely on industrial disclosures
and public data sources to develop plausible estimates.

 A large colocation provider reports an annual average WUE of 1.55~L/kWh for facilities using evaporative cooling in 2024 \cite{Equinix_SustainabilityReport_2024}. In comparison, a large technology company operating multiple data centers in the United States with a mix of water-intensive evaporative cooling and non-evaporative systems reports an estimated annual WUE of 1.13~L/kWh \cite{Google_SustainabilityReport_2025}. Note that excluding the four air-cooled U.S. data center sites (whose energy consumption is not disclosed) from the 23 sites in total \cite{Google_SustainabilityReport_2025} would increase the estimated annual WUE for evaporatively cooled sites to above 1.13~L/kWh, bringing it closer to the 1.55~L/kWh reported for evaporative cooling by \cite{Equinix_SustainabilityReport_2024}. 
These company-wide WUE values provide empirical U.S./global benchmarks for water efficiency in large-scale data centers using evaporative cooling (mostly cooling towers). 

The measured daily peaking factor  (or estimated based on monthly values) for data center cooling towers 
is approximately 2.2 (Section~\ref{sec:high_peaking}). 
For planning purposes, we adopt a conservative average WUE of 
1.2~L/kWh and assume a peaking factor of 2.5 to reflect 
 operational safety margins. 
This implies a peak daily \pwue of 3.0~L/kWh.
Assuming a consumptive ratio of 0.75, consistent with 
\cite{Water_Planning_WashingtonMetro_2050Prediction_DataCenter_PeakingFactor_3point62_Report_2025}, 
an evaporative cooling system supporting a 100~MW IT load 
would require a peak water withdrawal capacity of approximately 
2.5~MGD. In hotter climates, this requirement could be 
substantially higher, even exceeding 5.0 MGD. For example, measured monthly peak WUE 
values exceeding 9~L/kWh have been reported for data centers 
in Arizona \cite{Water_DataCenterEnergy_Tradeoff_Arizona_Real_Measurement_WUE_Monthly_2022_KARIMI2022106194}.

Next, we examine dry cooling with evaporative assistance. 
While this approach significantly reduces annual WUE compared to conventional cooling towers, 
peak water demand can still be substantial. 
 As a real example, consider a leading technology company's data center campus in  Iowa. 
In 2022, the reported annual WUE was 0.19~L/kWh,  \cite{Microsoft_DataCenter_019WUE_2022_Iowa}. 
The peak daily WUE (\pwue) can be estimated by multiplying the monthly peaking factor of approximately 3 in 2022 (as shown in Figure~\ref{fig:microsoft_monthly_water}) 
by a minimum adjustment factor of 1.5 based on regulatory guidance \cite{Water_DrinkingWaterRegulation_California_File_2025}, 
yielding a \pwue of 0.855~L/kWh. This is substantially higher than the annual average. 
Applying the default consumptive ratio of 0.75 \cite{Water_Planning_WashingtonMetro_2050Prediction_DataCenter_PeakingFactor_3point62_Report_2025} 
results in a water withdrawal intensity of 1.14~L/kWh.
Thus,
for this Iowa data center campus, cooling a 100~MW IT load requires approximately 0.72~MGD, which also needs to be further adjusted upward to approximately 1.0 MGD for planning to maintain operational safety margins in practice.

The same technology company also operates a large data center campus in Leesburg, Virginia, 
where it invested \$25~million to upgrade local water treatment and sewage infrastructure, 
ensuring that the associated costs are not passed on to local ratepayers 
\cite{Microsoft_CommunityFirst_AI_Infrastructure_Blog_2026}.
Public county records \cite{Water_Microsoft_point64MGD_Expansion_Leesburg_FinalOfficial_2024,Water_Microsoft_point59MGD_Intitial_Leesburg_Consulting_2023} indicate that, as of April 2024, the Town of Leesburg, Virginia, allocates 1.23 MGD of water capacity to the data center campus. This campus is also approved for approximately 314 MW of total diesel generator capacity by the Virginia Department of Air Quality as of March 2025 \cite{Virginia_AirPermitsDataCenter_Microsoft_Leesburg_Website}. 
Following the study \cite{Water_Planning_WashingtonMetro_2050Prediction_DataCenter_PeakingFactor_3point62_Report_2025} to apply the same ``2N'' redundancy configuration and an 80\% IT load utilization factor, 
a 314 MW diesel generator capacity corresponds to an effective IT load of approximately 126 MW. When compared against the 1.23 MGD water capacity allocation, this implies roughly 9,762 gallons of water withdrawal per MW of IT load per day, corresponding
to a peak water withdrawal intensity of 1.540 L/kWh.
Thus, cooling a 100 MW IT load can need a water capacity of approximately 0.98 MGD.
 With a consumptive ratio of 0.75 as considered in \cite{Water_Planning_WashingtonMetro_2050Prediction_DataCenter_PeakingFactor_3point62_Report_2025}, the peak water withdrawal intensity  corresponds
to a \pwue of 1.155 L/kWh. By comparison, the reported annual WUE for the company's data center fleet across Northern Virginia is 0.14~L/kWh as of 2023 (which may be lower due to efficiency improvements in 2025)~\cite{Shaolei_Water_AI_Thirsty_CACM}. This comparison suggests that the peaking factor for the Leesburg data center campus is approximately 8 or higher.
While future load growth without additional water capacity allocation may be feasible and could reduce the effective \pwue, the current \pwue estimate more closely reflects the planned design value, which accounts for worst case operating conditions as well as anticipated future expansion.
In colder climates, the peaking factor can even be substantially higher (e.g., reaching 30 \cite{AI_Water_Microsoft_PeakingFactor_30_MountPleasant_Wisconsin_News_WPR_2025}), because evaporative cooling is utilized only on a limited number of days each year.

We now consider company-wide average WUE values measured and reported by leading data centers that primarily employ evaporative assistance. 
These values typically range from 0.1 to 0.2~L/kWh (Table~\ref{tab:dc_metrics_2024}). 
Assuming 0.15 L/kWh and accounting for a peaking factor of 6.5 (Figure~\ref{fig:microsoft_monthly_peakingfactor}), the peak daily WUE (\pwue) can be roughly 1.0~L/kWh, 
which corresponds to a peak water withdrawal intensity of approximately 1.3~L/kWh under the default consumptive ratio of 0.75 
\cite{Water_Planning_WashingtonMetro_2050Prediction_DataCenter_PeakingFactor_3point62_Report_2025}. 
For planning purposes, a higher value is appropriate, suggesting that a 1~MGD water capacity allocation is a reasonable estimate for cooling a 100~MW IT load.

While the benefits of water evaporative cooling remain (e.g., a recent \$400 million investment to upgrade public water infrastructure to support
data centers hosting AI and cloud services \cite{Amazon_Water_400m_25_35per_PeakPowerReduction_AI_Cloud_Louisiana_PressRelease_2026}),
the adoption of liquid-cooled servers can tolerate higher server-level temperature setpoints, allowing air-cooled heat rejection to be used for a greater portion of the year. 
This can further reduce total, and potentially peak, water usage. 
To account for this effect, as well as regional climate variations, 
we apply a 50\% reduction to the water capacity need of 1.0~MGD 
for evaporatively cooling a 100~MW IT load, 
based on the current planning data from the leading technology company. 
This yields an adjusted water capacity need of roughly 0.5~MGD.

In summary, cooling a 100~MW IT load with evaporative cooling typically requires 
approximately 0.5--2.5~MGD of water capacity, depending on the cooling system design 
and operational configuration, while excluding extreme climate conditions.

\section{Water Allocation in an Innovation District in Boone County, Indiana}\label{appendix:water_allocation_leap_indiana}

The Limitless Exploration/Advanced Pace (LEAP) Lebanon Innovation District in Boone County, Indiana, is a state-led economic development initiative designed to attract high-tech, advanced manufacturing, and research businesses. To support anticipated water demand, the local water utility is upgrading infrastructure to supply up to an additional 25~MGD of potable water capacity and expand wastewater treatment capacity by an additional 15~MGD. These upgrades are estimated to cost approximately \$1~billion, including \$560~million (financed through Indiana’s Drinking Water State Revolving Fund) for new water supply, \cite{Water_Lebanon_LEAP_WaterOverview_560million_2025},
\$225~million for upgrades to existing infrastructure of the host town (Lebanon, Indiana), and \$350~million for wastewater facility expansion \cite{Water_Lebanon_LEAP_DataCenter_Meta_Final_2025}. The full 25~MGD is expected to be delivered by 2031.

We review disclosed water allocation agreements and pre-agreements across several projects spanning different categories to further examine how data centers use water differently. Unless otherwise specified, all water allocation values correspond to the projected full buildout capacity.

\begin{itemize}
    \item Project 1 (Hospital) \cite{Water_Lebanon_LEAP_Hospital_Parkview}: Peak water supply capacity of 34{,}200~gallons per day (GPD) and average demand of 11{,}400~GPD; peak wastewater capacity of 27{,}000~GPD and average flow of 7{,}000~GPD. The peaking factors for water and wastewater are 3.00 and 3.86, respectively.

    \item Project 2 (Housing) \cite{Water_Lebanon_LEAP_Housing_ProjectSpringCreek_Housing}: Water and wastewater allocations are not directly disclosed, except for 387 equivalent dwelling units (EDUs) and a peaking factor of 1.49. In the innovation district, each EDU corresponds to an average water demand of 500~GPD \cite{Water_Lebanon_LEAP_NewCold_Industry}, resulting in an estimated average demand of 193,500~GPD and peak capacity of approximately 288,315~GGD for water supply.

    \item Project 3 (Industry) \cite{Water_Lebanon_LEAP_NewCold_Industry}: Peak water supply capacity of 135{,}000~GPD and average demand of 100{,}000 GPD; peak wastewater capacity of 40{,}000~GPD and average flow of 33{,}500~GPD. The peaking factors for water and wastewater are 1.35 and 1.19, respectively.

    \item Project 4 (Pharmaceutical) \cite{Water_Lebanon_LEAP_Pharmaceutical_EliLilly_LP1X}: Peak water supply capacity of 1{,}675{,}529~GPD and average demand of 1{,}581{,}933~GPD; peak wastewater capacity of 1{,}350{,}000~GPD and average flow of 864{,}000~GPD. The peaking factors for water and wastewater are 1.06 and 1.56, respectively.

    \item Project 5 (Pharmaceutical) \cite{Water_Lebanon_LEAP_Pharmaceutical_EliLilly_MedicineFoundry}: Peak water supply capacity of 749{,}970~GPD and average demand of 513{,}482~GPD; peak wastewater capacity of 290{,}970~GPD and average flow of 205{,}232~GPD. The peaking factors for water and wastewater are 1.46 and 1.42, respectively.

    \item Project 6 (Data Center) \cite{Water_Lebanon_LEAP_DataCenter_Meta_Final_2025}: Peak water supply capacity of 8{,}000{,}000~GPD and average demand of 4{,}000{,}000~GPD; peak wastewater capacity of 4{,}000{,}000~GPD and average flow of 2{,}000{,}000~GPD. This is a final executed agreement.  The peaking factors for both water and wastewater are 2.00 based on the disclosed average and peak water use. However, the operator's existing U.S. data centers (15 different locations) report an average water withdrawal intensity of 0.20~L/kWh as of 2024 \cite{meta_environmental_data_index}. Considering that the planned ``state-of-the-art data center'' located in the Innovation District has a capacity of 1~GW and will use a closed-loop liquid cooling system that recirculates water and ``will use zero water for a majority of the year'' \cite{Meta_Lebanon_Water_AI_NoWaterMajority_100million_2026}, it is highly likely that the new facility will achieve an average water withdrawal intensity below 0.20~L/kWh. This would correspond to an average water demand of no more than 1.27~MGD, even under the assumption that all the full 1~GW capacity corresponds to IT load. Given the peak water capacity of 8~MGD, this implies a planned daily peaking factor of at least 6.31, consistent with other large technology operators using similar cooling systems (Appendix~\ref{appendix:peaking_factor}). Therefore, the specified average water demand in the agreement likely serves primarily as an administrative placeholder with limited relevance to the data center's actual operational needs, ensuring that sufficient withdrawal volumes remain authorized if the facility's design or cooling configuration changes in the future. Nonetheless, although peak water capacity allocations are also often requested above expected operating levels to hedge against worst-case scenarios such as extreme heatwaves or unanticipated IT load growth, securing such peak capacity is costly and requires substantial financial commitment, and thus more closely reflects both the anticipated operational demand and necessary reliability margins. 
\end{itemize}

Importantly, based
on the operator's existing U.S. data center water efficiency,
it is highly likely that the planned peaking factor is  substantially higher than those of other typical water users, indicating more variability between average and peak demand.

Moreover,
the data center's peak water allocation alone accounts for 32\% of the new 25~MGD water capacity
and 27\% of the expanded 15~MGD wastewater capacity 
supplied to the innovation district, making it likely the single largest user of the added capacity. Considering \$560 million for the new water capacity of 25~MGD
\cite{Water_Lebanon_LEAP_WaterOverview_560million_2025} 
and \$350 million for wastewater capacity expansion of 15~MGD
\cite{Water_Lebanon_LEAP_DataCenter_Meta_Final_2025}, 
the total estimated investment for the data center's water infrastructure is approximately \$270 million. 
We exclude the anticipated \$225 million funding required to upgrade existing water infrastructure \cite{Water_Lebanon_LEAP_DataCenter_Meta_Final_2025}.

Our cost estimate is intended to represent the scale of infrastructure investment to serve the data center's water capacity need and should \emph{not} be interpreted as the actual financial obligation of the data center operator,
which may vary depending on contractual arrangements, accounting rules, local rate structures, among others.

\section{Water-Power Tradeoff for Data Center Cooling}\label{appendix:water_power_tradeoff}

Data center cooling and heat rejection strategies present a fundamental tradeoff between water use and (peak) electricity consumption, particularly during summer ambient conditions. At the facility level, dry air-cooled heat rejection systems (a.k.a., dry coolers) can avoid water use entirely but generally use more energy compared to evaporative cooling methods, which use water to reduce the temperature of the air entering the condenser or facility loop. According to industry disclosures, during peak summer conditions, evaporative-based/-assisted coolers can use 10–35\% less electricity than equivalent air-cooled systems \cite{Amazon_AI_Water_EnergyWater_Tradeoff_Video,Google_Water_10percent_Less_Energy_CliamteCounscious_Blog}.
 Notably, multiple leading technology companies have recently signed agreements with local water utilities to secure substantial water capacity (e.g., up to 8~MGD at full buildout) for cooling their newest AI and cloud data center campuses across various states \cite{Water_Lebanon_LEAP_DataCenter_Meta_Final_2025,Meta_Lebanon_Water_AI_NoWaterMajority_100million_2026,Meta_Water_Lousiana_DataCenter_News_Feb_2026,Amazon_Water_400m_25_35per_PeakPowerReduction_AI_Cloud_Louisiana_PressRelease_2026,Google_Water_8MGD_Vrginia_WaterUtilityAgreement_2026},
with some water capacity expected to be delivered in 2031 due to required infrastructure upgrades \cite{Water_Lebanon_LEAP_DataCenter_Meta_Final_2025}.

The basic thermodynamic distinction that underlies facility‑level cooling choices is that evaporative cooling leverages the ambient wet‑bulb temperature, whereas dry cooling leverages the ambient dry‑bulb temperature. The dry‑bulb temperature is the routine measure of air temperature and does not depend on humidity. By contrast, the wet‑bulb temperature reflects the lowest temperature that air can reach by evaporative cooling, and is always less than or equal to the dry‑bulb temperature except at saturation (100\%  relative humidity) where they coincide. This is because evaporation removes sensible heat from the air until equilibrium is reached. 

For a given target temperature setpoint required by server‑level cooling, the cooling energy scales with the difference between that setpoint and the ambient reference temperature. When employing purely dry cooling, the system can reject heat only down toward the dry‑bulb temperature; therefore, dry cooling is sufficient to maintain a facility loop (e.g., 29\degree C for air-cooled servers or around 40\degree C for liquid-cooled servers) without mechanical or evaporative assistance only when the dry‑bulb temperature is a few degrees lower. In contrast, evaporative cooling can achieve effective heat rejection even when the dry‑bulb temperature exceeds the loop setpoint, provided the ambient wet‑bulb temperature is a few degrees lower than the setpoint. This evaporative supplement enables cooling at conditions where the dry‑bulb temperature alone would be too high without other mechanically-/energy-intensive cooling methods. 

Consequently, adjusting the facility-loop temperature setpoint can impact
the water-power tradeoff: higher setpoints can decrease the energy savings achievable through evaporation, while lower setpoints increase power reduction benefits of evaporative cooling but require more water. Effectively, the optimal balance between water and electricity depends on site-specific constraints, including local water infrastructure capacity, electricity cost, and peak power capacity.

The introduction of liquid cooling for high-density AI servers at the server level further complicates the design. Liquid-cooled servers allow the facility loop to operate at a higher supply temperature setpoint while still maintaining safe component temperatures. This higher loop setpoint enables a greater fraction of the year to be handled by dry coolers without additional water usage. However, even in these scenarios, evaporative cooling continues to provide energy benefits: by lowering the effective temperature seen by the dry cooler, evaporative assistance reduces pump and fan power (which typically changes in a cubic manner with the fan speed), particularly during peak summer conditions in which the PUE for data centers with dry coolers can be significantly higher than the annualized average PUE \cite{PUE_Peak_1point27_Microsoft_SanJose_Project_2025,PUE_Peak_1point3_GSOBGF_Project_2024}. Consequently, evaporative cooling remains a valuable tool for mitigating peak electricity demand, even when higher loop temperatures are permissible with liquid cooling \cite{Microsoft_Water_Design_EnergyTradeoff_2024,Equinix_WaterEnergy_Tradeoff_split_20_80_Blog_2025}. In other words, dry coolers supplemented with evaporative cooling effectively operate as if the ambient air were cooler than the actual summer conditions. This benefit is especially significant in hotter climates. 
For example, a leading technology company' state-of-the-art data center under construction to host AI and core products 
employs a ``closed-loop, liquid-cooled system that recirculates the same water'' \cite{Meta_Lebanon_Water_AI_NoWaterMajority_100million_2026},
but still needs up to 8~MGD for facility-level (evaporative) cooling assistance during the hottest
days of the year~\cite{Water_Lebanon_LEAP_DataCenter_Meta_Final_2025}.

While the high power density of AI server racks often necessitates liquid cooling 
(e.g., direct-to-chip and immersion cooling) because of its superior heat removal capability \cite{Water_HybridCooling_LiquidCooling_NREL_TechReport_2018_sickinger2018thermosyphon,Equinix_WaterEnergy_Tradeoff_YouTube_2025},  they still have approximately 20\% or more loads cooled
by air that have lower temperature setpoint than liquid-cooled IT loads \cite{Equinix_WaterEnergy_Tradeoff_split_20_80_Blog_2025,Microsoft_Water_10Percent_AI_DataCenter_Blog_2025}. This means that evaporative assistance
can have significant benefits of peak power reduction for the air-cooled portion of IT loads. For example, to balance the peak power demand  and the host community's limited available water capacity, a large data center
uses dry coolers for its liquid-cooled AI servers while evaporative assitance for other servers \cite{Microsoft_Water_Design_EnergyTradeoff_2024}.
Indeed, a leading technology company acknowledges that ``water is the most efficient means of cooling in many places'' \cite{Google_SustainabilityReport_2025}.

It is important to note that batteries in data centers are typically deployed to support IT loads and provide uninterruptible power during outages before backup generators are activated, but using large-scale energy storage to directly reduce facility-level cooling power remains both technically challenging and economically costly relative to the achievable peak power reduction. As a result, data centers generally do not rely on battery-based solutions for cooling power reduction. Instead, many facilities employ evaporative cooling assistance, particularly during hot periods, to reduce peak energy consumption for heat rejection. Nonetheless, the limited availability of public water capacity is increasingly emerging as a binding—and often under-recognized—constraint \cite{Water_DataCenter_FAQ_NoSurplus_SRBC_2025}, which can play a more decisive role in shaping the future water–power tradeoff for data center cooling.

\section{Additional Recommendations}\label{appendix:recommendation_additional}

We provide additional recommendations to complement those in Section~\ref{sec:recommendation_discussion} for comprehensively addressing the growing water demand of data centers.

\paragraph{Additional Recommendation 1: Waste Heat Recovery.} 
Waste heat from data centers represents a largely untapped resource that can be leveraged to improve overall energy and water efficiency, particularly as liquid-cooled AI servers increasingly operate at elevated temperatures exceeding 40\degree C. By capturing and repurposing this excess thermal energy, data centers can supply heating to buildings, district heating networks, or industrial processes, thereby reducing dependence on additional energy- and water-intensive cooling systems~\cite{Shaolei_CloudHeat_Datacenter_WasteHeatRecovery_TOMPECS_2018_10.1145/3199675}. Although significant engineering challenges exist---such as the relatively low ``quality'' of the heat and the spatial separation from heat demand centers---waste heat recovery remains a promising strategy. It can not only deliver tangible benefits to surrounding communities but also help lower the water demand of data center operations.

\paragraph{Additional Recommendation 2: Water-Aware Computing.} Water storage tanks can serve as a buffer to smooth hourly water demand peaks. However, it is challenging to use storage to fully offset multi-day or even month-long peak water demand during prolonged heatwaves, as this requires costly 
large-scale tanks, sanitation, maintenance, and other operational considerations. Data center workloads often exhibit substantial scheduling flexibility \cite{DataCenter_FlexibilityInitiative_EPRI_WhitePaper_2024,Google_CarbonAwareComputing_PowerSystems_2023_9770383}, which, when combined with multi-region site deployment, may effectively reduce peak water demand. Therefore, ``water-aware'' computing in combination with water tanks presents a promising approach to lowering peak water demand and strengthening public water system resilience.

\paragraph{Additional Recommendation 3: Water-AI Nexus.}
Technologies such as advanced AI have the potential to substantially enhance the operation, efficiency, and resilience of public water systems \cite{Shaolei_WaterInfrastructure_Decarbonizing_eEnergy_2025}. For instance, AI-driven leak detection and predictive maintenance can identify losses in real time, reducing water waste and lowering operational costs~\cite{Water_AI_LeakDetection_ACM_Trans_2025_10.1145/3729431}. These interventions are especially important given that water losses in distribution networks constitute a significant inefficiency, analogous to transmission and distribution losses in U.S. power grids, which  exceeded the total energy consumption of all U.S. data centers in 2023~\cite{Electricity_TransmissionDistributionLoss_5percent_EIA_2023,DoE_DataCenter_EnergyReport_US_2024}. By mitigating such losses, AI can effectively increase available water supply without new infrastructure investment. We therefore recommend development and deployment of AI and related technologies to assist public water systems in managing increasingly complex infrastructure networks, improving resilience, and simultaneously accommodating the growing water demands of data centers.

\section{Economic Potential of Evaporative Cooling for Peak Power Reduction}\label{appendix:water_power_cost_comparison}

We now perform an order-of-magnitude comparison to highlight the potential benefits of evaporative cooling as an economically competitive approach for shaving peak power demand.

Assuming an average power capacity utilization of 50\% \cite{DoE_DataCenter_EnergyReport_US_2024} 
and a peak PUE reduction of 0.15 as a \emph{reference} case, using evaporative cooling for a 100~MW IT load can effectively reduce peak power demand by 30~MW, 
while requiring an estimated water capacity of 0.5--2.5~MGD (Appendix~\ref{appendix:capacity_need_100MW_IT_load}). 
Based on the valuation of \$10--40~million per MGD capacity (Appendix~\ref{appendix:cost_water_10_40_mgd}),
we assume the mid valuation at \$25 million per MGD,
yielding a total cost range of \$12.5--62.5~million of water capacity to cool a 100 MW IT load. 
By comparison, installing power generators to supply 30~MW, excluding long-distance transmission costs, 
would cost approximately \$37.2~million in the U.S. South and \$70.9~million in the U.S. Northeast, 
according to the most recent 2023 capacity-weighted construction cost estimates from the U.S. Energy Information Administration \cite{EIA_ElectricityGeneratorConstructionCost_Data2023_Published2025}. 

Although it does not account for other factors such as environmental reviews and construction lead times, this simplified orders-of-magnitude comparison highlights the potential of evaporative cooling as an economically competitive approach to shaving peak power demand,
emphasizing the need for coordinated water-power planning to support growing data center loads.
It also aligns with a leading technology company's statement that ``water is the most efficient means of cooling in many places'' \cite{Google_SustainabilityReport_2025}, and that some data centers 
need up to 
8 MGD \cite{Water_Lebanon_LEAP_DataCenter_Meta_Final_2025,Google_Water_8MGD_Vrginia_WaterUtilityAgreement_2026} of water capacity (for peak power shaving) with capital costs for the necessary water infrastructure upgrades reaching hundreds of millions of dollars \cite{Google_Water_8MGD_300million_30x_BottlingCompany_Virginia_News_2025,Amazon_Water_400m_25_35per_PeakPowerReduction_AI_Cloud_Louisiana_PressRelease_2026,Water_Lebanon_LEAP_DataCenter_Meta_Final_2025}.
Nonetheless, overall water availability at regional watersheds remains important, and actual peak power reductions can vary depending on local climate conditions and cooling system design. 

\section{Water Footprint of Animal Products}\label{appendix:animal_water}

The study \cite{Water_Beef_1point1_0point8_Municipal_UNESCO_2010_Mekonnen2010} offers a comprehensive assessment of the water footprint associated with farm animals and their products. To make our study self-contained and for clarity, we summarize the key findings in \cite{Water_Beef_1point1_0point8_Municipal_UNESCO_2010_Mekonnen2010} related to animal  products.

In hydrological accounting, the total water footprint is divided into three distinct categories: green water, which is the volume of rainwater consumed by crops that is stored in the soil; blue water, which refers to surface and groundwater sourced from aquifers and rivers for 
agricultural irrigation, livestock consumption, and industrial processes,
including but not limited to municipal public water supply; and grey water, which represents the volume of freshwater required to dilute pollutants, such as fertilizer runoff, to meet water quality standards. 

The total water footprint of global animal production
was roughly 2,400~Gm$^3$ of water annually
during the period 1996-2005. The vast majority of the water footprint is green water (rainfall stored in soil), with smaller portions coming from blue (surface and groundwater) and grey (polluted) water sources. Beef production alone accounts for about one-third of the total water footprint, while dairy production contributes 19\%. The overwhelming portion of the water, i.e., approximately 98\%, is the indirect water used to grow feed for the animals, whereas direct water use make up only a small fraction of the total. More specifically,
drinking water for the animals, service water and feed mixing water account only for 1.1\%, 0.8\% and 0.03\%, respectively. 

Even assuming that all the direct water use originates from public or municipal water systems, it accounts for less than 2\% of the total water footprint. Therefore, comparing the total water footprint of animal products with data centers' Scope~1 municipal water use is uninformative and obscures the actual challenges faced by public water systems.

\section{Water Use of U.S. Golf Courses}\label{appendix:golf_course_water}

\begin{figure}[!t]
	\centering
	\includegraphics[trim = 0cm 0cm 0cm 0cm, clip, width=0.5\linewidth]{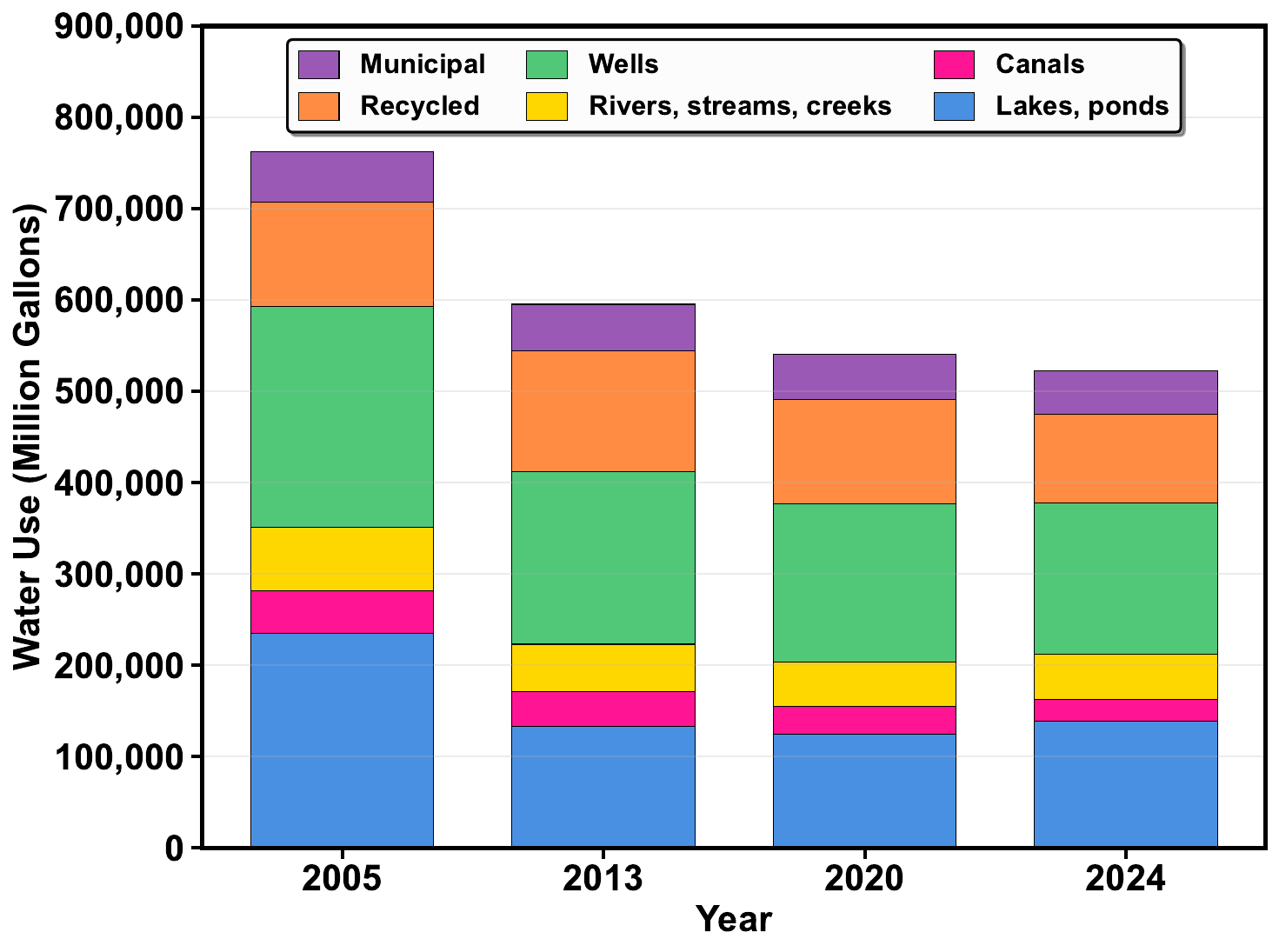}
	\caption{Statistics of the total water use of U.S. golf courses \cite{Water_GolfCourse_Statistics_US_2025_Shaddox-2025-horttech-35.5}}
	\label{fig:golf_absolute}
\end{figure}

We summarize the results of a national water use survey of U.S. golf courses reported in \cite{Water_GolfCourse_Statistics_US_2025_Shaddox-2025-horttech-35.5}. Although golf courses represent a substantial water user in aggregate when measured by total volumetric withdrawals, the reliance on municipal drinking water is relatively limited, representing approximately 7-9\% of their total applied water nationwide between 2005 and 2024. The limited use of municipal water is partly attributable to state and local regulations that restrict or discourage the use of potable water for landscape irrigation, particularly in water-stressed regions \cite{EPA_Water_California_Landscaping_Reuse_Website}.
In addition, total projected irrigation demand for U.S. golf courses declined by approximately 31\% nationally between 2005 and 2024, reflecting changes in course management practices and efficiency improvement.
Region-wise results and further breakdowns are provided in the original paper \cite{Water_GolfCourse_Statistics_US_2025_Shaddox-2025-horttech-35.5}. The summary of nation-level water use statistics is presented in Fig.~\ref{fig:golf_absolute}.
The numbers refer to ``applied water,'' representing all the 
water sprayed, dripped, or otherwise applied to turf and landscaped areas.

Additionally, in the context of golf course irrigation, the daily peaking factor is typically less than 3.0 \cite{Water_Golf_PeakingFactorMax3_AWWA2009}. 
Golf course irrigation is generally concentrated during nighttime or the early morning, taking place before golfers begin their rounds, despite fluctuations in seasonal water requirements.
Even when supplied by public water systems, 
a typical fully-irrigated golf course uses up to 0.5 MGD \cite{Water_DataCenter_FAQ_NoSurplus_SRBC_2025}.

\section{Capacity Planning and Management in Public Water Systems}\label{appandix:capacity_expansion}

In the United States, public water system capacity planning is governed by a combination of federal capacity development requirements under the Safe Drinking Water Act (SDWA) \cite{EPA_Water_SafeDrinkingWaterAct_Summary_Website}
and state-level implementation through permitting, financing, and utility regulation. While federal law does not prescribe explicit numerical utilization thresholds at which capacity expansion must occur, states typically operationalize capacity planning through a combination of engineering design criteria, financing rules, and legally enforceable limits on service commitments. 
Here, we use California as a representative and well-documented example \cite{Water_DrinkingWaterRegulation_California_Website}, which separates planning triggers based on actual utilization from legal limits based on allocated capacity.

\paragraph{Planning triggers based on utilization.}
Capacity expansion planning is typically initiated well before a system reaches its physical design limits. This expectation is reflected in  Drinking Water State Revolving Fund (DWSRF) policies administered by the California State Water Resources Control Board \cite{ca_dwsrf_capacity}. Specifically, under California's DWSRF rules, capacity expansions are generally limited to approximately 10\% above existing MDD. This is intended to ensure engineering reliability, maintain service to existing customers, and reduce financial and operational risk.

As a result, when observed or projected demand approaches roughly 80--90\% of existing design capacity, utilities are expected to begin formal planning activities, including demand forecasting, engineering evaluation, environmental review, and financing arrangements. The utilization range of 80--90\% serve as planning references rather than enforceable legal thresholds, and are intended to ensure that infrastructure expansion occurs proactively rather than in response to emergency conditions.

\paragraph{Allocated capacity and legal limits on new connections.}
Distinct from the actual operational utilization is the concept of \emph{allocated capacity}, which accounts for existing demand together with approved but possibly not completed service commitments. In California, once allocated capacity equals a system's demonstrated or permitted supply capacity, additional service connections may be legally restricted even if actual water usage still remains below physical limits.

Specifically, for regulated water utilities,  Section~2708 of the California Public Utilities Code empowers regulators to prohibit new or additional service connections when further commitments would impair service reliability or violate capacity constraints \cite{cpuc_2708}. Importantly, this determination is based on the system's ability to serve its committed load, not necessarily on its actual average usage.
Thus, regulatory decisions require
utilities that cannot demonstrate adequate source, treatment, storage, or distribution capacity to comply with or request a service connection moratorium until additional capacity is constructed or otherwise secured \cite{cpuc_moratorium_decisions}:
``\emph{A system's facilities shall have the capacity to meet the system's MDD, PHD plus any required fire flow in the system as a whole and in each individual pressure zone. If, at any time, the system does not have this capacity, the system shall request a service connection moratorium until such time as it can demonstrate the source capacity has been increased to meet system requirements}.'' As a result, a public water system may be considered legally ``at capacity'' once its allocated commitments fully exhaust approved capacity, even if actual demand remains, for example, near or lower than 70\% of the physical design capacity.

\paragraph{Implications for new service connections.}
Combined together, these mechanisms illustrate a general framework for public water system infrastructure management. First, infrastructure planning and investment are triggered as systems approach high utilization levels but below full design capacity. Second, legally binding restrictions on new service connections may apply once allocated capacity is fully committed, regardless of the actual water use. This structure aligns long-term infrastructure planning with public health  protection and service reliability while preventing over-extension of public water system resources.

Consequently, a public water system's ability to accommodate new service connection requests is governed by its available capacity, decided by (the more restrictive of) the remaining physical capacity and the uncommitted capacity (i.e., total physical capacity minus already allocated capacity). The system's full design capacity is therefore not the primary determinant in decisions regarding new connections.
Even a large public water system serving multiple existing large users may deny a new service connection request if its available capacity is limited, even when the requested capacity is substantially smaller than that of some existing users.

This practice is also reflected in recent data center projects. For example, a major technology company stated that its new data centers in Louisiana will draw water exclusively from the host community's ``verified surplus water'' capacity, aiming to ensure ``no strain on local water supplies''~\cite{Amazon_Water_400m_25_35per_PeakPowerReduction_AI_Cloud_Louisiana_PressRelease_2026}.

\end{document}